\documentclass[12pt,english,a4]{article}
\usepackage[T1]{fontenc}
\usepackage[latin1]{inputenc}
\usepackage{geometry}
\makeatletter

\usepackage{color}
\usepackage{graphicx}
\usepackage{setspace}

\@addtoreset{equation}{section}
\renewcommand{\theequation}{\thesection.\arabic{equation}}
\makeatletter
\renewcommand\@makefnmark{\@textsuperscript{\normalfont(\@thefnmark)}}
\makeatother

\usepackage{babel}
\makeatother
\begin{document}
\vspace*{0.8cm}

\thispagestyle{empty}

\begin{center}\textbf{\huge Radiative Corrections, New Physics Fits
and $e^{+}e^{-}\rightarrow?\rightarrow f\bar{f}$.}\end{center}{\huge \par}

\begin{center}{\huge \vspace*{1cm}}\end{center}{\huge \par}

\begin{center}{\large K.M.Hamilton}%
\footnote{email: k.hamilton1@physics.ox.ac.uk%
}\end{center}{\large \par}

\begin{center}\emph{\large $\textrm{Denys Wilkinson Building, Keble Road, Oxford OX1 3RH, U.K.}$}\end{center}{\large \par}

\begin{center}\vspace*{0.5cm}\end{center}

\begin{center}{\large November 2003}\end{center}{\large \par}

{\large \par{}}{\large \par}

\begin{center}\vspace*{1cm}\end{center}

\begin{center}\textbf{\large Abstract}\end{center}{\large \par}

{\large \par{}}{\large \par}

\noindent This document reviews the general approach to correcting
the process $e^{+}e^{-}\rightarrow X^{0}\rightarrow f\bar{f}$ for
radiative effects, where $X^{0}$ represents an exchanged boson arising
from some new physics. The validity of current methods is discussed
in the context of the differential cross section. To this end the
universality of the dominant QED radiative corrections to such processes
is discussed and an attempt is made to quantify it. The paper aims
to justify, as much as possible, the general approach taken by $e^{+}e^{-}$
collider experiments to the issue of how to treat the dominant radiative
corrections in fitting models of new physics using inclusive and exclusive
cross section measurements. We conclude that in all but the most pathological
of new physics models the dominant radiative corrections (QED) to
the tree level processes of the standard model can be expected to
hold well for new physics. This argument follows from the fact that
the phase space of indirect new physics searches is generally restrictive
(high $\sqrt{s^{\prime}}$ events) in such a way that the factorization
of radiative corrections is expected to hold well and generally universal
infrared corrections should be prevalent.

\newpage

\section{Introduction.}

In this document we discuss how radiative corrections which significantly
affect the standard model difermion process maybe exploited such that
they can be used with analogous difermion processes containing new
physics components, \emph{i.e.} we discuss the universality of radiative
corrections to the difermion process. To this end we must derive again
the significant QED corrections to the difermion process. The results
of our derivations agree with those in the literature. It has also
long been known that QED radiative corrections can exhibit soft behavior
which is independent of the underlying process, the so called \emph{hard
process} (see \emph{e.g.} \cite{Renton:td,Peskin:ev}), it is this behaviour
that we wish to find and study. The known radiative corrections generally
manifest themselves as long and complicated analytic expressions and
one generally has no clue as to whether they are significant or universal
or, ideally, both. In reviewing and {}``dismantling'' known results
we are able to thoroughly check the dependence of radiative corrections
on the hard scattering process thus highlighting, as much as possible,
universal parts of radiative corrections which are relevant to more
general difermion processes.

\section{Fitting Beyond the Standard Model Physics.}

Currently there are a large number of models involving physics beyond
the standard model which may be constrained from measurements of the
difermion process $e^{+}e^{-}\rightarrow f\bar{f}$. New physics models
appear frequently in the literature which propose modifications to
these observables. It is common for these models to involve the exchange
of a new gauge boson mimicking the exchange of a $Z^{0}$ or photon
in difermion processes. Let us generically denote these particles
$X^{0}$. The result of such exchanges is to increase the measured
cross sections above that which one would expect for difermion processes
as defined above, the differential cross section will also be modified.

The new physics contribution to the measured difermion cross section
is always given as a function of some parameters of the theory, typically
the energy scale of the new physics. The total new physics contribution
due to exchange of $X^{0}$ bosons comprises of a term due to interference
of $X^{0}$ exchange amplitudes with each of the $Z^{0}$ and $\gamma$
exchange amplitudes, as well as a pure $X^{0}$ exchange term. Given
that new physics effects are weak the largest contribution from them
comes from the interference term. By adding this new physics prediction
to the prediction for genuine difermion events and fitting the result
to the data for various values of the new physics parameters one can
obtain $\chi^{2}$ as a function of the parameters. From this $\chi^{2}$
function one can obtain confidence limits on the parameters.

Due to the fact that the standard model generally agrees well with
experimental measurements any new physics effect which may be occurring
in current $e^{+}e^{-}$ collider experiments must be small. Generally
one is trying to set the most stringent constraints on the new physics
model one can. It is for these reasons that when selecting events
with which to set limits one should not include events for which $\sqrt{s^{\prime}}\ll\sqrt{s}$%
\footnote{$\sqrt{s^{\prime}}$ is the so-called \emph{reduced invariant mass}.
$\sqrt{s^{\prime}}<\sqrt{s}$ due to bremsstrahlung emitted from the
initial state particles. Generally $\sqrt{s^{\prime}}$ is the invariant
mass of the exchanged boson, it is defined in appendix \ref{sec:Appendix-A}.%
} as previous runs of the experiment will have directly probed these
energies (and found nothing) so including such effects will essentially
\emph{dilute} the effects of a new physics contribution resulting
in a more relaxed limit. Obviously there is a trade off between the
number of events in the sample and the amount by which new physics
would be diluted. At LEP2 limits are set on the various models of
new physics using only samples of events for which $\sqrt{s^{\prime}}\geq0.85\sqrt{s}$.

Given that radiative corrections have striking consequences for difermion
events it would be naive to think that they are inconsequential for
new physics processes. New physics cross sections and differential
cross sections that appear in the literature are only ever given to
tree level, to obtain realistic limits for the parameters of the model
one should try to correct these predictions for radiative effects.
Effective theories are generally non-renormalizable, so the idea of
computing radiative corrections to processes derived from them may
not seem sensible. Nevertheless the idea of fitting the born level
predictions is certainly not correct. Finally, in the event that explicit
calculations of radiative corrections were possible, to do this would
be unfeasible given the current rate at which new physics scenarios
are conceived.

The conventional method employed by experiments to improve the tree
level new physics predictions of an observable $O_{NP}$, to include
radiative effects $\left.O^{\prime}\right|_{NP}$, is to multiply
the tree level prediction by a simple factor of the standard model
prediction including radiative corrections $\left.O^{\prime}\right|_{SM}$
divided by the standard model prediction at tree level $\left.O\right|_{SM}$,\begin{equation}
\left.O^{\prime}\right|_{NP}=\left.O\right|_{NP}\times\frac{\left.O^{\prime}\right|_{SM}}{\left.O\right|_{SM}}.\label{eq:5.1.1}\end{equation}
 This correction factor embodies the aggregated effects of all of
the \emph{standard model} radiative corrections. This approach amounts
to claiming that all standard model radiative corrections \emph{factorize}
trivially from the born level quantity, that radiative corrections
are independent of the born level process. In the rest of the paper
we investigate whether the approach of equation \ref{eq:5.1.1} is
at all valid.

\section{Radiator Functions: A {}``Black Box'' Empirical Study\label{sec:Radiator-Functions:-A}.}

The effect of bremsstrahlung on $e^{+}e^{-}$ collisions is highly
significant, the data show that the cross-sections for events with
$\sqrt{s^{\prime}}\geq75\textrm{ GeV}$ is larger than those events
with $\sqrt{\frac{s^{\prime}}{s}}\geq0.85$ by a factor of two or
more. If one imagines a single bremsstrahlung correction to a difermion
process, in which one of the colliding particles emits a photon, one
effect will be to lower the invariant mass of the basic difermion
reaction we are trying to measure from $\sqrt{s}$ to $\sqrt{s^{\prime}}$
(or $\sqrt{t}$ to $\sqrt{t^{\prime}}$ for $t$-channel processes).
Assuming that the processes of emitting a photon and undergoing a
difermion reaction are otherwise independent, that is to say they
\emph{factorize}, one can think of the cross section $\sigma\left(s\right)$measured
at some value of $\sqrt{s}$, in some range $s_{min}\leq s^{\prime}\leq s$
as being comprised of a weighted sum over all of the tree level cross
sections $\sigma_{Tree}\left(s^{\prime}\right)$ in that $\sqrt{s^{\prime}}$
range,\begin{equation}
\sigma\left(s\right)=\int_{s_{min}}^{s}\textrm{d}s^{\prime}\textrm{ }R\left(s,s^{\prime}\right)\sigma_{Tree}\left(s^{\prime}\right).\label{eq:5.2.1}\end{equation}
 $R\left(s,s^{\prime}\right)$ is the weighting function, it is known
in the literature as the \emph{radiator} \emph{function}. Given that
the cross section is proportional to the number of events in the sample
we could write\begin{equation}
N\left(s\right)=\int_{s_{min}}^{s}\textrm{d}s^{\prime}\textrm{ }R\left(s,s^{\prime}\right)N_{Tree}\left(s^{\prime}\right),\label{eq:5.2.2}\end{equation}
 which tells us that $\textrm{d}s^{\prime}\textrm{ }R\left(s,s^{\prime}\right)N_{Tree}\left(s^{\prime}\right)$
is the number of tree level difermion events in the range $s^{\prime}\rightarrow s^{\prime}+\textrm{d}s'$
entering the measured sample of $N\left(s\right)$. Therefore $\textrm{d}s^{\prime}\textrm{ }R\left(s,s^{\prime}\right)$
is the fraction of the tree level cross section $\sigma_{Tree}\left(s^{\prime}\right)$
that goes into making $\sigma\left(s\right)$. What is being described
here is essentially the \emph{structure} \emph{function} approach,
this is the idea that the electron and positron are not to be thought
of as fundamental but rather they should be thought of as objects
\emph{containing} truly fundamental electrons, positrons and photons.
With this in mind one can define a structure function for the electron
$D_{e}\left(x\right)$. $D_{e}\left(x\right)$ is a probability density
function of $x$ which is the \emph{fraction} of the longitudinal
component of the electrons momentum that it has when it undergoes
the tree level hard scattering process. $D_{e}\left(x\right)\textrm{d}x$
is therefore the probability that the electron collides with the positron
with a fraction $x$ of its original longitudinal four momentum. The
positron structure function is the same as the electron structure
function. Neglecting transverse momenta for the time being we have\begin{equation}
s^{\prime}=x_{1}x_{2}s.\label{eq:5.2.3}\end{equation}
 With these definitions the cross section for an electron and positron
to undergo a difermion interaction such that $s^{\prime}\geq s_{min}^{\prime}$
is\begin{equation}
\sigma\left(s\right)=\int_{\frac{s_{min}}{s}}^{1}\textrm{d}x_{1}\int_{\frac{s_{min}}{x_{1}s}}^{1}\textrm{d}x_{2}\textrm{ }D_{e}\left(x_{1}\right)D_{e}\left(x_{2}\right)\sigma_{Tree}\left(x_{1}x_{2}s\right).\label{eq:5.2.4}\end{equation}
 Defining $1-x=x_{1}x_{2}$ and demanding it be greater than $\frac{s_{min}}{s}$
this can be written\begin{equation}
\begin{array}{rcl}
\sigma\left(s\right) & = & \int_{0}^{1-\frac{s_{min}}{s}}\textrm{d}x\left[\int_{1-x}^{1}\textrm{d}x_{2}\textrm{ }\frac{1}{x_{2}}\textrm{ }D_{e}\left(\frac{1-x}{x_{2}}\right)D_{e}\left(x_{2}\right)\right]\sigma_{Tree}\left(\left(1-x\right)s\right)\\
 & = & \int_{0}^{1-\frac{s_{min}}{s}}\textrm{d}x\textrm{ }H\left(x,s\right)\sigma_{Tree}\left(\left(1-x\right)s\right)\end{array},\label{eq:5.2.5}\end{equation}
 allowing an identification between the radiator function and the
structure functions. Clearly $H\left(x,s\right)$ and $R\left(s,s^{\prime}\right)$
are related by a factor.

The most general new physics distribution to fit is the differential
cross section. In this case the quantity $\left.O\right|_{NP}$ to
be corrected via equation \ref{eq:5.1.1} is the value of the differential
cross section in a given angular bin. Conventionally one would then
use a number of correction factors determined using the aforementioned
standard model cross sections in that bin to correct the bins of the
new physics tree level angular distribution. It is however hard to
attach a physical meaning to these correction factors, they do not
lend themselves to a description in terms of a radiator function or
structure functions. Ideally one would like to employ a radiator function
approach to correcting new physics observables and so correct them
by folding the given tree level quantities with the radiator function.
This raises two questions, does a radiator function approach exist
for the differential cross section and if it does exist how should
it be implemented?

Radiator functions for the differential cross section do exist and
are present inside $e^{+}e^{-}\rightarrow f\bar{f}$ simulation programs.
There are a variety of such programs which are freely available. There
are two kinds of $e^{+}e^{-}\rightarrow f\bar{f}$ simulation packages.
Semi-analytic simulations such as \texttt{TOPAZ0} \cite{Montagna:1998kp}
and \texttt{ZFITTER} \cite{Bardin:1999yd} produce predictions of
observables like cross-sections and differential cross sections. They
contain analytic expressions for these quantities corrected for a
myriad of radiative corrections all in the context of the standard
model. Some outputs of these programs involve numerical integrations
hence {}``Semi-analytical''. In addition to the semi-analytic packages
there are difermion Monte Carlo programs such as ${\cal {KK}}\textrm{2f}$
\cite{Ward:2002qq} which simulate actual difermion events \emph{i.e.}
there output includes four vectors for the various final state particles.
At the heart of the Monte Carlo is a random number generator, as the
number of generated events tends to infinity the cross sections \emph{etc}
that are calculable with the generated events tend to those of the
standard model \emph{i.e.} random fluctuations die away. Naturally
the semi-analytic packages have the advantage of requiring much less
computational time than the Monte Carlo programs and the \texttt{ZFITTER}
package appears to be the most popular of these in the experimental
community.

In \cite{Bardin:ak} (and the \texttt{ZFITTER} manual \cite{Bardin:1999yd})
the differential cross section for difermion production in the presence
of ISR is given in the form%
\footnote{The form of equation \ref{eq:5.2.6} in \cite{Bardin:1999yd} differs
slightly from what we quote which is a simplified version. The only
difference of note is that \cite{Bardin:1999yd} split the born level
cross section into a part which is symmetric in $\cos\theta$ and
an asymmetric part, each of which is convoluted with a slightly different
radiator function. Equation \ref{eq:5.2.6} is appropriate for the
time being. %
}\begin{equation}
\frac{{\rm {d}}\sigma\left(s,\cos\theta\right)}{{\rm {d}}\cos\theta}=\int_{s_{min}^{\prime}}^{s}{\rm {d}}s^{\prime}\textrm{ }R\left(s,s^{\prime},\cos\theta\right)\sigma_{Tree}\left(s^{\prime}\right).\label{eq:5.2.6}\end{equation}
 Note crucially that \emph{all} angular dependence is contained within
the radiator function. Let us denote the cross section for a new physics
process incorporating radiative corrections by $\sigma_{NP}$ and
the corresponding tree level cross section $\sigma_{Tree,NP}$. Assuming
that it is possible to extract $R\left(s,s^{\prime},\cos\theta\right)$
from the \texttt{ZFITTER} package\begin{equation}
\frac{{\rm {d}}\sigma_{NP}\left(s,\cos\theta\right)}{{\rm {d}}\cos\theta}=\int_{s_{min}^{\prime}}^{s}{\rm {d}}s^{\prime}\textrm{ }R\left(s,s^{\prime},\cos\theta\right)\sigma_{Tree,NP}\left(s^{\prime}\right),\label{eq:5.2.7}\end{equation}
 will not be the correct differential cross section for a new physics
process, obviously there is absolutely no reference to the angular
distribution of new physics in the above formula, only to that of
the standard model. Studying references \cite{Bardin:ak} and \cite{Bardin:1999yd}
in a more detail one finds that in fact the radiator function which
exactly factorizes at the level of the integrated cross section (by
integrated we mean integrated over the full solid angle), as shown
in \ref{eq:5.2.6}, also \emph{largely} factorizes from the differential
cross section \emph{i.e.} \ref{eq:5.2.6} is of the form\begin{equation}
\begin{array}{rcl}
\frac{{\textrm{{d}}}\sigma\left(s,\cos\theta\right)}{{\textrm{{d}}}\cos\theta} & = & \int_{s_{min}^{\prime}}^{s}{\textrm{{d}}}s^{\prime}\textrm{ }\tilde{R}_{1}\left(s,s^{\prime}\right)\frac{{\rm {d}}\sigma_{Tree}\left(s,\cos\theta\right)}{{\rm {d}}\cos\theta}+\int_{s_{min}^{\prime}}^{s}{\textrm{{d}}}s^{\prime}\textrm{ }\tilde{R}_{2}\left(s,s^{\prime},\cos\theta\right)\sigma_{Tree}\left(s^{\prime}\right)\\
 & = & \int_{s_{min}^{\prime}}^{s}{\textrm{{d}}}s^{\prime}\textrm{ }\tilde{R}_{1}\left(s,s^{\prime}\right)\frac{{\rm {d}}\sigma_{Tree}\left(s,\cos\theta\right)}{{\rm {d}}\cos\theta}+\textrm{Non-factorizable.}\end{array}.\label{eq:5.2.8}\end{equation}
 In \ref{eq:5.2.8} $\textrm{ }R\left(s,s^{\prime},\cos\theta\right)$
has been split into a factorizable part%
\footnote{Henceforth we use the word factorizable to mean that the radiative
corrections can be represented by a radiator function convoluted with
the \emph{differential cross section} $\textrm{d}\sigma$.%
} $\tilde{R}_{1}\left(s,s^{\prime}\right)\frac{{\rm {d}}\sigma_{0}\left(s,\cos\theta\right)}{{\rm {d}}\cos\theta}$
and a non-factorizable part $\tilde{R}_{2}\left(s,s^{\prime},\cos\theta\right)$.
If the parts of the above which do not factorize into the differential
cross section are negligible then it appears that the radiator function
does not depend on the tree level differential cross section, in fact
it does not depend on angles at all. If this is true then $\tilde{R}_{1}\left(s,s^{\prime}\right)$
would appear to be independent of the tree level physics analogous
to a structure function (the electron structure function is discussed
in the next section), then one would appear to have a recipe for making
new physics angular distribution corrected for radiative effects\begin{equation}
\frac{{\rm {d}}\sigma_{NP}\left(s,\cos\theta\right)}{{\rm {d}}\cos\theta}=\int_{s_{min}^{\prime}}^{s}\textrm{d}s^{\prime}\textrm{ }\tilde{R}_{1}\left(s,s^{\prime}\right)\frac{{\rm {d}}\sigma_{T,NP}\left(s,\cos\theta\right)}{{\rm {d}}\cos\theta}.\label{eq:5.2.9}\end{equation}
 Assuming it is true that the non-factorizable parts of $R\left(s,s^{\prime},\cos\theta\right)$
are negligible we then wish to obtain $\tilde{R}_{1}\left(s,s^{\prime}\right)$.

\begin{figure}
\begin{center}\includegraphics[%
  width=0.30\paperwidth,
  height=0.35\paperwidth]{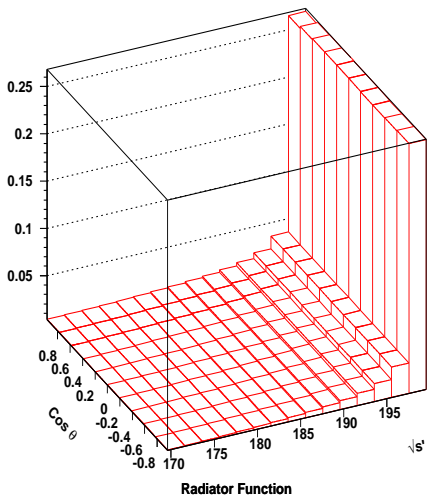}\end{center}

\caption{\label{cap:nointf}The radiator function $\tilde{R}_{1}\left(s,s^{\prime}\right)$
with $s=\left(200\textrm{ GeV}\right)^{2}$, obtained from the \texttt{ZFITTER}
package by extracting it separately for each angular bin for the $\mu^{+}\mu^{-}$
final state.}
\end{figure}

The \texttt{ZFITTER} package contains just this information though
as much as the function is convoluted with the standard model in the
above equations it is convoluted to an greater extent with the standard
model in terms of the actual coding of the package. A different, more
practical means of deconvolving the radiator function is required.
Denoting $\sigma\left(s,c_{b}\right)$ as the cross section for the
difermion process occurring such that the final state fermion goes
into the $\cos\theta$ bin in the range $\cos\theta_{b}\leq\cos\theta\leq\cos\theta_{b+1}$
denoted by $c_{b}$ we have\[
\begin{array}{rcl}
\sigma\left(s,c_{b}\right) & = & \int_{s_{min}^{\prime}}^{s}\textrm{d}s^{\prime}\textrm{ }\left\{ \tilde{R}_{1}\left(s,s^{\prime}\right)\sigma_{T}\left(s,c_{b}\right)+\int_{\cos\theta_{b}}^{\cos\theta_{b+1}}{\rm {d}}\cos\theta\textrm{ }\int_{s_{min}^{\prime}}^{s}\textrm{d}s^{\prime}\textrm{ }\tilde{R}_{2}\left(s,s^{\prime},\cos\theta\right)\sigma_{T}\left(s^{\prime}\right)\right\} \\
 & = & \int_{s_{min}^{\prime}}^{s}\textrm{d}s^{\prime}\textrm{ }\left\{ \tilde{R}_{1}\left(s,s^{\prime}\right)\sigma_{T}\left(s,c_{b}\right)+\int_{\cos\theta_{b}}^{\cos\theta_{b+1}}{\rm {d}}\cos\theta\textrm{ Non-factorizable}\right\} \end{array},\]
 where here {}``Non-factorizable'' is the same as in \ref{eq:5.2.8}.
The method by which we extract the radiator function assumes that
the non-factorizable part is zero. If this assumption is true one
should get the same function $\tilde{R}\left(s,s^{\prime}\right)$
regardless of what binning one uses in the equation above. The fact
that the non-factorizable part has a complex angular form (this can
be checked by reference to \cite{Bardin:1999yd}) will mean that the
$\tilde{R}_{1}\left(s,s^{\prime}\right)$ which one extracts from
the program could look very different depending on which angular bin
is used and, importantly, the size of the non-factorizable contributions.
If the non-factorizable component really is small then the $\tilde{R}_{1}\left(s,s^{\prime}\right)$
which is extracted from the program should not be a function of the
bin used to extract it, it should be flat when plotted against the
bin used to extract it. The author has devised a means of extracting
a numerical version of $\tilde{R}_{1}\left(s,s^{\prime}\right)$ from
the \texttt{ZFITTER} package by making finite approximations to the
integral over $s^{'}$. The resulting numerical radiator function
is shown in figure \ref{cap:nointf}, it is, as hoped, flat across
the angular region shown.%

\section{Structure Functions, Leading Log Approximation\label{sec:Structure-Functions,-Leading}.}

The purpose of this section is to add weight to the empirical observations
of the last section by deriving the electron structure function. The
idea for a structure function for the electron originates from the
paper of Gribov and Lipatov \cite{Gribov:rt}. In the papers of Kuraev,
Fadin, Jadach and Skrzypek \cite{Skrzypek:1990qs,Kuraev:hb} the authors
consider the process of $e^{+}e^{-}$ annihilation as being a Drell-Yan
process and exploit the perturbative nature of QED to calculate the
electron structure function analytically using the Gribov-Lipatov
Altarelli-Parisi equations. This structure function approach represents
a \emph{leading logarithmic} \emph{approximation} (\emph{LLA}), this
corresponds to assuming near collinear emission of photons from the
incoming $e^{+}e^{-}$. The radiator function in the literature is
by no means unique. The structure function approach / the leading
log approximation, is one way to take into account the effects of
ISR. The structure function approach is intuitive, simple and it produces
a cross section to better than $0.015$\% \cite{Skrzypek:1992vk}%
\footnote{This number is quoted in relation to the accuracy of the LLA around
the $Z^{0}$ peak at LEP1 however in LEP2 studies \cite{Boudjema:1996qg}
the LEP1 and higher order LEP2 structure functions have been shown
to produce cross sections agreeing to better than 0.1\% for cuts like
those used in obtaining our non-radiative class of events. %
}, this alone is a good reason to discuss it. For clarity and completeness
we review the results of Kuraev and Fadin (KF) as well as Jadach and
Skrzypek (JS) with the help of a combination of papers \cite{Skrzypek:1990qs,Skrzypek:1992vk,Kuraev:hb,Gribov:ri,Nicrosini:1986sm}.
In doing this we explicitly confirm (as noted in the introduction
of \cite{Skrzypek:1992vk}) that no element of the underlying hard
scattering process enters this description of ISR%
\footnote{It had been a cause of concern that, contrary to QCD, the perturbative,
calculable nature of this structure function may somehow allow for
some specifically process dependent corrections to enter it, as is
suggested by the explicit angular dependence in the radiator functions
of \cite{Bardin:1999yd}.%
} and check whether or not it may be applied to the differential cross
section as well as the integrated cross section.

The \emph{non-singlet} structure function $D^{NS}\left(x,s\right)$
corresponds to the probability of finding an electron with longitudinal
momentum fraction $x$ `inside' the original on-shell electron at
energy scale $\sqrt{s}$. The generic starting point for determining
the structure function is Lipatov's equation (see \emph{e.g.} \cite{Peskin:ev})\begin{equation}
D^{NS}\left(x,s\right)=\delta\left(1-x\right)+\int_{m_{e}^{2}}^{s}\frac{{\rm {d}}s_{1}}{s_{1}}\frac{\alpha\left(s_{1}\right)}{2\pi}\int_{x}^{1}\frac{{\rm {d}}x_{1}}{x_{1}}\textrm{ }P\left(x_{1}\right)D^{NS}\left(\frac{x}{x_{1}},s_{1}\right).\label{eq:5.3.1}\end{equation}
 where $P\left(z\right)$ is the regularized splitting function\begin{equation}
P\left(x_{1}\right)=\frac{1+x_{1}^{2}}{1-x_{1}}-\delta\left(1-x_{1}\right)\int_{0}^{1}\textrm{d}z\textrm{ }\frac{1+z^{2}}{1-z}.\label{eq:5.3.2}\end{equation}
 For our purposes a more useful form of the regularized splitting
function involves introducing a cut off at $z=1-\epsilon$ where $\epsilon\ll1$
is very small\begin{equation}
-\int_{0}^{1}\textrm{d}z\textrm{ }\frac{1+z^{2}}{1-z}=\lim_{\epsilon\rightarrow0}2\left(\frac{3}{4}+\log\epsilon\right)-2\epsilon+\frac{1}{2}\epsilon^{2}.\label{eq:5.3.3}\end{equation}
 In this way we can define a different regularized splitting function\begin{equation}
\textrm{ }P\left(x_{1}\right)=\textrm{ }2\delta\left(1-x_{1}\right)\left(\frac{3}{4}+\log\epsilon\right)+\theta\left(1-\epsilon-x_{1}\right)\frac{1+x_{1}^{2}}{1-x_{1}}.\label{eq:5.3.4}\end{equation}
 which is finite when integrated over in the limit $\epsilon\rightarrow0$.
Clearly this form is the natural extension of \ref{eq:5.3.2}, in
the limit $\epsilon\rightarrow0$ \ref{eq:5.3.2} and \ref{eq:5.3.3}
are identical. Henceforth we omit writing $\lim_{\epsilon\rightarrow0}$
and drop all terms from our calculations which are vanishing in the
limit $\epsilon\rightarrow0$.

Lipatov's equation \ref{eq:5.3.1} has an obvious iterative solution,
to see this it is convenient to first rewrite the second integral
using\begin{equation}
\int_{0}^{1}\textrm{d}x_{1}\textrm{d}x_{2}\textrm{ }\delta\left(x-x_{1}x_{2}\right)P\left(x_{1}\right)D^{NS}\left(x_{2},s_{1}\right)=\int_{x}^{1}\frac{\textrm{d}x_{1}}{x_{1}}\textrm{ }P\left(x_{1}\right)D^{NS}\left(\frac{x}{x_{1}},s_{1}\right).\label{eq:5.3.6}\end{equation}
 In addition we approximate the running coupling constant to a constant
$\alpha\left(s\right)=\alpha$ and define $\eta\equiv\eta\left(s\right)=\frac{2\alpha}{\pi}\log\frac{s}{m_{e}^{2}}$,
$\eta_{i}\equiv\eta\left(s_{i}\right)$ therefore\begin{equation}
\textrm{d}\eta_{i}=\frac{\textrm{d}s_{i}}{s_{i}}\frac{2\alpha}{\pi}.\label{eq:5.3.6}\end{equation}
 With these substitutions Lipatov's equation \ref{eq:5.3.1} becomes,\begin{equation}
D^{NS}\left(x,\eta\right)=\delta\left(1-x\right)+\frac{1}{4}\int_{0}^{\eta}\textrm{d}\eta_{1}\int_{0}^{1}\textrm{d}x_{1}\textrm{d}x_{2}\textrm{ }\delta\left(x-x_{1}x_{2}\right)P\left(x_{1}\right)D^{NS}\left(x_{2},\eta_{1}\right)\label{eq:5.3.7}\end{equation}
 Substituting relation \ref{eq:5.3.6} into Lipatov's equation \ref{eq:5.3.1}
and iterating it \emph{i.e.} substituting $D\left(x_{i},\eta_{i}\right)$as
defined by \ref{eq:5.3.1} back into \ref{eq:5.3.1} we have an expansion
in $\eta$;\begin{equation}
\begin{array}{rcl}
D^{NS}\left(x,\eta\right) & = & \delta\left(1-x\right)\\
 & + & \frac{1}{4}\int_{0}^{\eta}{\rm {d}}\eta_{1}\int_{0}^{1}{\rm {d}}x_{1}{\rm {d}}x_{2}\textrm{ }\delta\left(x-x_{1}x_{2}\right)P\left(x_{1}\right)\delta\left(1-x_{2}\right)\\
 & + & \frac{1}{4}\int_{0}^{\eta}{\rm {d}}\eta_{1}\int_{0}^{1}{\rm {d}}x_{1}{\rm {d}}x_{2}\textrm{ }\delta\left(x-x_{1}x_{2}\right)P\left(x_{1}\right)\frac{1}{4}\int_{0}^{\eta_{1}}{\rm {d}}\eta_{2}\int_{0}^{1}{\rm {d}}x_{3}{\rm {d}}x_{4}\textrm{ }\delta\left(x_{2}-x_{3}x_{4}\right)P\left(x_{3}\right).....\\
 & + & {\mathcal{{O}}}\left(\eta^{3}\right)\end{array}.\label{eq:5.3.8}\end{equation}
 The fact that the integrals $\int_{0}^{\eta_{i}}\textrm{d}\eta_{i}$
are \emph{nested}, that is to say the upper limit of one integrand
is the dummy variable of the one preceding it, results in a sequence
of exponential factors $\frac{1}{i!}$ multiplying the terms above.
\ref{eq:5.3.7} is condensed by simple delta function manipulations\begin{equation}
\begin{array}{rl}
 & D^{NS}\left(x,\eta\right)\\
= & \delta\left(1-x\right)\\
+ & 1\times\left(\frac{\eta}{4}\right)^{1}P\left(x\right)\\
+ & \frac{1}{2}\times\left(\frac{\eta}{4}\right)^{2}\int_{0}^{1}{\rm {d}}x_{1}{\rm {d}}x_{2}\textrm{ }\delta\left(x-x_{1}x_{2}\right)P\left(x_{1}\right)P\left(x_{2}\right)\\
+ & \frac{1}{6}\times\left(\frac{\eta}{4}\right)^{3}\int_{0}^{1}{\rm {d}}x_{1}{\rm {d}}x_{2}\textrm{ }\delta\left(x-x_{1}x_{2}\right)P\left(x_{1}\right)\int_{0}^{1}{\rm {d}}x_{3}{\rm {d}}x_{4}\textrm{ }\delta\left(x_{2}-x_{3}x_{4}\right)P\left(x_{3}\right)P\left(x_{4}\right)\\
+ & {\mathcal{{O}}}\left(\eta^{4}\right)\end{array}.\label{eq:5.3.9}\end{equation}
 Denoting the convolution integrals above by\begin{equation}
\int_{0}^{1}\textrm{d}x_{i}\textrm{d}x_{j}\textrm{ }\delta\left(x-x_{i}x_{j}\right)f\left(x_{i}\right)g\left(x_{j}\right)=f\left(.\right)\otimes g\left(.\right)\left(x\right),\label{eq:5.3.10}\end{equation}
 equation \ref{eq:5.3.8} becomes\begin{equation}
D^{NS}\left(x,s\right)=\delta\left(1-x\right)+\sum_{i=1}^{\infty}\frac{1}{i!}\left(\frac{\eta}{4}\right)^{i}P^{\otimes i}\left(x\right),\label{eq:5.3.11}\end{equation}
 where\begin{equation}
P^{\otimes3}\left(x\right)=\int_{0}^{1}\textrm{d}x_{1}\textrm{d}x_{2}\textrm{ }\delta\left(x-x_{1}x_{2}\right)P\left(x_{1}\right)\int_{0}^{1}\textrm{d}x_{3}\textrm{d}x_{4}\textrm{ }\delta\left(x_{2}-x_{3}x_{4}\right)P\left(x_{3}\right)P\left(x_{4}\right)\label{eq:5.3.12}\end{equation}
 \emph{etc}.

The ${\mathcal{{O}}}\left(\eta\right)$ term is simply the regularized
splitting function, $\frac{1}{4}\eta P\left(x\right)$. In calculating
the ${\mathcal{{O}}}\left(\eta^{2}\right)$ term first we calculate
the \emph{theta} term, we proceed essentially in an identical way
to that in which we calculated the regularized splitting function.
We calculate a regularized version of this term by taking $x$ to
be less than a cut off $x<1-\epsilon$.\begin{equation}
\begin{array}{rl}
 & \theta\left(1-x-\epsilon\right)\int_{0}^{1}{\rm {d}}x_{1}{\rm {d}}x_{2}\textrm{ }\delta\left(x-x_{1}x_{2}\right)P\left(x_{1}\right)P\left(x_{2}\right)\\
= & \theta\left(1-x-\epsilon\right)4\left(\frac{3}{4}+\log\epsilon\right)^{2}\int_{0}^{1}{\rm {d}}x_{1}{\rm {d}}x_{2}\textrm{ }\delta\left(x-x_{1}x_{2}\right)\delta\left(1-x_{1}\right)\delta\left(1-x_{2}\right)\\
+ & \theta\left(1-x-\epsilon\right)2\left(\frac{3}{4}+\log\epsilon\right)\int_{0}^{1-\epsilon}{\rm {d}}x_{1}\int_{0}^{1}{\rm {d}}x_{2}\textrm{ }\delta\left(x-x_{1}x_{2}\right)\delta\left(1-x_{2}\right)\frac{1+x_{1}^{2}}{1-x_{1}}\\
+ & \theta\left(1-x-\epsilon\right)2\left(\frac{3}{4}+\log\epsilon\right)\int_{0}^{1}{\rm {d}}x_{1}\int_{0}^{1-\epsilon}{\rm {d}}x_{2}\textrm{ }\delta\left(x-x_{1}x_{2}\right)\textrm{ }\delta\left(1-x_{1}\right)\frac{1+x_{2}^{2}}{1-x_{2}}\\
+ & \theta\left(1-x-\epsilon\right)\int_{0}^{1}{\rm {d}}x_{1}\int_{0}^{1}{\rm {d}}x_{2}\delta\left(x-x_{1}x_{2}\right)\textrm{ }\theta\left(1-\epsilon-x_{1}\right)\theta\left(1-\epsilon-x_{2}\right)\frac{1+x_{1}^{2}}{1-x_{1}}\frac{1+x_{2}^{2}}{1-x_{2}}\end{array}\label{eq:5.3.14}\end{equation}
 The first term is zero due to the cut off we impose on $x$, the
other terms are also straightforward to evaluate. As stated earlier
small terms ${\mathcal{{O}}}\left(\epsilon\right)$ and above, vanishing
in the limit $\epsilon\rightarrow0$, are dropped. The total theta
term obtained with the cut off $x<1-\epsilon$ is \begin{equation}
\begin{array}{rl}
 & \theta\left(1-\epsilon-x\right)\int_{0}^{1}{\rm {d}}x_{1}{\rm {d}}x_{2}\textrm{ }\delta\left(x-x_{1}x_{2}\right)P\left(x_{1}\right)P\left(x_{2}\right)\\
= & \theta\left(1-x-\epsilon\right)2\left(\frac{1+x^{2}}{1-x}\left(2\log\left(1-x\right)-\log x+\frac{3}{2}\right)+\frac{1}{2}\left(1+x\right)\log x-1+x\right)\end{array}\label{eq:5.3.17}\end{equation}
 The \emph{delta} term, analogous to the delta function term in the
regularized splitting function \ref{eq:5.3.4}, {}``soaks up'' the
divergent part of the term above arising from the limit $x\rightarrow1$
$\left(\epsilon\rightarrow0\right)$\begin{equation}
\begin{array}{rl}
 & -\lim_{\epsilon\rightarrow0}\int_{0}^{1-\epsilon}{\rm {d}}x\textrm{ }2\left(\frac{1+x^{2}}{1-x}\left(2\log\left(1-x\right)-\log x+\frac{3}{2}\right)+\frac{1}{2}\left(1+x\right)\log x-1+x\right)\\
= & 4\left(\left(\frac{3}{4}+\log\epsilon\right)^{2}-\zeta\left(2\right)\right)\end{array}.\label{eq:5.3.18}\end{equation}
 Finally the full regularized ${\mathcal{{O}}}\left(\eta^{2}\right)$
term is \begin{equation}
\begin{array}{rl}
 & \frac{1}{2}\left(\frac{\eta}{4}\right)^{2}\int_{0}^{1}{\rm {d}}x_{1}{\rm {d}}x_{2}\textrm{ }\delta\left(x-x_{1}x_{2}\right)P\left(x_{1}\right)P\left(x_{2}\right)\\
= & \left(\frac{\eta}{4}\right)^{2}\delta\left(1-x\right)\left(2\left(\frac{3}{4}+\log\epsilon\right)^{2}-2\zeta\left(2\right)\right)\\
+ & \left(\frac{\eta}{4}\right)^{2}\theta\left(1-x-\epsilon\right)\left(\frac{1+x^{2}}{1-x}\left(2\log\left(1-x\right)-\log x+\frac{3}{2}\right)+\frac{1}{2}\left(1+x\right)\log x-1+x\right)\end{array}.\label{eq:5.3.19}\end{equation}
 The third order term, ${\mathcal{{O}}}\left(\eta^{3}\right)$, is
simply\begin{equation}
\frac{1}{6}\left(\frac{\beta}{4}\right)^{3}\int_{0}^{1}{\rm {d}}x_{1}{\rm {d}}x_{2}\textrm{ }\delta\left(x-x_{1}x_{2}\right)P\left(x_{1}\right)\int_{0}^{1}{\rm {d}}x_{3}{\rm {d}}x_{4}\textrm{ }\delta\left(x_{2}-x_{3}x_{4}\right)P\left(x_{3}\right)P\left(x_{4}\right).\label{eq:5.3.20}\end{equation}
 Clearly the $\int_{0}^{1}\textrm{d}x_{3}\textrm{d}x_{4}$ part of
this term is given by simply substituting in the ${\mathcal{{O}}}\left(\eta^{2}\right)$
term with the replacement, $x\rightarrow x_{2}$. Noting this simplification
the resulting calculation is significantly simplified and a long but
straightforward sequence of integrations gives the ${\mathcal{{O}}}\left(\eta^{3}\right)$
theta and delta terms%
\footnote{The following dilogarithm identity is required, $\textrm{Li}_{2}\left(x\right)=\frac{\pi^{2}}{6}-\log\left(x\right)\log\left(1-x\right)-\textrm{Li}_{2}\left(1-x\right)$
to obtain the term in the form of \cite{Skrzypek:1990qs}.%
} of \cite{Skrzypek:1990qs}.

The Lipatov equation can also be solved \emph{analytically} in the
soft limit $x\rightarrow0$ by means of a Mellin transformation. The
Mellin transform is defined as,\begin{equation}
\tilde{D}^{NS\left(z\right)}\left(\eta\right)=\int_{0}^{1}{\rm {d}}x\textrm{ }x^{z-1}D^{NS}\left(x,\eta\right)\label{eq:5.3.25}\end{equation}
 for which the inverse is,\begin{equation}
D^{NS}\left(x,\eta\right)=\frac{1}{2\pi i}\int_{c-i\infty}^{c+i\infty}{\rm {d}}z\textrm{ }x^{-z}\tilde{D}^{\left(z\right)}\left(\eta\right).\label{eq:5.3.26}\end{equation}
 The transform exists if $\int_{0}^{1}\textrm{d}x\textrm{ }x^{z-1}\left|D^{NS}\left(x,\eta\right)\right|$
is bounded for some $z>0$ in which case the inverse exists for $c>z$.

Taking the Mellin transform of Lipatov's equation (\ref{eq:5.3.7})
gives\begin{equation}
\tilde{D}^{NS\left(z\right)}\left(\eta\right)=1+\frac{1}{4}\int_{0}^{\eta}\textrm{d}\eta_{1}\tilde{P}^{\left(z\right)}\tilde{D}^{NS\left(z\right)}\left(\eta_{1}\right)\label{eq:5.2.27}\end{equation}
 Differentiating the above with respect to $\eta$ gives, \begin{equation}
\frac{\textrm{d}}{\textrm{d}\eta}\tilde{D}^{NS\left(z\right)}\left(\eta\right)=\frac{1}{4}\tilde{P}^{\left(z\right)}\tilde{D}^{NS\left(z\right)}\left(\eta\right).\label{eq:5.3.28}\end{equation}
 Integrating this equation one finds a simple expression for the Mellin
moments of the structure function in terms of those of the splitting
function\begin{equation}
\tilde{D}^{NS\left(z\right)}\left(\eta\right)=\exp\left(\frac{1}{4}\eta\tilde{P}^{\left(z\right)}\right)\label{eq:5.3.29}\end{equation}
 where we have used as an initial condition $\tilde{D}^{NS\left(z\right)}\left(m_{e}^{2}\right)=1$.
The Mellin moments for the splitting function $\tilde{P}^{\left(z\right)}$
are\begin{equation}
\tilde{P}^{\left(z\right)}=2\left(\frac{3}{4}+\log\epsilon\right)-\int_{0}^{1-\epsilon}\textrm{d}x\textrm{ }\frac{1-x^{z-1}}{1-x}+\frac{1-x^{z+1}}{1-x}-\frac{2}{1-x}\label{eq:5.3.30}\end{equation}
 Using the Taylor expansion $\frac{x^{z}}{1-x}=\sum_{n=z}^{\infty}x^{n}$
the integral in \ref{eq:5.3.30} is trivial, as before we drop terms
vanishing in the limit $\epsilon\rightarrow0$. With a little manipulation
of the summations $\tilde{P}^{\left(z\right)}$ becomes,\begin{equation}
\begin{array}{rcl}
\tilde{P}^{\left(z\right)} & = & 2\left(\frac{3}{4}+\log\epsilon\right)+\int_{0}^{1-\epsilon}\textrm{d}x\textrm{ }x^{z-1}+x^{z}+\frac{2}{1-x}-2\sum_{n=0}^{z}x^{n}\\
 & = & -\sum_{n=3}^{z+1}\textrm{ }\frac{1}{n}-\sum_{n=1}^{z-1}\textrm{ }\frac{1}{n}\end{array}.\label{eq:5.3.31}\end{equation}
 It is worth pointing out that in dropping the terms vanishing in
the limit $\epsilon\rightarrow0$ the cut off $\epsilon$ has now
disappeared altogether from the calculation. The Euler function is
defined as\begin{equation}
\psi\left(x\right)=\frac{\textrm{d}}{\textrm{d}x}\log\Gamma\left(x\right)\label{eq:5.3.32}\end{equation}
 hence,\begin{equation}
\psi\left(x+1\right)-\psi\left(x\right)=\frac{1}{x}\label{eq:5.3.33}\end{equation}
 ($\Gamma\left(x\right)$ is the Gamma function). In terms of the
Euler function,\begin{equation}
\tilde{P}^{\left(z\right)}=-\psi\left(z+2\right)-\psi\left(z\right)+\psi\left(3\right)+\psi\left(1\right).\label{eq:5.3.34}\end{equation}
 Using the identity \ref{eq:5.3.33} again and expanding $\Gamma\left(x\right)$
around $x=1$ we find\begin{equation}
\tilde{P}^{\left(z\right)}=-\psi\left(z+2\right)-\psi\left(z\right)+\frac{3}{2}-2\gamma.\label{eq:5.3.35}\end{equation}
 Finally, substituting \ref{eq:5.3.35} into \ref{eq:5.3.29}, we
have an explicit expression for the structure function in terms of
its Mellin transform (\ref{eq:5.3.26})\begin{equation}
D^{NS}\left(x,\eta\right)=\frac{1}{2\pi i}\exp\left(\frac{1}{2}\eta\left(\frac{3}{4}-\gamma\right)\right)\int_{c-i\infty}^{c+i\infty}\textrm{d}z\textrm{ }x^{-z}\exp\left(\frac{1}{4}\eta\left(-\psi\left(z+2\right)-\psi\left(z\right)\right)\right).\label{eq:5.3.37}\end{equation}
 An analytic expression for the integral above is not known. The integral
may be performed in the \emph{soft} \emph{limit} $x\rightarrow1$,
this is known as the \emph{Gribov} \emph{approximation}. The factor
$x^{-z}$ shows that the integral is dominated by large values of
$z$. The Euler Gamma function is defined,\begin{equation}
\Gamma\left(z\right)=\left(z-1\right)!\label{eq:5.3.38}\end{equation}
 in the large $z$ limit one can approximate the factorial operation
by Stirling's formula $n!\approx\sqrt{2\pi}z^{z+\frac{1}{2}}\exp\left(-z\right)$.
In the large $z$ limit we have $\psi\left(z\right)\approx\log z$\begin{equation}
\Rightarrow D^{NS}\left(x,\eta\right)=\frac{1}{2\pi i}\exp\left(\frac{1}{2}\eta\left(\frac{3}{4}-\gamma\right)\right)\int_{c-i\infty}^{c+i\infty}\textrm{d}z\textrm{ }x^{-z}z^{-\frac{1}{2}\eta}.\label{eq:5.3.41}\end{equation}
 By substituting $y=z\log x$,\begin{equation}
D^{NS}\left(x,\eta\right)=\frac{1}{2\pi i}\exp\left(\frac{1}{2}\eta\left(\frac{3}{4}-\gamma\right)\right)\left(\log x\right)^{\frac{1}{2}\eta-1}\int_{c^{\prime}+i\infty}^{c^{\prime}-i\infty}\textrm{d}y\textrm{ }y^{-\frac{1}{2}\eta}\exp\left(-y\right).\label{eq:5.3.42}\end{equation}
 After the transformation the limits have changed $c+i\infty\rightarrow c\log x+i\infty\log x=c^{\prime}-i\infty$.
Importantly $c^{\prime}$ is negative as $\log x<0$ and $c>z>0$
is required for the Mellin transform to exist. The integrand has two
obvious singularities, one at $y=0$ and another at $\textrm{Re}\left(y\right)\rightarrow-\infty$
$\left(\textrm{Re}\left(z\right)\rightarrow\infty\right)$. For $\left|y\right|\rightarrow\infty$
and $\textrm{Re}\left(y\right)\ne-\infty$ the integrand is zero,
this being the case we can \emph{close} the integration contour by
joining up the two ends at $c^{\prime}+i\infty$ and $c^{\prime}-i\infty$
such that it becomes a hemisphere of radius $\left|y\right|=\infty$
in the $\textrm{Re}\left(y\right)>c^{\prime}$ region of the complex
$y-$plane, as the integrand is zero along this addition to the contour.
Given the contour is closed in the complex plane we can deform it
as we please so long has we keep all poles inside it as its value
is given by $2\pi i$ times the residue at the pole, in this case
the only pole is at $y=0$. The \emph{Hankel} \emph{contour} ${\mathcal{{C_{H}}}}$
in the complex plane is an open contour which comes in along just
under the real axis from $+\infty$ goes around the origin and back
out to $-\infty$ just above the real axis. We can clearly deform
our contour to Hankel's contour as it contains the origin and no other
poles and because our integrand is zero at $\left|y\right|=\infty$
and $\textrm{Re}\left(y\right)\ne-\infty$ we can open the contour
again at $\left|y\right|=\infty$ making it exactly the Hankel contour.
The integral can then be brought into the form of Hankel's integral
representation of the Gamma function\begin{equation}
\Gamma\left(\frac{1}{2}\eta\right)^{-1}=-\frac{1}{2\pi i}\int_{{\mathcal{{C_{H}}}}}\textrm{d}y\textrm{ }\left(-y\right)^{-\frac{1}{2}\eta}\exp\left(-y\right).\label{eq:5.3.43}\end{equation}
 Inserting $1=\left(-1\right)^{-\frac{1}{2}\eta}\left(-1\right)^{+\frac{1}{2}\eta-1}\left(-1\right)^{+1}$
and \ref{eq:5.3.43} into \ref{eq:5.3.42} gives\begin{equation}
D_{Gribov}^{NS}\left(x,\eta\right)=\exp\left(\frac{1}{2}\eta\left(\frac{3}{4}-\gamma\right)\right)\frac{\left(1-x\right)^{\frac{1}{2}\eta-1}}{\Gamma\left(\frac{1}{2}\eta\right)}.\label{eq:5.3.44}\end{equation}
 (we have used $\lim_{x\rightarrow1}-\log\left(x\right)=\lim_{x\rightarrow1}1-x$).

The result of the Gribov approximation constitutes the perturbative
result derived earlier extended to \emph{all} \emph{orders} with the
caveat that it is only valid for the limit $x\rightarrow1$. To quote
Gribov and Lipatov \cite{Gribov:ri} the approximation above {}``can
be considered as a generalization of the Sudakov form factor''. The
perturbative structure function calculated earlier has no such constraint
on it but is nonetheless a finite order perturbative result. Ideally
one wants a single expression for the structure function which tends
to the Gribov result at in the soft limit and to the perturbative
result away from $x\rightarrow1$, the desired expression will interpolate
between the perturbative structure function and the Gribov approximation.
Such interpolation constitutes what is known as \emph{exponentiation}.
Integrating \ref{eq:5.3.44} over the soft phase space gives,\begin{equation}
\int_{1-\epsilon}^{1}\textrm{d}x\textrm{ }D_{Gribov}^{NS}\left(x,\eta\right)=\exp\left(\frac{1}{2}\eta\left(\frac{3}{4}-\gamma\right)\right)\frac{\epsilon^{\frac{1}{2}\eta}}{\Gamma\left(1+\frac{1}{2}\eta\right)}.\label{eq:5.3.45}\end{equation}
 We denote this $D_{Soft}^{NS}\left(\epsilon,\eta\right)$ and rewrite
the $\epsilon$ term $\exp\left(\frac{1}{2}\log\epsilon\right)$,\begin{equation}
D_{Soft}^{NS}\left(\epsilon,\eta\right)=\frac{\exp\left(-\frac{1}{2}\eta\gamma\right)}{\Gamma\left(1+\frac{1}{2}\eta\right)}\exp\left(\frac{1}{2}\eta\left(\frac{3}{4}+\log\epsilon\right)\right).\label{eq:5.3.46}\end{equation}
 The soft part of the structure function $D_{Soft}^{NS}\left(\epsilon,\eta\right)$
gives an expansion in powers of $\eta$ with coefficients exactly
the same as the coefficients of $\eta$ in the delta term obtained
earlier by iterating the Lipatov equation, so justifying the earlier
quote of Gribov and Lipatov. We have checked this explicitly to ${\mathcal{{O}}}\left(\beta^{3}\right)$.
Had we calculated the ${\mathcal{{O}}}\left(\beta^{4}\right)$ delta
term with Lipatov's equation we would see its coefficient is the same
as the coefficient of the ${\mathcal{{O}}}\left(\beta^{4}\right)$
term in the expansion of $D_{Soft}^{NS}\left(\epsilon,\eta\right)$.
In the perturbative solution the delta terms give the structure function
in the $1-\epsilon<x\leq1$ region. If we were to integrate the delta
terms over $1-\epsilon<x\leq1$ and expand in $\eta$ we would get
the same result as we would just expanding $D_{Soft}^{NS}\left(\epsilon,\eta\right)$
up to the same power in $\eta$. Denoting the delta term in the perturbative
expansion calculated to ${\mathcal{{O}}}\left(\eta^{i}\right)$ as
$\delta\left(1-x\right)\Delta^{\left(i\right)}\left(\epsilon,\eta\right)$
we have,\begin{equation}
\int_{1-\epsilon}^{1}\textrm{d}x\textrm{ }\delta\left(1-x\right)\Delta^{\left(i\right)}\left(\epsilon,\eta\right)-\int_{1-\epsilon}^{1}\textrm{d}x\textrm{ }D_{Gribov}^{NS}\left(x,\eta\right)\sim{\mathcal{{O}}}\left(\eta^{i+1}\right)\label{eq:5.3.48}\end{equation}
 this suggests that we modify the structure function by hand to include
higher order soft effects by replacing the delta term altogether with
$D_{Gribov}^{NS}\left(x,\eta\right)$,\begin{equation}
\delta\left(1-x\right)\Delta^{\left(i\right)}\left(\epsilon,\eta\right)\rightarrow D_{Gribov}^{NS}\left(x,\eta\right).\label{eq:5.3.49}\end{equation}
 How does this affect the theta term \emph{etc}? With this replacement
we can safely take the already implied limit $\epsilon\rightarrow0$,
the theta function for the theta term is then replaced by $1$. Then
the most obvious thing to do is to demand that, up to the same order
in $\eta$, the new structure function is the same as the old purely
perturbative one was in the region $0<x<1-\epsilon$ and not care
about the terms that are higher order in $\eta$. Expanding in $\eta$
we have\begin{equation}
\begin{array}{rcl}
D_{Gribov}^{NS}\left(x,\eta\right) & = & \frac{2}{1-x}\left(\frac{\eta}{4}\right)^{1}\\
 & + & \frac{3+4\log\left(1-x\right)}{1-x}\left(\frac{\eta}{4}\right)^{2}\\
 & + & \frac{\frac{9}{4}-\frac{3}{4}\pi^{2}+6\log\left(1-x\right)+4\log^{2}\left(1-x\right)}{1-x}\left(\frac{\eta}{4}\right)^{3}\\
 & + & {\mathcal{{O}}}\left(\eta^{4}\right)\end{array}.\label{eq:5.3.49b}\end{equation}
 Recall that the Gribov solution is valid in the large $x$ limit.
The perturbative solution obtained order by order in $\eta$ by iterating
solutions through Lipatov's equation requires no such approximation.
Consequently one expects that the perturbative solution and the Gribov
approximation should agree in the limit $x\rightarrow1$. This is
indeed the case, it is easy to verify that the $x\rightarrow1$ limit
of the theta terms obtained in the perturbative case are equal to
those obtained in the non-perturbative case. If we denote the $ith$
order in $\eta$ of a quantity by appending superscript $\left(i\right)$
to it we can define an improved structure function\begin{equation}
D^{\prime NS\left(i\right)}\left(x,\eta\right)=\left\{ \begin{array}{ll}
D_{Gribov}^{NS\left(i\right)}\left(x,\eta\right)+\left(\left.D^{NS\left(i\right)}\left(x,\eta\right)\right|_{\theta-term}-D_{Gribov}^{NS\left(i\right)}\left(x,\eta\right)\right) & 0<i\leq3\\
D_{Gribov}^{NS\left(i\right)}\left(x,\eta\right) & i>3\end{array}\right.\label{eq:5.3.50}\end{equation}
 \emph{i.e.}\begin{equation}
D^{\prime NS}\left(x,\eta\right)=D_{Gribov}^{NS}\left(x,\eta\right)+\sum_{i=0}^{3}\left(\left.D^{NS\left(i\right)}\left(x,\eta\right)\right|_{\theta-term}-D_{Gribov}^{NS\left(i\right)}\left(x,\eta\right)\right).\label{eq:5.3.50b}\end{equation}
 This is the exponentiation prescription of Kuraev and Fadin \cite{Kuraev:hb}.
The condition on $i$ relates to the order to which the perturbative
solution was found, in our case $3$. Applying prescription \ref{eq:5.3.50}
to the perturbative theta terms obtained so far is trivial albeit
tedious subtraction, we find agreement with the results in \cite{Skrzypek:1990qs,Skrzypek:1992vk}.
It is of note that the integral of this function over the range $1-\epsilon\rightarrow\epsilon$
is finite in the limit $\epsilon\rightarrow0$. The integral is finite
because, as stated above, in the divergent $x\rightarrow1$ limit
the theta terms are equal to the corresponding Gribov terms inside
the summation, in addition the integral over the all orders Gribov
approximation to the left of the sum is finite (see equation \ref{eq:5.3.46}).

An alternative exponentiation prescription is that of Jadach and Ward
called YFS exponentiation. Here the only difference is that the Gribov
term is extracted from the iterative solution as a \emph{factor} rather
than subtracted as in the Kuraev Fadin exponentiation \emph{viz}\begin{equation}
D^{\prime NS\left(i\right)}\left(x,\eta\right)=\left\{ \begin{array}{ll}
\sum_{j+k=i}D_{Gribov}^{NS\left(j\right)}\left(x,\eta\right)\Delta^{\left(k\right)}\left(x,\eta\right) & 0<i\leq3\\
D_{Gribov}^{NS\left(i\right)}\left(x,\eta\right)\Delta^{\left(k\right)}\left(x,\eta\right) & i>3\end{array}\right.,\label{eq:5.3.51}\end{equation}
 where $\Delta^{\left(k\right)}\left(x,\eta\right)$ is defined by\begin{equation}
\left.D^{NS\left(i\right)}\left(x,\eta\right)\right|_{\theta-term}=\sum_{j+k=i}D_{Gribov}^{NS\left(j\right)}\left(x,\eta\right)\Delta^{\left(k\right)}\left(x,\eta\right)\textrm{ }\left(1\leq j+k\leq3,\textrm{ }0\leq k\leq2\right).\label{eq:5.3.52}\end{equation}
 This is a system of linear equations that is easily solved for the
$\Delta^{\left(k\right)}\left(x,\eta\right)$,\begin{equation}
\begin{array}{rcl}
\Delta^{\left(0\right)}\left(x,\eta\right) & = & \frac{\left.D^{NS\left(1\right)}\left(x,\eta\right)\right|_{\theta-term}}{D_{Gribov}^{NS\left(1\right)}\left(x,\eta\right)}\\
\Delta^{\left(1\right)}\left(x,\eta\right) & = & \frac{\left.D^{NS\left(2\right)}\left(x,\eta\right)\right|_{\theta-term}}{D_{Gribov}^{NS\left(1\right)}\left(x,\eta\right)}-\frac{D_{Gribov}^{NS\left(2\right)}\left(x,\eta\right)}{D_{Gribov}^{NS\left(1\right)}\left(x,\eta\right)}\Delta^{\left(0\right)}\left(x,\eta\right)\\
\Delta^{\left(2\right)}\left(x,\eta\right) & = & \frac{\left.D^{NS\left(3\right)}\left(x,\eta\right)\right|_{\theta-term}}{D_{Gribov}^{NS\left(1\right)}\left(x,\eta\right)}-\frac{D_{Gribov}^{NS\left(3\right)}\left(x,\eta\right)}{D_{Gribov}^{NS\left(1\right)}\left(x,\eta\right)}\Delta^{\left(0\right)}\left(x,\eta\right)-\frac{D_{Gribov}^{NS\left(2\right)}\left(x,\eta\right)}{D_{Gribov}^{NS\left(1\right)}\left(x,\eta\right)}\Delta^{\left(1\right)}\left(x,\eta\right)\end{array}\label{eq:5.3.53}\end{equation}
 giving\begin{equation}
\begin{array}{rl}
 & D^{\prime NS}\left(x,\eta\right)\\
= & D_{Gribov}^{NS}\left(x,\eta\right)\times\left(\frac{1}{2}\left(1+x\right)^{2}+\frac{1}{16}\left(-2\left(1-x\right)^{2}-\left(1+3x^{2}\right)\log x\right)\eta\right.\\
+ & \left.\frac{1}{384}\left(12\left(1-x\right)^{2}+\left(6\left(1-4x+3x^{2}\right)+\left(1+7x^{2}\right)\log x\right)\log x-12\left(1-x\right)^{2}\textrm{Li}_{2}\left(1-x\right)\right)\eta^{2}\right)\end{array}.\label{eq:5.3.54}\end{equation}
 This is the result given in \cite{Skrzypek:1990qs,Skrzypek:1992vk}.

To summarize, we have re-derived in this section again the structure
functions of Kuraev and Fadin \ref{eq:5.3.50} and Jadach and Skrzypek.
At no point is the underlying hard scattering process referred to,
with this approach folding the radiator function with New Physics
is just as valid as folding it with the standard model. In addition,
within the context of Lipatov's equation (\emph{i.e.} for the case
of near collinear emission of radiation) the calculations suggest
that the factorization of the corrections holds at the level of the
differential cross section not just the total cross section. We are
seeing this effect in the flatness of the radiator functions we have
derived in figure \ref{cap:nointf}. In fact the third order Kuraev-Fadin
structure function \ref{eq:5.3.50} is folded with itself \ref{eq:5.2.5}
and present within \texttt{ZFITTER} as the default setting for the
flag higher \texttt{FOT2} which governs the implementation of higher
order radiative corrections. Crucially at this level of approximation
the hard scattering interaction is not referred to, this is clearly
good news for fits to new physics. Approximations are made in the
leading log approximations \emph{e.g.} the finite order perturbative
result nevertheless they have been found to agree very well (KF better
than 0.2\%, YFS better than \textasciitilde{}0.01\%) with \emph{exact}
numerical solutions of the Lipatov equation. This would seem to vindicate
the use of ISR corrections, the numerical radiator function, derived
from \texttt{ZFITTER} and other such packages.

\section{The Breakdown of the Simple LLA Structure Function Approach.}

\begin{figure}
\begin{center}\includegraphics[%
  width=0.30\paperwidth,
  height=0.35\paperwidth]{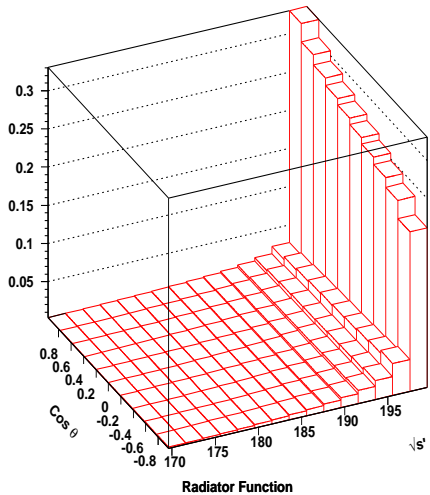}\end{center}

\caption{\label{cap:interference}The plot above shows the radiator functions
$\tilde{R}_{1}\left(s,s^{\prime}\right)$ with $s=\left(200\textrm{ GeV}\right)^{2}$,
obtained from the \texttt{ZFITTER} package by extracting it separately
for each angular bin just as in figure \ref{cap:nointf} but this
time including initial state-final state interference effects.}
\end{figure}

In general the standard model predictions obtained from \texttt{ZFITTER}
and the other standard model difermion packages allow many more sophisticated
radiative corrections to be applied to the process in question besides
just ISR. In particular initial state-final state interference (ISR{*}FSR)
effects corresponding to interference between diagrams with bremsstrahlung
emitted from initial state particles and diagrams with bremsstrahlung
emitted from final state particles, as well as box diagrams resulting
from emission and re-absorption of photons, have a significant (\textasciitilde{}10\%)
effect at the level of the differential cross section despite having
only a negligible (<1\%) effect on the total cross section. The radiator
functions derived from \texttt{ZFITTER} with these corrections included
are shown in figure \ref{cap:interference} where they have acquired
a significant angular dependence that was not seen in figure \ref{cap:nointf}.

This angular dependence would seem to ruin the argument put forward
in \ref{sec:Radiator-Functions:-A}, which says that these corrections
should be independent of the polar angle if in fact they are independent
of the hard physics process. It also highlights the weak point in
the leading log approximation, the leading log approximation starts
to break down for exclusive observables \emph{i.e.} highly differential
quantities and measurements obtained with strong cuts \cite{Skrzypek:1990qs}.
ISR{*}FSR interference corrections clearly fall outside the mandate
of the structure function / leading log approach which describes ISR
effects so well.

In fact the cross section obtained with angular cuts requires \emph{two}
radiator functions. The $\cos\theta$ symmetric and $\cos\theta$
asymmetric parts of the tree level cross section are convoluted separately
with different radiator functions and the result added \cite{Bardin:1989cw,Bardin:1990fu,Bardin:1999yd}.
The cross section in an angular bin $i$ with edges at $c=\cos\theta=c_{i},c_{i+1}$
$\sigma_{i}\left(s,c_{i},c_{i+1}\right)$ is\begin{equation}
\sigma\left(s,c_{i},c_{i+1}\right)=\int_{c_{i}}^{c_{i+1}}\textrm{d}c\textrm{ }\frac{{\rm {d}}\sigma}{{\rm {d}}c}.\label{eq:5.3.55}\end{equation}
 Defining the symmetric and asymmetric cross sections $\sigma_{T}\left(s,-c_{i},c_{i}\right)$
and $\sigma_{FB}\left(s,-c_{i},c_{i}\right)$ respectively in the
region $-c_{i}<c\leq c_{i}$ as\begin{equation}
\begin{array}{rcl}
\sigma_{T}\left(s,-c_{i},c_{i}\right) & = & \frac{1}{2}\left[\int_{0}^{c_{i}}\textrm{d}c\textrm{ }\frac{{\rm {d}}\sigma}{{\rm {d}}c}+\int_{-c_{i}}^{0}\textrm{d}c\textrm{ }\frac{{\rm {d}}\sigma}{{\rm {d}}c}\right]\\
\sigma_{FB}\left(s,-c_{i},c_{i}\right) & = & \frac{1}{2}\left[\int_{0}^{c_{i}}\textrm{d}c\textrm{ }\frac{{\rm {d}}\sigma}{{\rm {d}}}-\int_{-c_{i}}^{0}\textrm{d}c\textrm{ }\frac{{\rm {d}}\sigma}{{\rm {d}}c}\right]\end{array},\label{eq:5.3.56}\end{equation}
 the cross section in the angular bin is\begin{equation}
\sigma\left(s,c_{i},c_{i+1}\right)=\sigma_{T}\left(s,-c_{i+1},c_{i+1}\right)+\sigma_{FB}\left(s,-c_{i+1},c_{i+1}\right)-\sigma_{T}\left(s,-c_{i},c_{i}\right)-\sigma_{FB}\left(s,-c_{i},c_{i}\right).\label{eq:5.3.57}\end{equation}
 In terms of a cross section corrected for radiative effects we have\begin{equation}
\begin{array}{rcl}
\sigma_{T}\left(s,-c_{i},c_{i}\right) & = & \int_{s_{min}^{\prime}}^{s}\textrm{d}s^{\prime}\textrm{ }R_{T}\left(s,s^{\prime},c_{i}\right)\sigma_{Tree}\left(s^{\prime}\right)\\
\sigma_{FB}\left(s,-c_{i},c_{i}\right) & = & \int_{s_{min}^{\prime}}^{s}\textrm{d}s^{\prime}\textrm{ }R_{FB}\left(s,s^{\prime},c_{i}\right)\sigma_{Tree}\left(s^{\prime}\right)\end{array}.\label{eq:5.3.58}\end{equation}

So far we have assumed \begin{equation}
R_{T}\left(s,s^{\prime},c_{i}\right)=R_{FB}\left(s,s^{\prime},c_{i}\right),\label{eq:5.3.59}\end{equation}
 and that to good approximation (\ref{eq:5.2.8}),\begin{equation}
\sigma\left(s,c_{i},c_{i+1}\right)=\int_{s_{min}^{\prime}}^{s}\textrm{d}s^{\prime}\textrm{ }\tilde{R}_{1}\left(s,s^{\prime}\right)\sigma_{Tree}\left(s^{\prime},c_{i},c_{i+1}\right)\label{eq:5.3.60}\end{equation}
 where, recalling section \ref{sec:Radiator-Functions:-A}, $\tilde{R}\left(s,s^{\prime}\right)$
is a universally applicable radiator function. According to Bardin
\emph{et al} \cite{Bardin:1990fu} {}``Strictly speaking an ansatz
like \ref{eq:5.3.59} is wrong.'' In \cite{Bardin:1990fu} the authors
go on to say that near the Z peak the ansatz \ref{eq:5.3.59} nonetheless
gives {}``excellent agreement with the correct result''. The authors
say that this effect is due to the dominant corrections being from
soft photons. They go on to say that if an $s^{\prime}$ cut is in
use the agreement is even better. We appeal to this soft photon dominance
ansatz due to the restrictive $s^{\prime}$ cuts used in obtaining
samples to fit with, $\sqrt{s^{\prime}}\geq0.85\sqrt{s}$, the degree
to which it holds is born out by the flatness in $\cos\theta$ of
the corrections we derived, $\tilde{R}\left(s,s^{\prime}\right)$
in figure \ref{cap:nointf}. If the corrections were significantly
different for the $\cos\theta$ symmetric and asymmetric parts of
the differential cross section $\tilde{R}\left(s,s^{\prime}\right)$would
not be flat in $\cos\theta$ as we essentially are just dividing the
corrected cross section in each angular bin by the born cross section
in each angular bin, bin by bin in $s^{\prime}$, the born cross section
could not possibly \emph{factor out}. Prior to considering initial
state-final state interference effects there seems to be nothing wrong
in using the same radiator function for $c$ even and $c$ odd parts
of the cross section, the structure functions we derived describe
the radiator function very well.

As you can see in figure \ref{cap:interference} the addition of initial
state-final state interference effects seem to ruin all our previous
hypotheses. Ideally one would hope that just because the radiative
corrections have acquired this angular dependence they are still nonetheless
essentially universal like the other (ISR) corrections or at least
largely process independent in our phase space. Given that we are
dealing with box diagrams such factorization seems like wishful thinking.

\section{ISR{*}FSR Universality.}

\begin{figure}
\begin{center}\includegraphics[%
  width=0.30\textwidth,
  height=0.20\textwidth]{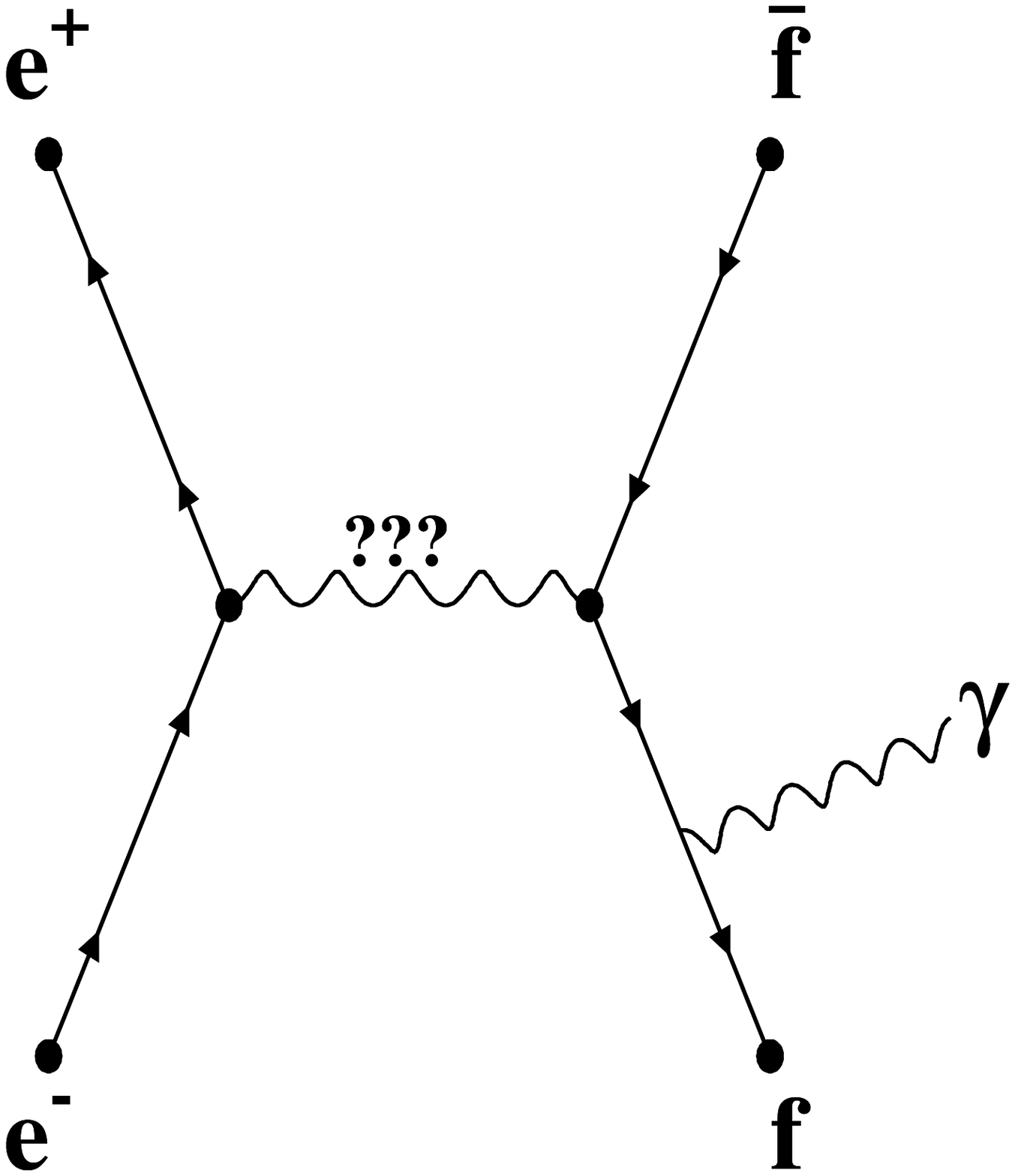}\includegraphics[%
  clip,
  width=0.9cm,
  height=0.20\textwidth]{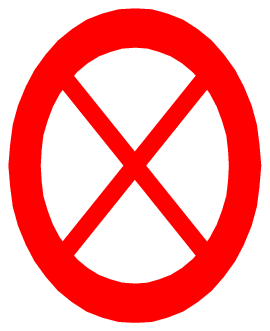}\includegraphics[%
  width=0.30\textwidth,
  height=0.20\textwidth]{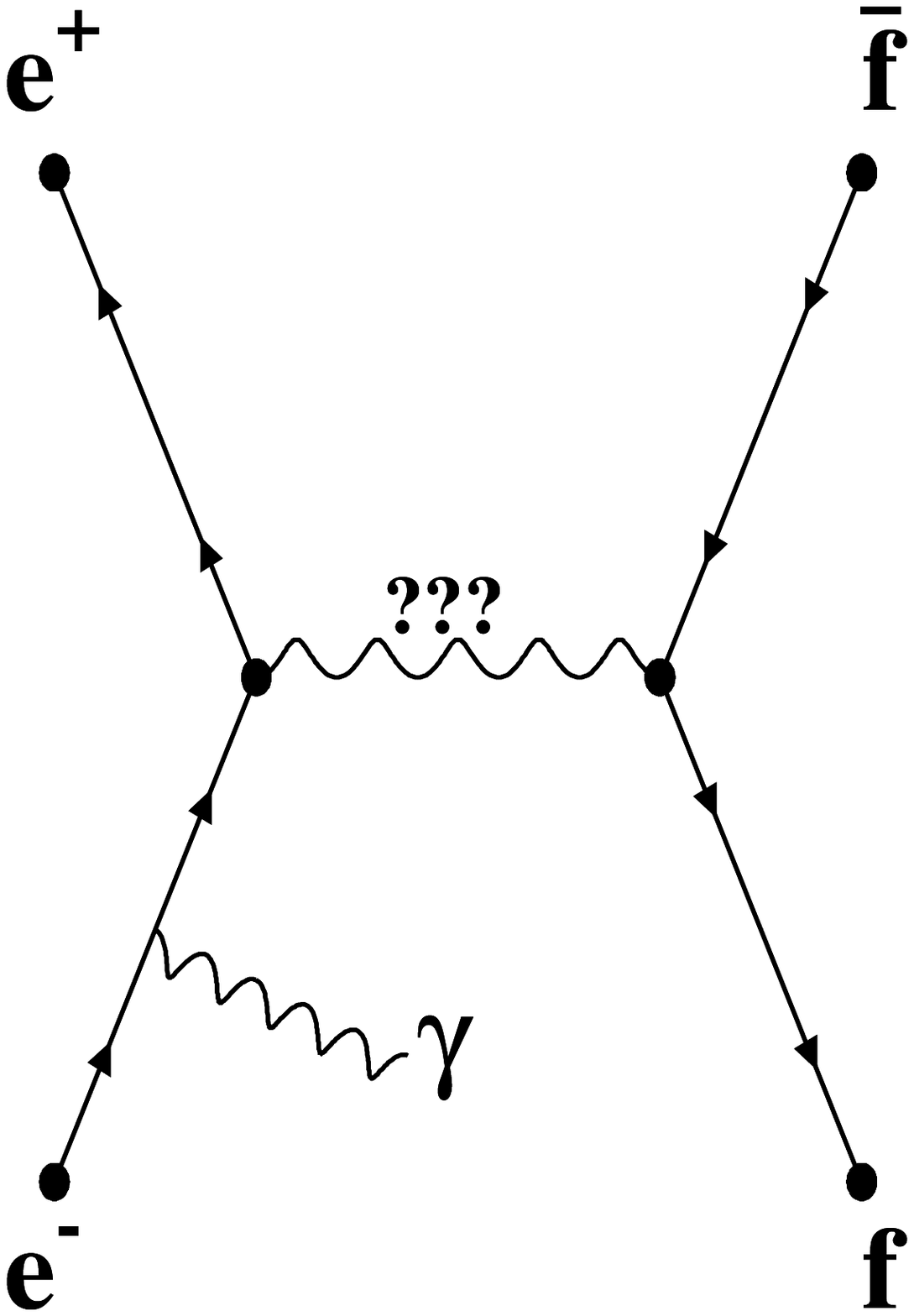}\end{center}

\caption{\label{cap:ISR*FSR-brem}ISR{*}FSR terms in the cross section originate
from diagrams with photons emitted from the initial legs and final
state legs interfering. }
\end{figure}

\begin{figure}
\begin{center}\includegraphics[%
  width=0.30\textwidth,
  height=0.20\textwidth]{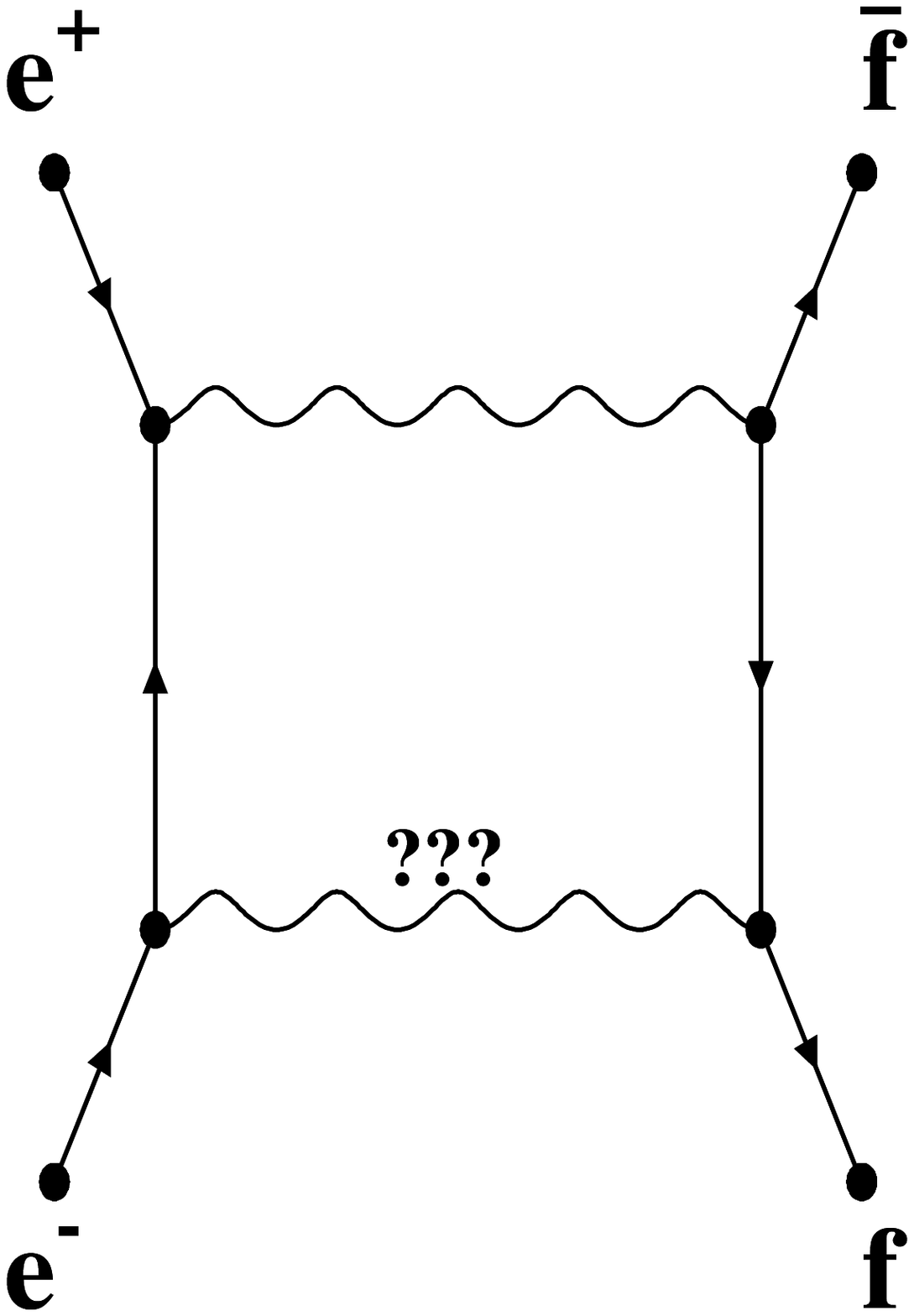}\includegraphics[%
  clip,
  width=0.9cm,
  height=0.20\textwidth]{redcross.eps}\includegraphics[%
  width=0.30\textwidth,
  height=0.20\textwidth]{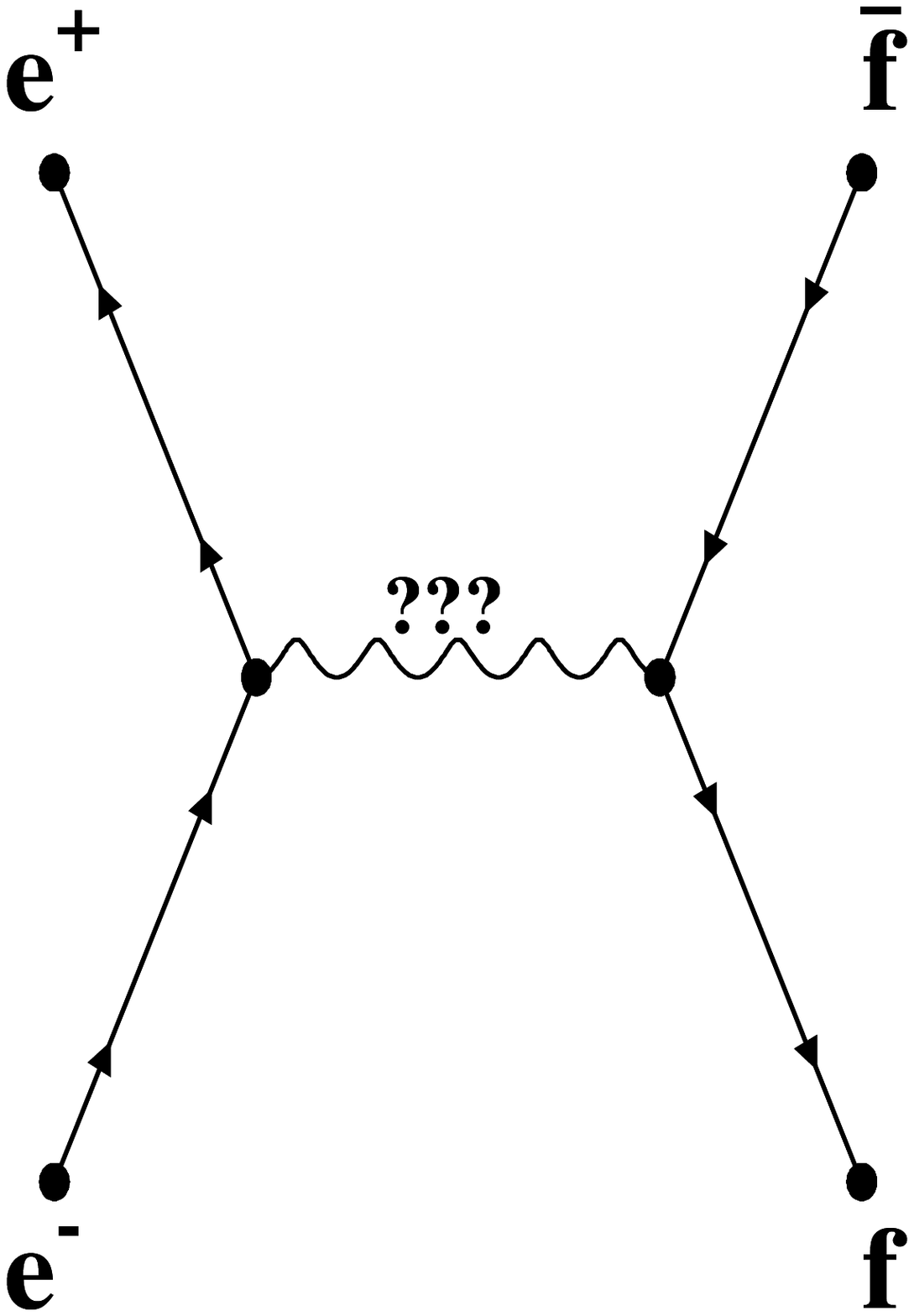}\end{center}

\caption{\label{cap:ISR*FSR-box}In addition to the bremsstrahlung diagrams
in figure \ref{cap:ISR*FSR-brem}, at the same order in $\alpha$,
ISR{*}FSR terms in the cross section also originate from diagrams
with photons box diagrams interfering with the lowest order diagram. }
\end{figure}

In this section we describe how the ISR{*}FSR interference corrections,
despite giving the naive radiator function an important angular dependence,
is nonetheless mostly a universal correction. The ISR{*}FSR interference
corrections for some underlying hard scattering involving the exchange
of an unknown particle (???) may be represented at ${\mathcal{{O}}}\left(\alpha^{3}\right)$
by the Feynman diagrams in figures \ref{cap:ISR*FSR-brem} and \ref{cap:ISR*FSR-box}.%

In subsections \ref{sub:Oa3brem} and \ref{sub:Oa3Virtual} we will
use the Feynman rules of \cite{Peskin:ev} and the kinematic notations
of appendix A. In these subsections we will try to maintain an explicit
notation. The matrix elements of the two fermion events under discussion
at this order have four fermions and hence four spinors in them, $\bar{\nu}_{e}\left(p_{+}\right),\textrm{ }u_{e}\left(p_{-}\right)$
represent the incoming $e^{+}$ and $e^{-}$, $\nu_{f}\left(q_{+}\right),\textrm{ }\bar{u}_{f}\left(q_{-}\right)$
represent the outgoing antifermion and fermion respectively. Given
that we are trying to discuss radiative corrections in such a way
as to find and highlight any universal features of them it will therefore
be useful to denote the born level process generically as\begin{equation}
{\mathcal{{A}}}_{Born}=\bar{\nu}_{e}\left(p_{+}\right)...Hard...\nu_{f}\left(q_{+}\right)\label{eq:5.5.1}\end{equation}
 where $...Hard...$ is defined to contain the electron and fermion
spinors and more importantly, the propagator of the exchanged particle
and its couplings to the incoming and outgoing particles (see also
equation \ref{eq:5.5.1}). For example in the case of QED we have
simply\begin{equation}
...Hard...=iQ_{e}e\gamma^{\mu}u_{e}\left(p_{-}\right)\frac{-ig_{\mu\nu}}{\left(p_{-}+p_{+}\right)^{2}}\bar{u}_{f}\left(q_{-}\right)iQ_{f}e\gamma^{\nu}.\label{eq:5.5.1b}\end{equation}

\subsection{${\mathcal{{O}}}\left(\alpha^{3}\right)$ Soft Bremsstrahlung\label{sub:Oa3brem}.}

\begin{figure}
\begin{center}\includegraphics[%
  width=0.30\textwidth,
  height=0.20\textwidth]{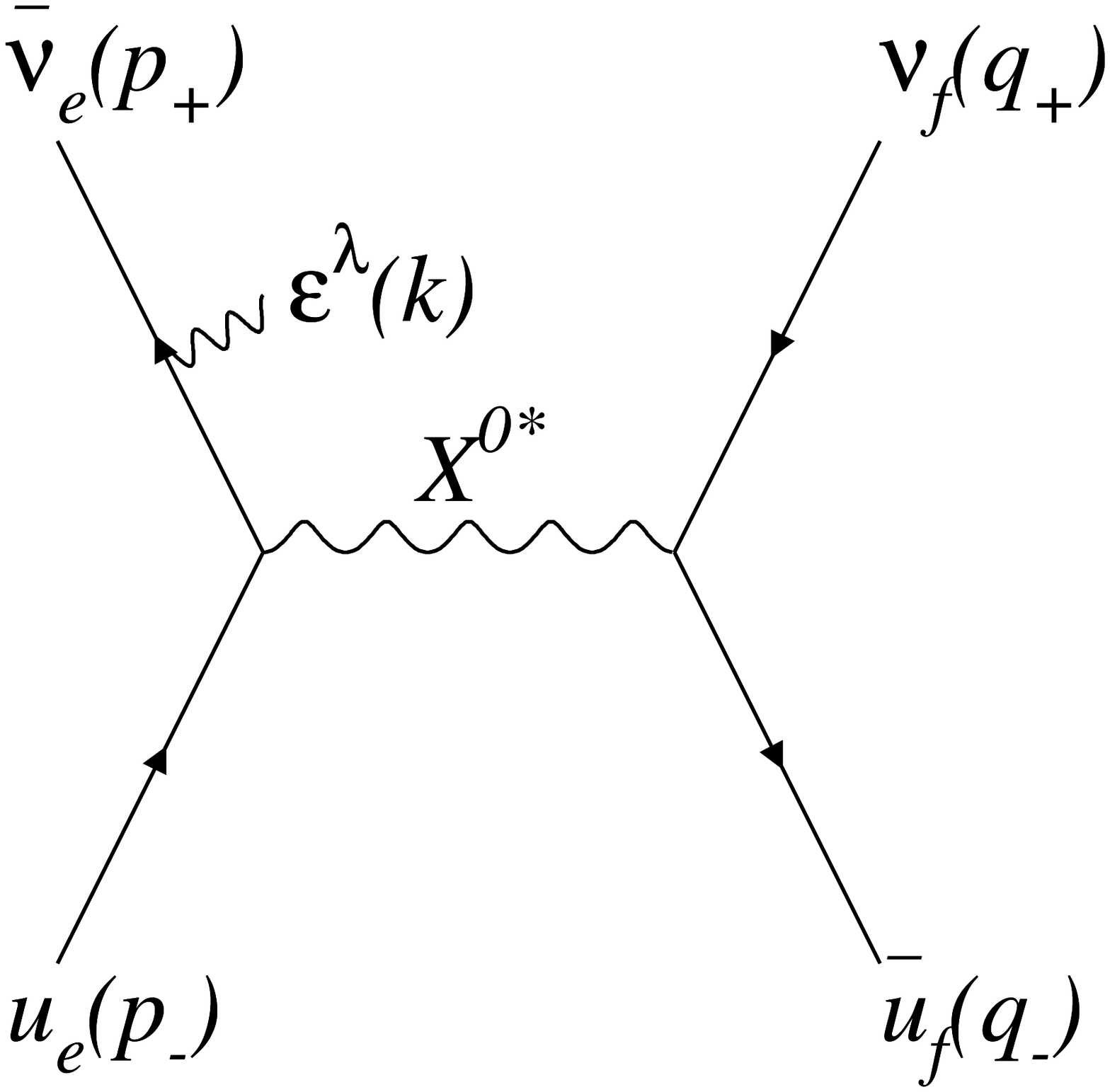}\includegraphics[%
  width=0.30\textwidth,
  height=0.20\textwidth]{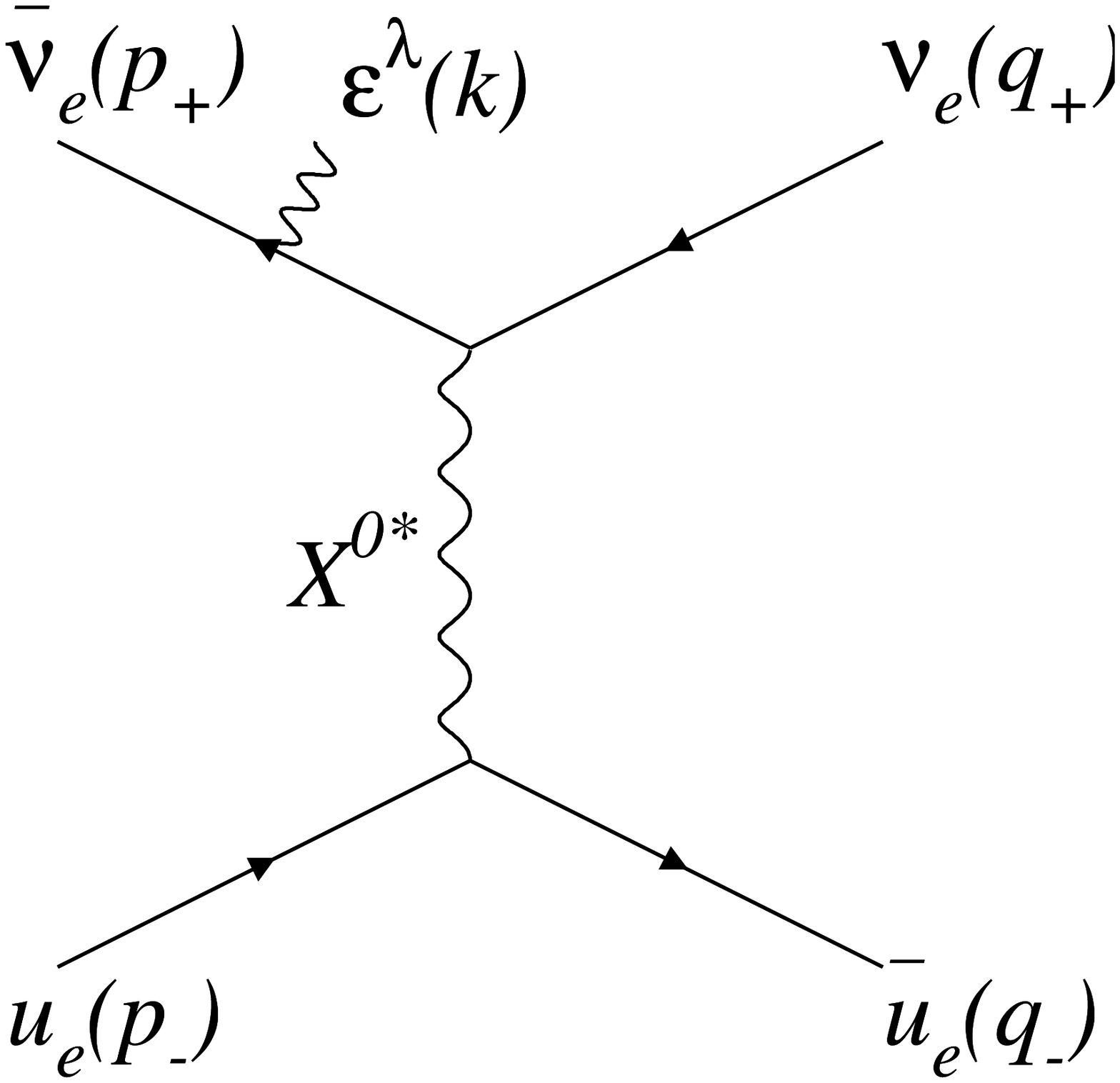}\end{center}

\caption{\label{cap:Some brem diags}Some ${\mathcal{{O}}}\left(\alpha^{3}\right)$
bremsstrahlung correction diagrams.}
\end{figure}

The contribution of soft bremsstrahlung corrections to 2-fermion processes
are greatly important. By soft bremsstrahlung we mean bremsstrahlung
which carries off an amount of energy smaller than the energy resolution
of the detector, soft photons carry off a negligible amount of energy.
Given that the detector cannot therefore tell the difference between
a genuine ${\mathcal{{O}}}\left(\alpha\right)$ process and an ${\mathcal{{O}}}\left(\alpha^{3}\right)$
soft bremsstrahlung correction to it, we must add such contributions
to our predicted cross section. The amplitude for a \emph{general
difermion process} with bremsstrahlung emitted from the external \emph{positron}
leg of the diagram (figure \ref{cap:Some brem diags}) will be of
the form%

\noindent \textcolor{black}{\begin{equation}
{\mathcal{{A}}}_{Brem}=\epsilon_{\lambda}^{*}\left(k\right)\bar{\nu}_{e}\left(p_{+}\right)\left\{ iQ_{e}e\gamma^{\lambda}\frac{i\left(-\left(\not p_{+}-\not k\right)+m_{e}\right)}{\left(p_{+}-k\right)^{2}-m_{e}^{2}}\right\} ...Hard...\nu_{f}\left(q_{+}\right),\label{eq:5.5.1}\end{equation}
 with $Q_{e}=-1$ for an electron (generally $Q_{f}$ is the charge
on the fermion/antifermion $f$ in units of $e$). If we now work
in the limit that $\left|k\right|\rightarrow0$, $s^{\prime}\rightarrow s$,
the} \textcolor{black}{\emph{soft photon approximation}}\textcolor{black}{,
we see that\begin{equation}
{\mathcal{{A}}}_{Soft}=-Q_{e}e\epsilon_{\lambda}^{*}\left(k\right)\bar{\nu}_{e}\left(p_{+}\right)\gamma^{\lambda}\frac{\not p_{+}-m}{2k.p_{+}}...Hard...\nu_{f}\left(q_{+}\right).\label{eq:5.5.2}\end{equation}
 Simple $\gamma$ matrix manipulation and the application of the Dirac
equation to \ref{eq:5.5.2} gives\begin{equation}
\begin{array}{rcl}
{\mathcal{{A}}}_{soft} & = & -\frac{1}{2k.p_{+}}Q_{e}e\epsilon_{\lambda}^{*}\left(k\right)\bar{\nu}_{e}\left(p_{+}\right)\left(2p_{+}^{\lambda}\right)...Hard...\nu_{f}\left(q_{+}\right)\\
 & = & -Q_{e}e\frac{\epsilon^{*}.p_{+}}{k.p_{+}}{\mathcal{{A}}}_{born}\end{array}.\label{eq:5.5.5}\end{equation}
 From this we see that in the soft photon approximation the bremsstrahlung
corrections to the born level process are independent of what that
born process is, we have extracted the bremsstrahlung correction as
an over all constant factor while staying blissfully ignorant of the
underlying process. Hopefully it is clear that these tricks can be
employed with the other external legs in considering bremsstrahlung
emitted from them in the soft photon approximation giving for bremsstrahlung
emitted from the $i$th external leg of the diagram\begin{equation}
{\mathcal{{A}}}_{Soft}^{\left(i\right)}=-Q_{i}e\frac{\epsilon^{*}.p_{i}}{k.p_{i}}{\mathcal{{A}}}_{Born}\label{eq:5.5.6}\end{equation}
 where $p_{i}$ denotes the momentum of the external leg. Consequently,
taking into account the bremsstrahlung emitted from all different
legs we find the matrix element\begin{equation}
{\mathcal{{A}}}_{Soft}^{\dagger}{\mathcal{{A}}}_{Soft}=\sum_{i,j}Q_{i}Q_{j}e^{2}\frac{\left(\epsilon.p_{i}\right)\left(\epsilon^{*}.p_{j}\right)}{\left(k.p_{i}\right)\left(k.p_{j}\right)}{\mathcal{{A}}}_{Born}^{\dagger}{\mathcal{{A}}}_{Born}.\label{eq:5.5.7}\end{equation}
 Summing fermion spins and polarizations of the photon $\left(\sum_{pols}\epsilon^{\mu}\epsilon^{*\nu}=-g^{\mu\nu}\right)$
and inserting the phase space and flux factors factors gives in the
$k\approx0$} \textcolor{black}{\emph{soft photon approximation}}\textcolor{black}{\begin{equation}
\textrm{d}\sigma=-\frac{\alpha}{4\pi^{2}}\sum_{i,j}Q_{i}Q_{j}p_{i}.p_{j}\frac{{\rm {d}}^{3}k}{k_{0}}\frac{1}{\left(k.p_{i}\right)\left(k.p_{j}\right)}\textrm{d}\sigma_{Born}.\label{eq:5.5.11}\end{equation}
 The soft photon cross section is the contribution to the cross section
from soft photons where soft photons were defined earlier as being
those low enough in energy to render the bremsstrahlung process indistinguishable
from the tree level process, we denote this energy $\omega$, the} \textcolor{black}{\emph{soft
photon cut-off}}\textcolor{black}{. Integrating over the final state
phase space of the soft photon we have\begin{equation}
\textrm{d}\sigma_{Soft}=-\frac{\alpha}{\pi}\sum_{i,j}Q_{i}Q_{j}p_{i}.p_{j}\left(\frac{1}{4\pi}\int_{\left|\vec{k}\right|<\omega}\frac{{\rm {d}}^{3}\vec{k}}{k_{0}}\frac{1}{\left(k.p_{i}\right)\left(k.p_{j}\right)}\right)\textrm{d}\sigma_{Born}.\label{eq:5.5.12}\end{equation}
 Note that this differential cross section contains contributions
from pairs of diagrams which constitute initial state-final state
interference. This would appear to be the desired result, the radiative
corrections are factorized from the hard scattering process at the
level of the differential cross section. }

On its own the result \ref{eq:5.5.12} arguably does not make much
sense as the phase space integral\begin{equation}
\frac{1}{4\pi}\int_{\left|\vec{k}\right|<\omega}\frac{{\rm {d}}^{3}\vec{k}}{k_{0}}\frac{1}{\left(k.p_{i}\right)\left(k.p_{j}\right)}\label{eq:5.5.13}\end{equation}
 is infrared divergent, as $k\rightarrow0$ $\textrm{d}\sigma\rightarrow\infty$.
The infrared divergence of the ISR{*}FSR bremsstrahlung contribution
actually cancels an infrared divergence from the ISR{*}FSR box diagrams.
The phase space integral must be expressed in a form whereby we can
add the bremsstrahlung and box contributions to cancel the divergences
before we can fully assess the universality of the ISR{*}FSR interference.
We have treated the divergence with dimensional regularization. We
replace all four dimensional dependence with the usual prescription
$4\rightarrow n=4+\epsilon^{\prime}$

\begin{equation}
\frac{1}{4\pi}\int_{\left|\vec{k}\right|<\omega}\frac{{\rm {d}}^{3}\vec{k}}{k_{0}}\frac{1}{\left(k.p_{i}\right)\left(k.p_{j}\right)}\rightarrow\frac{1}{4\pi}\mu^{N}\int_{\left|\vec{k}\right|<\omega}\frac{{\rm {d}}^{n-1}\vec{k}}{k_{0}}\frac{1}{\left(k.p_{i}\right)\left(k.p_{j}\right)}.\label{eq:5.5.14}\end{equation}
 The factor $\mu^{N}$ is introduced to keep the overall mass dimensions
of the integral the same as they were before extending to $N$ dimensions.
$\mu^{N}$ has mass dimensions $\left[\mu^{N}\right]=N$ \emph{i.e.}
$N=-\epsilon^{\prime}$ \emph{i.e.} \ref{eq:5.5.14} is\begin{equation}
\frac{1}{4\pi}\left(\frac{1}{\mu}\right)^{\epsilon^{\prime}}\int_{\left|\vec{k}\right|<\omega}\frac{{\rm {d}}^{\epsilon^{\prime}+3}\vec{k}}{k_{0}}\frac{1}{\left(k.p_{i}\right)\left(k.p_{j}\right)}.\label{eq:5.5.15}\end{equation}
 This photon phase space integral is technical and not so illuminating
for our present discussion so we shall confine it to appendix B. After
completing the integral the contribution to the differential cross
section from ${\mathcal{{O}}}\left(\alpha^{3}\right)$ soft bremsstrahlung
diagrams is found to be\begin{equation}
\textrm{d}\sigma_{Soft,Int}=\frac{2\alpha}{\pi}\left[\left(\frac{1}{\hat{\epsilon}}+\log\left(\frac{4\omega^{2}}{\mu^{2}}\right)\right)\log\left(\frac{t}{u}\right)-\textrm{Li}_{2}\left(1+\frac{s}{t}\right)+\textrm{Li}_{2}\left(1+\frac{s}{u}\right)\right]\textrm{d}\sigma_{Born}.\label{eq:5.5.45}\end{equation}

\textcolor{black}{This is the} \textcolor{black}{\emph{universal}} \textcolor{black}{soft
photon bremsstrahlung correction to the differential cross section
from ISR{*}FSR interference. Left as it is above, the correction to
the differential cross section is infrared divergent, there are terms
${\mathcal{{O}}}\left(\epsilon^{\prime-1}\right)$ and the limit $\epsilon^{\prime}\rightarrow0$
must be taken to return to four dimensional Minkowski space. For our
purposes the essential feature of this result is that the radiative
correction is factorized from / independent of the new physics exchange
process. }

\subsection{${\mathcal{{O}}}\left(\alpha^{3}\right)$ Virtual Diagrams\label{sub:Oa3Virtual}.}

\begin{figure}
\begin{center}\textcolor{black}{\includegraphics[%
  width=0.49\textwidth,
  height=0.40\textwidth]{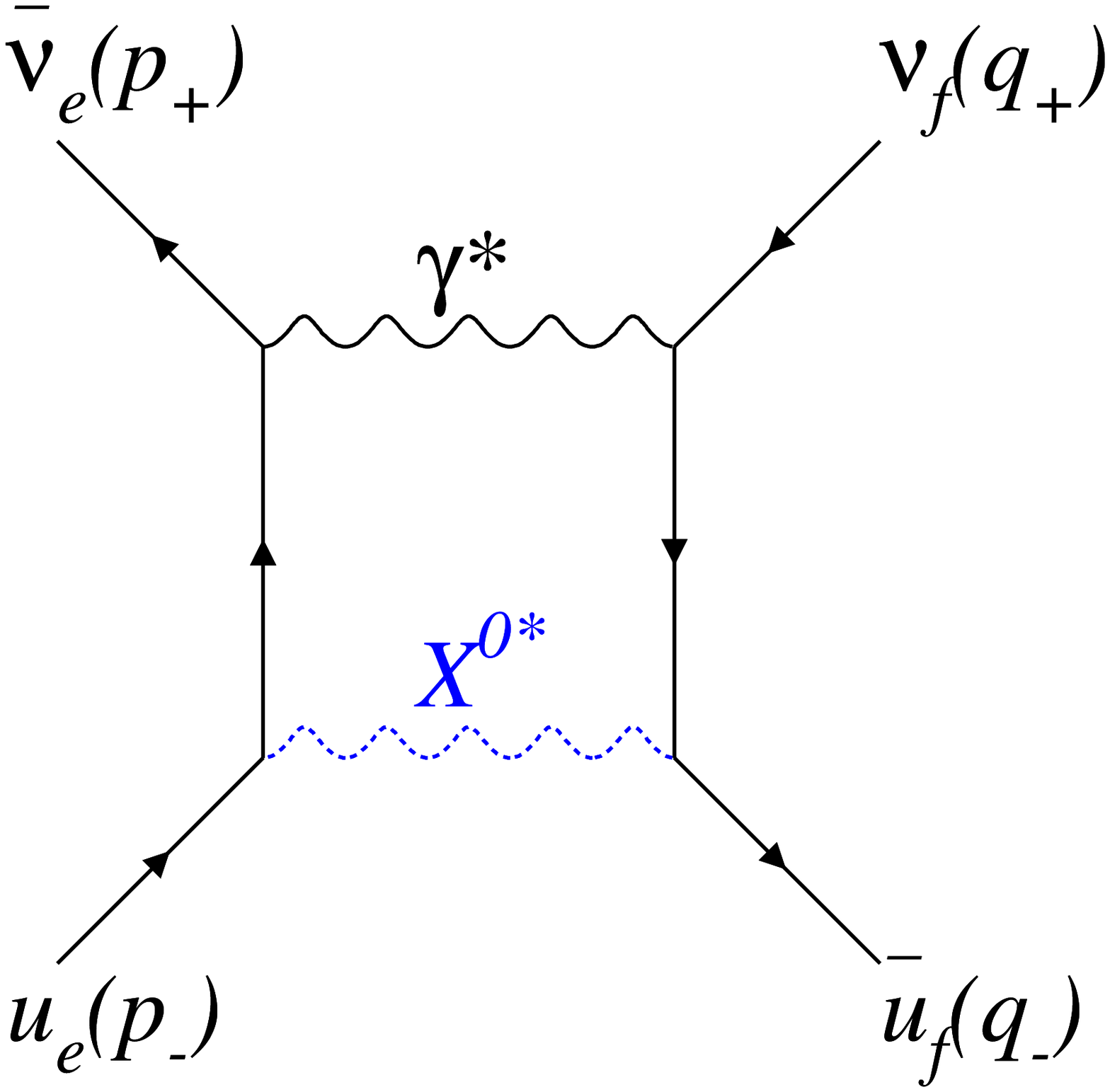}\includegraphics[%
  width=0.49\textwidth,
  height=0.40\textwidth]{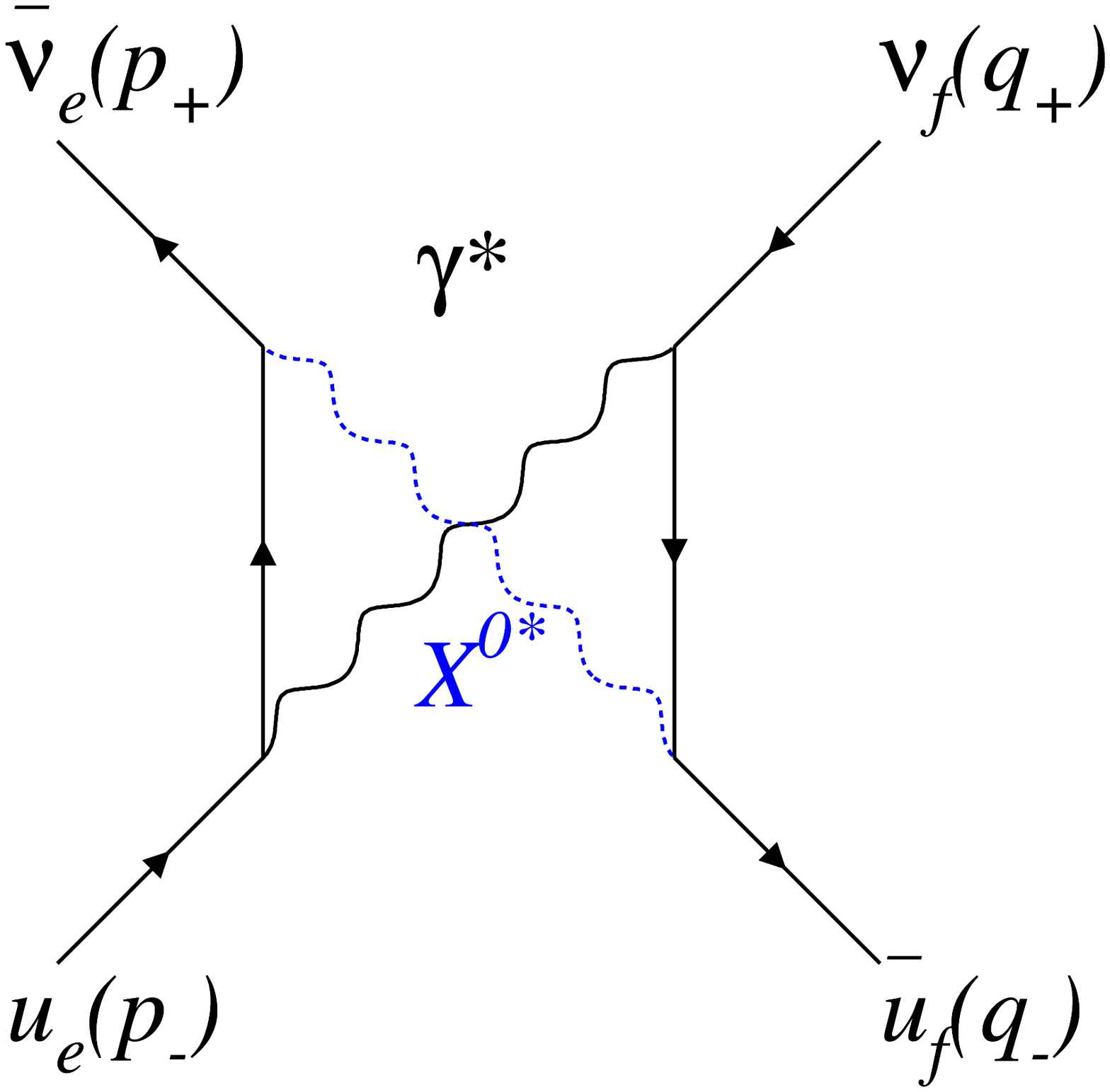}}\end{center}

\begin{center}\textcolor{black}{\includegraphics[%
  width=0.49\textwidth,
  height=0.40\textwidth]{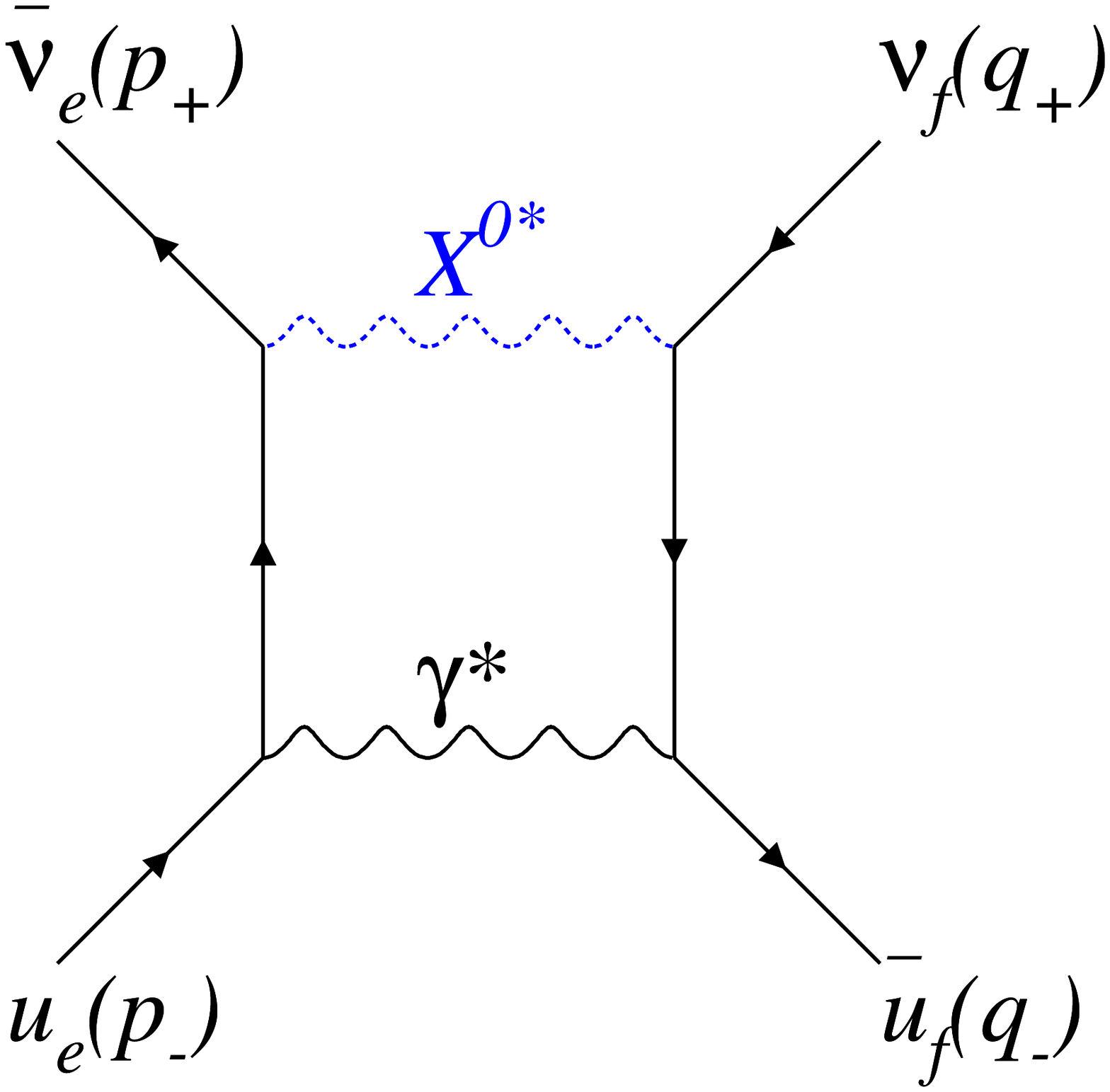}\includegraphics[%
  width=0.49\textwidth,
  height=0.40\textwidth]{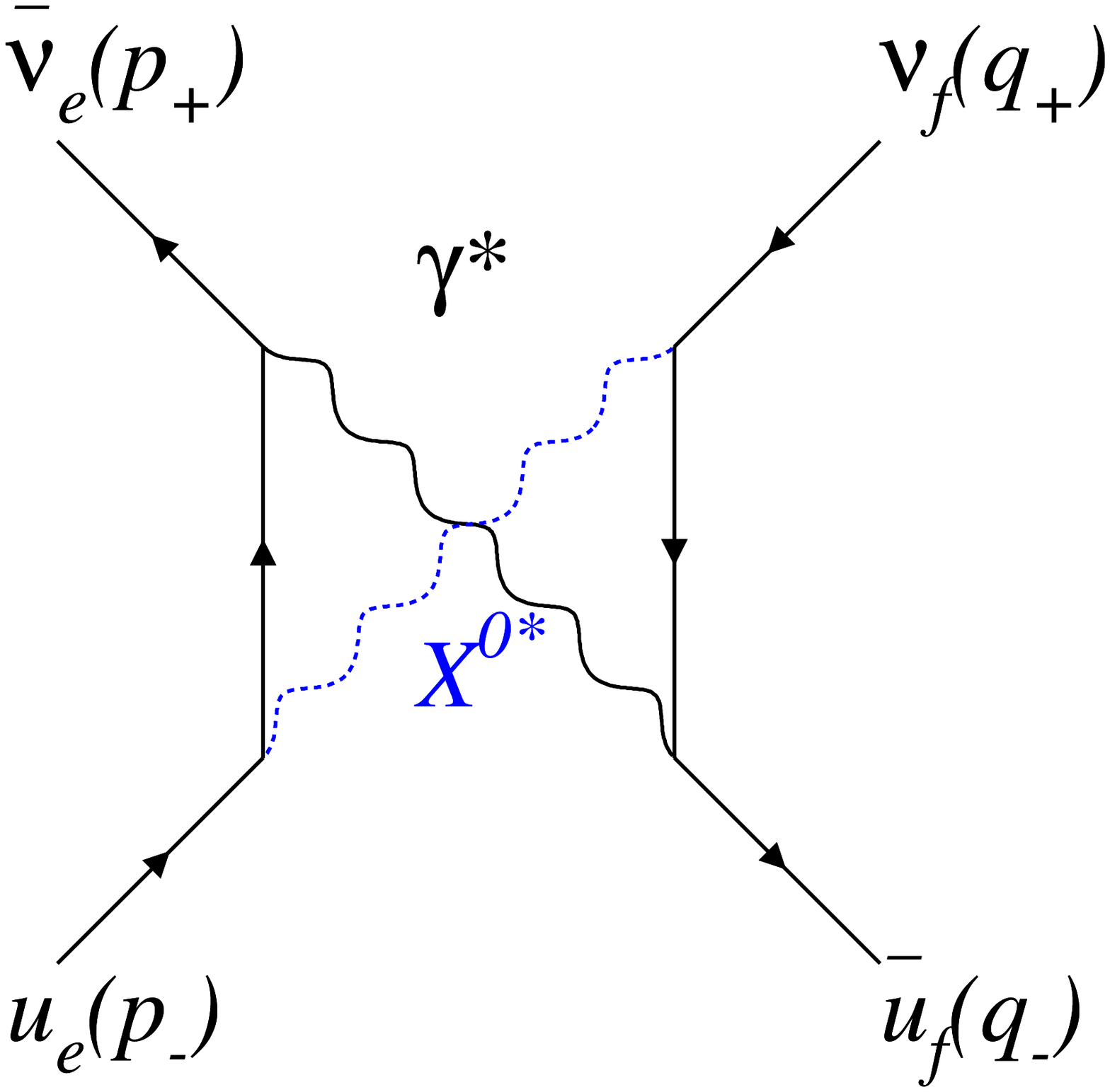}}\end{center}

\caption{\textcolor{black}{\label{cap:box diagrams}The box diagrams contributing
to the ISR{*}FSR interference cross section. On the left hand side
we have} \textcolor{black}{\emph{direct box diagrams}}\textcolor{black}{,
from top to bottom these are represented by amplitudes ${\mathcal{{A}}}_{D1}$
and ${\mathcal{{A}}}_{D2}$ respectively. On the right hand side we
have} \textcolor{black}{\emph{crossed box diagrams}}\textcolor{black}{,
from top to bottom these are represented by amplitudes ${\mathcal{{A}}}_{C1}$
and ${\mathcal{{A}}}_{C2}$ respectively. }}
\end{figure}

\textcolor{black}{The infrared divergences $\left(\frac{1}{\hat{\epsilon}}\right)$
in the soft bremsstrahlung result are well known to be } \textcolor{black}{\emph{physical}}\textcolor{black}{,
that is to say that they are a real part of the theory as opposed
to the ultraviolet divergences encountered in renormalization which
are a consequence of our ignorance of short distance physics, likewise
the infrared divergences cannot be subtracted} \textcolor{black}{\emph{ad
hoc}}\textcolor{black}{. The infrared divergences due to the emission
of soft photons do however cancel against the corresponding virtual
diagrams, it is a well known fact that the soft photon contribution
to the cross section from initial state bremsstrahlung cancels against
a corresponding infrared divergence from the corresponding vertex
diagram which involves emitting a photon from one initial state leg
and absorbing it on the other. Likewise the initial state-final state
interference cross section arising from the interference of initial
and final state bremsstrahlung diagrams does not make much sense on
its own, it must be added to the corresponding virtual graph. The
virtual graph for ISR{*}FSR bremsstrahlung is by analogy to the case
of initial state bremsstrahlung the one formed by taking the photon
emitted from an initial state fermion and joining it to a final state
fermion. This gives the two diagrams of figure \ref{cap:box diagrams}.}

\textcolor{black}{}%

\noindent \textcolor{black}{At ${\mathcal{{O}}}\left(\alpha^{3}\right)$
these diagrams only contribute by their interference with the tree
level (hard) process. Unlike the soft bremsstrahlung corrections the
ISR{*}FSR box diagrams do not simply represent multiplicative corrections
to the born level amplitude. These diagrams were studied in the the
hope that, though they do not fall under the banner of universal radiative
corrections, they may have some component which is universal. It is
quite easy to show that the box diagrams have universal, factorizable
corrections. The tricks previously applied to the soft bremsstrahlung
diagrams also work, to some degree, with box diagrams (provided there
is an internal photon). Let us write down the amplitude for the first
direct box diagram in figure \ref{cap:box diagrams}, were we feign
ignorance of the couplings and propagator of the $X^{0}$ boson}%
\footnote{\textcolor{black}{For now let us assume that $X^{0}$ has the generic
$\frac{1}{k^{2}}$ high energy boson behavior} \textcolor{black}{\emph{i.e.}} \textcolor{black}{the
diagram is UV finite. We discuss this point more later. }%
}\textcolor{black}{. We denote the couplings associated of $X^{0}$
some particle $A$ as ${\mathcal{{C}}}_{A}$ and propagator contributions
to the numerator as ${\mathcal{{P}}}_{X^{0}}$. We assume the denominator
of the propagator is the usual $\left(k^{2}-m_{X}^{2}\right)$ form
(for an $X^{0}$ particle four momentum $k$ and mass $m_{X}$). The
couplings can be considered to have Dirac and Lorentz indices, the
Dirac indices are contracted with the rest of the Dirac algebra in
the numerator and the Lorentz indices are contracted with those in
${\mathcal{{P}}}_{X^{0}}$. With these notations ${\mathcal{{P}}}_{X^{0}}...{\mathcal{{C}}}_{e}u\left(p_{-}\right)\bar{u}\left(p_{+}\right){\mathcal{{C}}}_{f}.../\left(s^{2}-m_{X}^{2}\right)$
corresponds to $...Hard...$ in the case of the soft bremsstrahlung.
Finally we shall assume the interaction is charge conjugation invariant} \textcolor{black}{\emph{i.e.}} \textcolor{black}{${\mathcal{{C}}}_{A}=-{\mathcal{{C}}}_{\bar{A}}$.
The amplitudes for the diagrams in figure \ref{cap:box diagrams}
are then }

\begin{equation}
\begin{array}{rcl}
{\mathcal{{A}}}_{D1} & = & -ie^{2}\int\frac{{\rm {d}}^{4}k}{\left(2\pi\right)^{4}}\textrm{ }{\mathcal{{P}}}_{X^{0}}\frac{\bar{\nu}\left(p_{+}\right)\gamma^{\mu}\left(-{\not k}-\not p_{+}+m_{e}\right){\mathcal{{C}}}_{e^{-}}u\left(p_{-}\right)\bar{u}\left(q_{-}\right){\mathcal{{C}}}_{f}\left(-{\not k}-\not q_{+}+m_{f}\right)\gamma_{\mu}\nu\left(q_{+}\right)}{k^{2}\left(k^{2}+2k.p_{+}\right)\left(k^{2}+2k.q_{+}\right)\left(k^{2}+s+2k.\left(p_{+}+p_{-}\right)-m_{X}^{2}\right)}\\
{\mathcal{{A}}}_{D2} & = & -ie^{2}\int\frac{{\rm {d}}^{4}k}{\left(2\pi\right)^{4}}\textrm{ }{\mathcal{{P}}}_{X^{0}}\frac{\bar{\nu}\left(p_{+}\right){\mathcal{{C}}}_{e^{+}}\left(\not k+\not p_{-}+m_{e}\right)\gamma^{\mu}u\left(p_{-}\right)\bar{u}\left(q_{-}\right)\gamma_{\mu}\left(\not k+\not q_{-}+m_{f}\right){\mathcal{{C}}}_{\bar{f}}\nu\left(q_{+}\right)}{k^{2}\left(k^{2}+2k.p_{-}\right)\left(k^{2}+2k.q_{-}\right)\left(k^{2}+s+2k.\left(p_{+}+p_{-}\right)-m_{X}^{2}\right)}\\
{\mathcal{{A}}}_{C1} & = & +ie^{2}\int\frac{{\rm {d}}^{4}k}{\left(2\pi\right)^{4}}\textrm{ }{\mathcal{{P}}}_{X^{0}}\frac{\bar{\nu}\left(p_{+}\right){\mathcal{{C}}}_{e^{+}}\left(\not k+\not p_{-}+m_{e}\right)\gamma^{\mu}u\left(p_{-}\right)\bar{u}\left(q_{-}\right){\mathcal{{C}}}_{f}\left(-{\not k}-\not q_{+}+m_{f}\right)\gamma_{\mu}\nu\left(q_{+}\right)}{k^{2}\left(k^{2}+2k.p_{-}\right)\left(k^{2}+2k.q_{+}\right)\left(k^{2}+s+2k.\left(p_{+}+p_{-}\right)-m_{X}^{2}\right)}\\
{\mathcal{{A}}}_{C2} & = & +ie^{2}\int\frac{{\rm {d}}^{4}k}{\left(2\pi\right)^{4}}\textrm{ }{\mathcal{{P}}}_{X^{0}}\frac{\bar{\nu}\left(p_{+}\right)\gamma^{\mu}\left(-{\not k}-\not p_{+}+m_{e}\right){\mathcal{{C}}}_{e^{-}}u\left(p_{-}\right)\bar{u}\left(q_{-}\right)\gamma_{\mu}\left(\not k+\not q_{-}+m_{f}\right){\mathcal{{C}}}_{\bar{f}}\nu\left(q_{+}\right)}{k^{2}\left(k^{2}+2k.p_{+}\right)\left(k^{2}+2k.q_{-}\right)\left(k^{2}+s+2k.\left(p_{+}+p_{-}\right)-m_{X}^{2}\right)}\end{array},\label{eq:5.5.2.1}\end{equation}
 where we have defined the loop momentum in each case such that it
flows counter-clockwise with the $X^{0}$ propagator carrying momentum
$k+p_{+}+p_{-}$\emph{i.e.} the propagator is the same in each diagram.
We have defined the loop momentum $k$ such that in each diagram it
is the photon four momentum flowing into the incoming antifermion
line. Firstly condense the numerator as much as possible. The ${\not p}$
terms in the numerator may be quickly got rid of by anticommuting
them so that the on-shell Dirac equation may be used. Considering
the left hand side of the numerator of ${\mathcal{{A}}}_{D1}$\begin{equation}
\begin{array}{rl}
 & \bar{\nu}\left(p_{+}\right)\gamma^{\mu}\left(-{\not k}-{\not p_{+}}+m_{e}\right)...\\
= & e\bar{\nu}\left(p_{+}\right)\left(-\gamma^{\mu}{\not k}+2p_{+}^{\mu}\right)...\end{array}.\label{eq:5.5.2.2}\end{equation}
 Using the same techniques on the right hand side of the various numerators
the amplitude can be simplified to

\begin{equation}
\begin{array}{rcl}
{\mathcal{{A}}}_{D1} & = & -ie^{2}\int\frac{{\rm {d}}^{4}k}{\left(2\pi\right)^{4}}\textrm{ }{\mathcal{{P}}}_{X^{0}}\frac{\bar{\nu}\left(p_{+}\right)\left(\gamma^{\mu}\not k+2p_{+}^{\mu}\right){\mathcal{{C}}}_{e^{-}}u\left(p_{-}\right)\bar{u}\left(q_{-}\right){\mathcal{{C}}}_{f}\left(\not k\gamma_{\mu}+2q_{+\mu}\right)\nu\left(q_{+}\right)}{k^{2}\left(k^{2}+2k.p_{+}\right)\left(k^{2}+2k.q_{+}\right)\left(k^{2}+s+2k.\left(p_{+}+p_{-}\right)-m_{X}^{2}\right)}\\
{\mathcal{{A}}}_{D1} & = & -ie^{2}\int\frac{{\rm {d}}^{4}k}{\left(2\pi\right)^{4}}\textrm{ }{\mathcal{{P}}}_{X^{0}}\frac{\bar{\nu}\left(p_{+}\right){\mathcal{{C}}}_{e^{-}}\left(\not k\gamma^{\mu}+2p_{-}^{\mu}\right)u\left(p_{-}\right)\bar{u}\left(q_{-}\right)\left(\gamma_{\mu}\not k+2q_{-\mu}\right){\mathcal{{C}}}_{f}\nu\left(q_{+}\right)}{k^{2}\left(k^{2}+2k.p_{-}\right)\left(k^{2}+2k.q_{-}\right)\left(k^{2}+s+2k.\left(p_{+}+p_{-}\right)-m_{X}^{2}\right)}\\
{\mathcal{{A}}}_{C1} & = & +ie^{2}\int\frac{{\rm {d}}^{4}k}{\left(2\pi\right)^{4}}\textrm{ }{\mathcal{{P}}}_{X^{0}}\frac{\bar{\nu}\left(p_{+}\right){\mathcal{{C}}}_{e^{-}}\left(\not k\gamma^{\mu}+2p_{-}^{\mu}\right)u\left(p_{-}\right)\bar{u}\left(q_{-}\right){\mathcal{{C}}}_{f}\left(\not k\gamma_{\mu}+2q_{+\mu}\right)\nu\left(q_{+}\right)}{k^{2}\left(k^{2}+2k.p_{-}\right)\left(k^{2}+2k.q_{+}\right)\left(k^{2}+s+2k.\left(p_{+}+p_{-}\right)-m_{X}^{2}\right)}\\
{\mathcal{{A}}}_{C2} & = & +ie^{2}\int\frac{{\rm {d}}^{4}k}{\left(2\pi\right)^{4}}\textrm{ }{\mathcal{{P}}}_{X^{0}}\frac{\bar{\nu}\left(p_{+}\right)\left(\gamma^{\mu}\not k+2p_{+}^{\mu}\right){\mathcal{{C}}}_{e^{-}}u\left(p_{-}\right)\bar{u}\left(q_{-}\right)\left(\gamma_{\mu}\not k+2q_{-\mu}\right){\mathcal{{C}}}_{f}\nu\left(q_{+}\right)}{k^{2}\left(k^{2}+2k.p_{+}\right)\left(k^{2}+2k.q_{-}\right)\left(k^{2}+s+2k.\left(p_{+}+p_{-}\right)-m_{X}^{2}\right)}\end{array}.\label{eq:5.5.2.3}\end{equation}
 Assuming that the couplings $\left({\mathcal{{C}}}_{e/f}\right)$and
numerator of the $X^{0}$ propagator ${\mathcal{{P}}}_{X^{0}}$ do
not depend on the loop momentum we can decompose the amplitudes into
\emph{tensor} $\left(T_{\alpha\beta}\right)$, \emph{vector} $\left(V_{\alpha}\right)$
and \emph{scalar} $\left(S\right)$ loop integrals according to the
number of powers of $k$ in the numerator

\begin{equation}
\begin{array}{rcl}
{\mathcal{{A}}}_{D1} & = & -ie^{2}\bar{\nu}\left(p_{+}\right)\gamma^{\mu}\gamma^{\alpha}...\gamma^{\beta}\gamma_{\mu}\nu\left(q_{+}\right)\times T_{\alpha\beta}\\
 &  & -ie^{2}\bar{\nu}\left(p_{+}\right)\left(\gamma^{\mu}\gamma^{\alpha}...2q_{+\mu}+2p_{+}^{\mu}...\gamma^{\alpha}\gamma_{\mu}\right)\nu\left(q_{+}\right)\times V_{\alpha}\\
 &  & -4ie^{2}\left(p_{+}.q_{+}\right)\bar{\nu}\left(p_{+}\right)...\nu\left(q_{+}\right)\times S\end{array},\label{eq:5.5.2.4}\end{equation}
 where\begin{equation}
\begin{array}{lcl}
T_{\alpha\beta} & = & \int\frac{{\rm {d}}^{4}k}{\left(2\pi\right)^{4}}\frac{k_{\alpha}k_{\beta}}{k^{2}\left(\left(k+p_{+}\right)^{2}-m_{e}^{2}\right)\left(\left(k+q_{+}\right)^{2}-m_{f}^{2}\right)\left(\left(k+p_{+}+p_{-}\right)^{2}-m_{X}^{2}\right)}\\
V_{\alpha} & = & \int\frac{{\rm {d}}^{4}k}{\left(2\pi\right)^{4}}\frac{k_{\alpha}}{k^{2}\left(\left(k+p_{+}\right)^{2}-m_{e}^{2}\right)\left(\left(k+q_{+}\right)^{2}-m_{f}^{2}\right)\left(\left(k+p_{+}+p_{-}\right)^{2}-m_{X}^{2}\right)}\\
S & = & \int\frac{{\rm {d}}^{4}k}{\left(2\pi\right)^{4}}\frac{1}{k^{2}\left(\left(k+p_{+}\right)^{2}-m_{e}^{2}\right)\left(\left(k+q_{+}\right)^{2}-m_{f}^{2}\right)\left(\left(k+p_{+}+p_{-}\right)^{2}-m_{X}^{2}\right)}\end{array}\label{eq:5.5.2.5}\end{equation}
 and ${\mathcal{{C}}}_{e^{-}}u\left(p_{-}\right)\bar{u}\left(q_{-}\right){\mathcal{{C}}}_{f}$
has been replaced with {}``$...$'' for brevity. We can see that
the scalar term in the amplitude of \ref{eq:5.5.2.4} is a factor
multiplied by the born scattering amplitude of the $X^{0}$particle\begin{equation}
2it\times e^{2}\times S\times\left(s-m_{X}^{2}\right)\times\left(\frac{\bar{\nu}\left(p_{+}\right){\mathcal{{C}}}_{e^{-}}u\left(p_{-}\right)\bar{u}\left(q_{-}\right){\mathcal{{C}}}_{f}\nu\left(q_{+}\right)}{s-m_{X}^{2}}\right)=2it\times e^{2}\times S\times\left(s-m_{X}^{2}\right)\times{\mathcal{{A}}}_{Born}.\label{eq:5.5.2.6}\end{equation}
 The numerator algebra has given a universal%
\footnote{Universal in the sense that we have not referred to the specific details
of the $X^{0}$ exchange process. The amplitudes corresponding to
the other diagrams $\left({\mathcal{{A}}}_{D2},{\mathcal{{A}}}_{C1},{\mathcal{{A}}}_{C2}\right)$
also give such factors the only difference being the replacement $t\leftrightarrow u$
for crossed box diagrams \emph{i.e.} the differences in the universal
factors are due to the topology of the diagrams rather than their
physics. %
} factor which multiplies the tree level numerator algebra $4\left(p_{+}.q_{+}\right)=-2t$
however the same is not true of the denominator. The factor $S\times\left(s-m_{X}^{2}\right)$
depends on the underlying hard scattering process through the dependence
of $S$ on $m_{X}$, the mass of the exchanged particle. Using the
on mass shell relations we can rewrite the denominator of the integrals
in the $\left|k\right|\rightarrow0$ limit as\begin{equation}
\begin{array}{rl}
 & \lim_{\left|k\right|\rightarrow0}\frac{1}{k^{2}\left(\left(k+p_{+}\right)^{2}-m_{e}^{2}\right)\left(\left(k+q_{+}\right)^{2}-m_{f}^{2}\right)\left(\left(k+p_{+}+p_{-}\right)^{2}-m_{X}^{2}\right)}\\
= & \frac{1}{k^{2}\left(2k.p_{+}\right)\left(2k.q_{+}\right)\left(s-m_{X}^{2}\right)}\end{array}.\label{eq:5.5.2.7}\end{equation}
 By trivially counting powers of $\left|k\right|$ we see that in
limit $\left|k\right|\rightarrow0$ the tensor integral goes as $\int\textrm{d}\left|k\right|\left|k\right|$,
the vector integral goes as $\int\textrm{d}\left|k\right|$ and the
scalar integral as $\int\textrm{d}\left|k\right|\left|k\right|^{-1}$.
Consequently the scalar integral is infrared divergent and the other
integrals are infrared finite. Simple power counting also shows that
all of the integrals are finite in the ultraviolet limit $\left|k\right|\rightarrow\infty$.
Expanding the last bracket in the denominator we can write,\begin{equation}
S=\int\frac{{\rm {d}}^{4}k}{\left(2\pi\right)^{4}}\frac{1}{k^{2}\left(\left(k+p_{+}\right)^{2}-m_{e}^{2}\right)\left(\left(k+q_{+}\right)^{2}-m_{f}^{2}\right)\left(k^{2}+s+2k.\left(p_{+}+p_{-}\right)-m_{X}^{2}\right)}.\label{eq:5.5.2.8}\end{equation}
 In the infrared divergent $\left|k\right|\rightarrow0$ limit we
can rewrite the last part of the denominator\begin{equation}
\left(k^{2}+s+2k.\left(p_{+}+p_{-}\right)-m_{X}^{2}\right)=\left(s-m_{X}^{2}\right),\label{eq:5.5.2.9}\end{equation}
 therefore the radiative correction factor in the scalar integral
term \begin{equation}
2it\times e^{2}\times S\times\left(s-m_{X}^{2}\right),\label{eq:5.5.2.10}\end{equation}
 has a divergent piece which is universal because the constant factor
$\left(s-m_{X}^{2}\right)$ in \ref{eq:5.5.2.10} cancels that which
appears in the divergent limit of \ref{eq:5.5.2.8}. We can perhaps
express this result in a better way writing\begin{equation}
\frac{\left(s-m_{X}^{2}\right)}{\left(k^{2}+2k.\left(p_{+}+p_{-}\right)+s-m_{X}^{2}\right)}=1-\frac{k^{2}+2k.\left(p_{+}+p_{-}\right)}{\left(k^{2}+2k.\left(p_{+}+p_{-}\right)+s-m_{X}^{2}\right)},\label{eq:5.5.2.11}\end{equation}
 which makes the scalar integral term equal to\begin{equation}
\begin{array}{rcl}
2it\times e^{2}\times S\times\left(s-m_{X}^{2}\right) & = & \int\frac{{\rm {d}}^{4}k}{\left(2\pi\right)^{4}}\frac{2it\times e^{2}}{k^{2}\left(\left(k+p_{+}\right)^{2}-m_{e}^{2}\right)\left(\left(k+q_{+}\right)^{2}-m_{f}^{2}\right)}\\
 & - & \int\frac{{\rm {d}}^{4}k}{\left(2\pi\right)^{4}}\frac{2it\times e^{2}\times\left(k^{2}+2k.\left(p_{+}+p_{-}\right)\right)}{k^{2}\left(\left(k+p_{+}\right)^{2}-m_{e}^{2}\right)\left(\left(k+q_{+}\right)^{2}-m_{f}^{2}\right)\left(k^{2}+s+2k.\left(p_{+}+p_{-}\right)-m_{X}^{2}\right)}\end{array}.\label{eq:5.5.2.12}\end{equation}
 The first term on the right hand side of \ref{eq:5.5.2.12} is universal,
the terms in the denominator correspond to the photon propagator and
two fermion propagators and the $-2t$ in the numerator came from
the universal part of the numerator algebra. The second term on the
right hand side of \ref{eq:5.5.2.12} clearly depends on the hard
scattering process (note the $m_{X}$ terms), it is actually a vector
and a tensor loop integral. The universal term goes as $\int\textrm{d}\left|k\right|\left|k\right|^{-1}$
in the $\left|k\right|\rightarrow0$ limit while the other term goes
as $\int\textrm{d}\left|k\right|$ \emph{i.e.} only the universal
part of the scalar integral is infrared divergent. The degree to which
the box diagram corrections are universal depends on how big the first
integral is in \ref{eq:5.5.2.12} relative to the other terms. If
we denote the universal part of the scalar integral coming from the
denominator of the amplitude \begin{equation}
S_{U}\left(p_{+},q_{+}\right)=\int\frac{{\rm {d}}^{4}k}{\left(2\pi\right)^{4}}\frac{1}{k^{2}\left(\left(k+p_{+}\right)^{2}-m_{e}^{2}\right)\left(\left(k+q_{+}\right)^{2}-m_{f}^{2}\right)},\label{eq:5.5.2.13}\end{equation}
 we can then rewrite the amplitude \ref{eq:5.5.2.4}

\begin{equation}
\begin{array}{rcl}
{\mathcal{{A}}}_{D1} & = & -ie^{2}\bar{\nu}\left(p_{+}\right)\gamma^{\mu}\gamma^{\alpha}...\gamma^{\beta}\gamma_{\mu}\nu\left(q_{+}\right)T_{\alpha\beta}-2ie^{2}tT_{\alpha}^{\alpha}{\mathcal{{A}}}_{Born}\\
 &  & -ie^{2}\bar{\nu}\left(p_{+}\right)\left(\gamma^{\mu}\gamma^{\alpha}...2q_{+\mu}+2p_{+}^{\mu}...\gamma^{\alpha}\gamma_{\mu}\right)\nu\left(q_{+}\right)V_{\alpha}-4ie^{2}tV.\left(p_{+}+p_{-}\right){\mathcal{{A}}}_{Born}\\
 &  & +2ie^{2}tS_{U}\left(p_{+},q_{+}\right){\mathcal{{A}}}_{Born}\end{array}.\label{eq:5.5.2.14}\end{equation}
 where {}``$...$'' represents ${\mathcal{{C}}}_{e^{-}}u\left(p_{-}\right)\bar{u}\left(q_{-}\right){\mathcal{{C}}}_{f}$.
Hopefully it is clear that the decomposition of such scalar integrals
into a universal part and non-universal part is as general as propagators
with denominators of the form $p^{2}-m^{2}$, that is to say no matter
what $X^{0}$is we will always find a term $S_{U}$ \ref{eq:5.5.2.13}.

We can repeat this process with the other diagrams (figure \ref{cap:box diagrams})
in exactly the same way, in each case we have the generic result that
the radiative correction from ISR{*}FSR box diagrams factorizes from
the tree level process in the infrared divergent part of the scalar
loop integral. Referring back to \ref{eq:5.5.2.3} we see that the
other topologies of the process give the following scalar terms\begin{equation}
\begin{array}{rcl}
\left.{\mathcal{{A}}}_{D1}\right|_{Scalar} & = & +2ie^{2}tS_{U}\left(p_{+},q_{+}\right){\mathcal{{A}}}_{Born}\\
\left.{\mathcal{{A}}}_{D2}\right|_{Scalar} & = & +2ie^{2}tS_{U}\left(p_{-},q_{-}\right){\mathcal{{A}}}_{Born}\\
\left.{\mathcal{{A}}}_{C1}\right|_{Scalar} & = & -2ie^{2}uS_{U}\left(p_{-},q_{+}\right){\mathcal{{A}}}_{Born}\\
\left.{\mathcal{{A}}}_{C2}\right|_{Scalar} & = & -2ie^{2}uS_{U}\left(p_{+},q_{-}\right){\mathcal{{A}}}_{Born}\end{array}.\label{eq:5.5.2.14a}\end{equation}

Again we have found what we are looking for, factorization of the
radiative corrections. The ideal result, namely that the differential
cross section due to ISR{*}FSR box diagrams interfering with the tree
level process is of the form\begin{equation}
\textrm{d}\sigma_{Box}=\textrm{Universal Factor}\times\textrm{d}\sigma_{Born},\label{eq:5.5.2.15}\end{equation}
 \emph{i.e.} \emph{exactly} \emph{factorizable} requires that the
vector and tensor integral terms as well as the non-universal parts
of $S$ are negligible relative to the infrared divergent, universal
piece of $S$. Clearly an exact factorization is impossible, factorization
will only ever occur to some degree. The question then becomes, is
a good degree of factorization possible and generic? Requiring a universally
good degree of factorization means that the factorizable part of the
amplitude must somehow dominate all of the other parts. It does not
seem improbable that this be the case as the universal correction
we are interested in \emph{always} corresponds to an infrared divergence,
in fact it corresponds to the only divergence, infrared or otherwise.
We shall briefly postpone the discussion regarding the relative sizes
of the various contributions to $\textrm{d}\sigma_{Box}$ to discuss
an initial simplifying assumption.

It is important to note that some of the new physics $X^{0}$ particles
which we hope to apply this analysis to will have momentum dependent
couplings \emph{e.g.} gravitons (gravitons couple to the energy momentum
tensor). Another, more pertinent point is that bosonic propagators
generically contribute polynomials in the momentum of the particle
they represent to the numerator as well as the denominator of the
amplitude. This does not change our current analysis except at the
point of decomposition into scalar vector and tensor integrals \ref{eq:5.5.2.4}
as gravitons will give ${\mathcal{{C}}}_{e^{-}},\textrm{ }{\mathcal{{C}}}_{f}$
and ${\mathcal{{P}}}_{X^{0}}$ then have a dependence on the loop
momentum. In this case the same decomposition can be done into scalar,
vector and tensor integrals and crucially one finds that the scalar
term is the same as in equation \ref{eq:5.5.2.14}. To see this all
one has to do is simply multiply out any $k$ dependence in ${\mathcal{{P}}}_{X^{0}},\textrm{ }{\mathcal{{C}}}_{e^{-}},\textrm{ }{\mathcal{{C}}}_{f}$
\emph{i.e.} decompose ${\mathcal{{P}}}_{X^{0}},\textrm{ }{\mathcal{{C}}}_{e^{-}},\textrm{ }{\mathcal{{C}}}_{f}$
into scalar, vector bits \emph{etc}. Another way to understand this
is to recall that the scalar part is infrared divergent and the momentum
in the photon in our diagrams is just $k$ the loop momentum, so naively
one can think of the photon line in the diagrams as vanishing in the
infrared divergent limit. Thus even in the case where there is a non-trivial
dependence of the couplings and propagator of $X^{0}$ on the loop
momentum the generic infrared term still occurs, essentially one just
has to expand in the loop momentum. In the case of momentum dependence
of the couplings \emph{etc} the tensor integrals will naturally involve
tensors of higher rank than just two \emph{i.e.} we will have tensor
integrals of the form\begin{equation}
\int\frac{\textrm{d}^{4}k}{\left(2\pi\right)^{4}}\frac{k_{\alpha}k_{\beta}k_{\gamma}...}{k^{2}\left(\left(k+p_{+}\right)^{2}-m_{e}^{2}\right)\left(\left(k+q_{+}\right)^{2}-m_{f}^{2}\right)\left(\left(k+p_{+}+p_{-}\right)^{2}-m_{X}^{2}\right)}.\label{eq:5.5.2.25}\end{equation}
 This causes us to rethink the divergent structure of the amplitude,
up to now only the term involving the scalar integral, the term that
represents the universal radiative correction was (infrared) divergent,
all other terms were ultraviolet and infrared finite. This was encouraging
from the point of view that this could be expected to make the universal
correction the dominant one. In the most general scenario momentum
dependent couplings and the numerator of the propagator could give
ultraviolet terms. Power counting shows that one requires four powers
of the loop momentum in the numerator to have an ultraviolet divergence.
In Giudice \emph{et al} \cite{Giudice:1998ck} the Feynman rules are
derived in the unitary gauge in which the graviton has a propagator
similar to that of the standard model gauge bosons in unitary gauge,
that is to say the high energy behavior goes as $\frac{1}{m^{2}}$.
We shall suppose that in this model (or at least in a fully consistent
theory of gravity) it is possible to work in a gauge analogous to
the the $R_{\xi}$ gauges in which the graviton propagator goes as
$\frac{1}{p^{2}}$ at high energy. With this assumption then we have
at most four powers of the loop momentum in the numerator of any loop
integral, two from the internal fermions and one from each coupling
of the graviton to the fermions. Such integrals diverge logarithmically.
Assuming we cut the integral off at $\left|k\right|=\Lambda$ they
should contribute logarithms $\sim\log\frac{\Lambda^{2}}{s}$ ($\Lambda$
and $\sqrt{s}$ are essentially the only two energy scales in the
diagram). The model of \cite{Giudice:1998ck} has a cut off of order
$1\textrm{ TeV}$. For $\Lambda=10\textrm{ TeV}$, $s=200\textrm{ GeV}$
we have $\log\frac{\Lambda^{2}}{s}=5$, this is something we should
be mindful of when considering the size of any infrared divergences.

The infrared divergent universal term must cancel the corresponding
infrared divergence from the bremsstrahlung calculated in the last
section. To see this we have to regularize the integral. We extend
the dimensionality of the integral from $4$ to $n=4+\epsilon^{\prime}$
dimensions as in section \ref{sub:Oa3brem}\begin{equation}
S_{U}\left(p_{+},q_{+}\right)=\frac{1}{\left(2\pi\right)^{4}}\lim_{\epsilon^{\prime}\rightarrow0}\left(\frac{1}{\mu}\right)^{\epsilon^{\prime}}\int\textrm{d}^{4+\epsilon^{\prime}}k\textrm{ }\frac{1}{k^{2}\left(\left(k+p_{+}\right)^{2}-m_{e}^{2}\right)\left(\left(k+q_{+}\right)^{2}-m_{f}^{2}\right)}.\label{eq:5.5.2.26}\end{equation}
 The denominators can be combined through the usual method of introducing
Feynman parameters\begin{equation}
\frac{1}{k^{2}\left(\left(k+p_{+}\right)^{2}-m_{e}^{2}\right)\left(\left(k+q_{+}\right)^{2}-m_{f}^{2}\right)}=2\int_{0}^{1}\textrm{d}y\textrm{ }y\int_{0}^{1}\textrm{d}x\textrm{ }\frac{1}{\left(k^{2}+2yk.\left(xp_{+}+\left(1-x\right)q_{+}\right)\right)^{3}}.\label{eq:5.5.2.27}\end{equation}
 To use the standard integrals we complete the square in the denominator
and shift the variable of integration by a constant amount, $k^{\mu}\rightarrow\tilde{k}^{\mu}=k^{\mu}+y\left(xp_{+}^{\mu}+\left(1-x\right)q_{+}^{\mu}\right)$
to give\begin{equation}
S_{U}\left(p_{+},q_{+}\right)=\frac{-i\pi^{2}}{\left(2\pi\right)^{4}}\lim_{\epsilon^{\prime}\rightarrow0}\left(\frac{1}{\mu}\right)^{\epsilon^{\prime}}\int_{0}^{1}\textrm{d}x\textrm{ }\int_{0}^{1}\textrm{d}y\textrm{ }y\textrm{ }\pi^{\frac{1}{2}\epsilon^{\prime}}\Gamma\left(1-\frac{1}{2}\epsilon^{\prime}\right)\left(\frac{1}{y^{2}\left(xp_{+}+\left(1-x\right)q_{+}\right)^{2}}\right)^{1-\frac{1}{2}\epsilon^{\prime}}.\label{eq:5.5.2.29}\end{equation}
 From now on we denote $\Delta\left(p_{+},q_{+}\right)=\left(xp_{+}+\left(1-x\right)q_{+}\right)^{2}$.
In the current form we can perform the integral over $y$,\begin{equation}
\int_{0}^{1}\textrm{d}y\textrm{ }y\left(\frac{1}{y^{2}}\right)^{1-\frac{1}{2}\epsilon^{\prime}}=\left[\frac{1}{\epsilon^{\prime}}y^{\epsilon^{\prime}}\right]_{0}^{1}=\frac{1}{\epsilon^{\prime}}.\label{eq:5.5.2.30}\end{equation}
 Substituting this into $S_{U}\left(p_{+},q_{+}\right)$ and expanding
the term in brackets about $\epsilon^{\prime}=0$ gives\begin{equation}
S_{U}\left(p_{+},q_{+}\right)=-\frac{i}{32\pi^{2}}\lim_{\epsilon^{\prime}\rightarrow0}\int_{0}^{1}\textrm{d}x\textrm{ }\frac{1}{\Delta}\left(\frac{1}{\hat{\epsilon}}-\log\left(\frac{\mu^{2}}{s}\right)+\log\left(\frac{\Delta}{s}\right)\right).\label{eq:5.5.2.32}\end{equation}
 Expanding $\Delta\left(p_{+},q_{+}\right)$ (keeping all masses)\begin{equation}
\Delta\left(p_{+},q_{+}\right)=\left(x^{2}t-x\left(t-m_{e}^{2}+m_{f}^{2}\right)+m_{f}^{2}\right),\label{eq:5.5.2.33}\end{equation}
 we see that $S_{U}\left(p_{+},q_{+}\right)$ is really a function
of the Mandelstam variable $t$. Likewise $\Delta\left(p_{+},q_{+}\right)$
is really a function of $t$, $\Delta\left(t\right)$. We can perform
the integral by completing the square in $\Delta\left(t\right)$\begin{equation}
\Delta\left(p_{+},q_{+}\right)=-t\left(k^{2}-y^{2}\right),\label{eq:5.5.2.34}\end{equation}
 where \begin{equation}
\begin{array}{rcl}
y & = & x-\frac{1}{2}\left(1+\delta\right)\\
k^{2} & = & \frac{1}{4}\left(1+\delta\right)^{2}-\frac{m_{f}^{2}}{t}\\
\delta & = & \frac{m_{f}^{2}-m_{e}^{2}}{t}\end{array}.\label{eq:5.5.2.35}\end{equation}
 In this case we have\begin{equation}
\int_{0}^{1}\textrm{d}x\textrm{ }\frac{1}{\Delta\left(t\right)}\log\left(\frac{\Delta\left(t\right)}{s}\right)=-\frac{1}{t}\int_{-\frac{1}{2}\left(1+\delta\right)}^{\frac{1}{2}\left(1-\delta\right)}\textrm{d}y\textrm{ }\frac{1}{k^{2}-y^{2}}\log\frac{-t}{s}\left(k^{2}-y^{2}\right).\label{eq:5.5.2.34}\end{equation}
 The integral is awkward and may be made easier by using the following
relation\begin{equation}
\frac{1}{k-y}\log\left(k+y\right)=\frac{\textrm{d}}{\textrm{d}y}\textrm{Li}_{2}\left(\frac{1}{2k}\left(k-y\right)\right)+\frac{1}{k-y}\log\left(2k\right).\label{eq:5.5.2.36}\end{equation}
 Substituting back in for $k^{2}$ $\left(k=\frac{1}{2}+\frac{1}{2}\delta-\frac{m_{f}^{2}}{t}\right)$
\emph{etc} and dropping terms ${O}\left(\delta^{2}\right)$ the integral
is found to be \begin{equation}
\begin{array}{rcl}
\int_{0}^{1}\textrm{d}x\textrm{ }\frac{1}{\Delta\left(t\right)}\log\left(\frac{\Delta\left(t\right)}{s}\right) & = & \frac{1}{4kt}\left[\log^{2}\left(\frac{m_{e}^{2}}{s}\right)+\log^{2}\left(\frac{m_{f}^{2}}{s}\right)-2\log^{2}\left(\frac{-t}{s}\right)\right]\\
 & + & \frac{1}{2kt}\left[\textrm{Li}_{2}\left(1+\frac{m_{f}^{2}}{t}\right)-\textrm{Li}_{2}\left(-\frac{m_{e}^{2}}{t}\right)\right]\\
 & + & \frac{1}{2kt}\left[\textrm{Li}_{2}\left(1+\frac{m_{e}^{2}}{t}\right)-\textrm{Li}_{2}\left(-\frac{m_{f}^{2}}{t}\right)\right]\end{array}.\label{eq:5.5.2.39}\end{equation}
 This simple step is illustrated to condense the structure of divergences
that are present in the limit $m_{e},m_{f}\rightarrow0$, the \emph{collinear
divergences}. A similar decomposition to that in equation \ref{eq:5.5.2.35}
gives the first two terms in \ref{eq:5.5.2.32} as \begin{equation}
\int_{0}^{1}\textrm{d}x\textrm{ }\frac{1}{\Delta}\left(\frac{1}{\hat{\epsilon}}-\log\left(\frac{\mu^{2}}{s}\right)\right)=\frac{1}{2kt}\left(\frac{1}{\hat{\epsilon}}-\log\left(\frac{\mu^{2}}{s}\right)\right)\left(\log\left(-\frac{m_{e}^{2}}{t}\right)+\log\left(-\frac{m_{f}^{2}}{t}\right)\right).\label{eq:5.5.2.40}\end{equation}
 Finally, taking $2k\approx1$ \emph{outside} logarithms gives {\small \begin{equation}
S_{U}\left(t\right)=-\frac{i}{32\pi^{2}}\lim_{\epsilon^{\prime}\rightarrow0}\left[\frac{1}{t}\left(\log\left(-\frac{m_{e}^{2}}{t}\right)+\log\left(-\frac{m_{f}^{2}}{t}\right)\right)\left(\frac{1}{\hat{\epsilon}}-\log\left(\frac{\mu^{2}}{s}\right)\right)+\int_{0}^{1}\textrm{d}x\textrm{ }\frac{1}{\Delta\left(t\right)}\log\left(\frac{\Delta\left(t\right)}{s}\right)\right].\label{eq:5.5.2.41}\end{equation}
}{\small \par}

In summary we have that in the most general case the box diagrams
under consideration have an amplitude which consists of a scalar IR
divergent term $2ie^{2}tS_{U}\left(t\right){\mathcal{{A}}}_{Born}$
($-2ie^{2}uS_{U}\left(u\right){\mathcal{{A}}}_{Born}$ for the crossed
box diagrams) along with other infrared finite terms the exact nature
of which depends on the physics of the exchanged $X^{0}$ boson. In
addition it is possible that ultraviolet divergences are present among
the other non-universal terms, in the case of the quantum gravity
model of Giudice \emph{et al} the amplitude is at most logarithmically
divergent and we expect that divergence enhances such a term by a
small factor. Adding the \emph{universal} \emph{contributions} of
the four diagrams \ref{cap:box diagrams} together we have\begin{equation}
\begin{array}{rl}
 & \left.{\mathcal{{A}}}_{Box}\right|_{Universal}\\
= & {\mathcal{{A}}}_{D1}+{\mathcal{{A}}}_{D2}+{\mathcal{{A}}}_{C1}+{\mathcal{{A}}}_{C2}\\
= & \frac{\alpha}{\pi}\lim_{\epsilon^{\prime}\rightarrow0}\left[\left(-\frac{1}{\hat{\epsilon}}+\log\left(\frac{\mu^{2}}{s}\right)\right)\log\left(\frac{t}{u}\right)+\frac{t}{2}\int_{0}^{1}\textrm{d}x\textrm{ }\frac{1}{\Delta\left(t\right)}\log\left(\frac{\Delta\left(t\right)}{s}\right)-\frac{u}{2}\int_{0}^{1}{\rm {d}}x\textrm{ }\frac{1}{\Delta\left(u\right)}\log\left(\frac{\Delta\left(u\right)}{s}\right)\right]{\mathcal{{A}}}_{Born}\end{array}.\label{eq:5.5.2.42}\end{equation}

\noindent The integral final integral in \ref{eq:5.5.2.42} is the
same as the one preceding it (shown in \ref{eq:5.5.2.39}) with the
replacement $t\rightarrow u$. These two integrals add to give terms
which are finite as $m_{e},m_{f}\rightarrow0$. It is known in the
literature that the initial state-final state interference corrections
contain no collinear divergences, here we see explicitly that this
result is true also for the universal correction as one would expect.
We can safely take the limit $m_{e},m_{f}\rightarrow0$ leaving

\noindent \begin{equation}
\left.{\mathcal{{A}}}_{Box}\right|_{Universal}=\frac{\alpha}{\pi}\lim_{\epsilon^{\prime}\rightarrow0}\left[\left(-\frac{1}{\hat{\epsilon}}+\log\left(\frac{\mu^{2}}{s}\right)\right)\log\left(\frac{t}{u}\right)+\frac{1}{2}\log^{2}\left(\frac{-u}{s}\right)-\frac{1}{2}\log^{2}\left(\frac{-t}{s}\right)\right].\label{eq:5.5.2.44}\end{equation}
 Denoting $\lim_{\hat{\epsilon}\rightarrow0}\left[...\right]$ as
${\mathcal{{F}}}$, the interference of the box diagrams with the
born amplitude may be written\begin{equation}
\Sigma_{spins}\left(A_{Born}A_{Box}^{\dagger}+A_{Box}A_{Born}^{\dagger}\right)=\frac{2\alpha}{\pi}{\mathcal{{F}}}\textrm{ }\Sigma_{spins}\textrm{ }A_{Born}^{\dagger}A_{Born},\label{eq:5.5.2.45}\end{equation}
 hence \begin{equation}
\left.\textrm{d}\sigma_{Box}\right|_{Universal}=\frac{2\alpha}{\pi}{\mathcal{{F}}}\textrm{d}\sigma_{Born}.\label{eq:5.5.2.46}\end{equation}
 If we now add the differential cross section due to box diagrams
interfering with the Born level amplitude to the differential cross
section due to soft bremsstrahlung diagrams we obtain the total differential
cross section for the \emph{universal part of} \emph{initial state-final
state interference} as\begin{equation}
\begin{array}{rcl}
\textrm{d}\sigma_{Universal} & = & \textrm{d}\sigma_{Soft,Int}+\left.\textrm{d}\sigma_{Box}\right|_{Universal}\\
 & = & \frac{2\alpha}{\pi}\left(\log\left(\frac{s}{4\omega^{2}}\right)\log\left(\frac{u}{t}\right)-\textrm{Li}_{2}\left(1+\frac{s}{t}\right)+\textrm{Li}_{2}\left(1+\frac{s}{u}\right)\right.\\
 &  & \left.+\frac{1}{2}\log^{2}\left(\frac{-u}{s}\right)-\frac{1}{2}\log^{2}\left(\frac{-t}{s}\right)\right)\textrm{d}\sigma_{Born}\end{array}.\label{eq:5.5.2.47}\end{equation}
 This is the universal contribution to the differential cross section
from initial state-final state interference contributions to some
new physics exchange process $\left(X^{0}\right)$ with tree level
differential cross section $\textrm{d}\sigma_{Born}$. For $\sqrt{s}\sim200\textrm{ GeV}$
while one may take the soft photon cut-off to be $\omega\sim50\textrm{ MeV}$
(\emph{i.e.} typical LEP experiments) this gives $\log\left(\frac{s}{4\omega^{2}}\right)=15$,
this constitutes a so-called \emph{large logarithm}. This large logarithm
$\log\left(\frac{s}{4\omega^{2}}\right)$ arises from the cancellation
of the infrared divergences in the box diagram and initial and final
state bremsstrahlung contributions. As noted earlier only the universal
scalar integral terms are infrared divergent thus \emph{no other terms
will undergo this large logarithmic enhancement}. These universal
factorizable corrections maybe resummed (exponentiated) as in the
conventional treatment of the Sudakov effect. Considering only terms
${\mathcal{{O}}}\left(\alpha\log\left(\frac{s}{4\omega^{2}}\right)\right)$
corresponds to the leading log approximation for initial state-final
state interference. We stress that terms which are not universal are
not infrared divergent and so they will not have the large logarithm
$\log\left(\frac{s}{4\omega^{2}}\right)$, they correspond to \emph{sub-leading
terms} ${\mathcal{{O}}}\left(\alpha\right)$. We conclude that initial
state-final state interference corrections are universal within the
leading log approximation and we anticipate, conservatively, that
the leading log (universal) terms dominate non-universal sub-leading
terms by a factor of 10.

It is of note that the theoretical study of the ISR{*}FSR interference
by the author began with the calculation of the QED box diagram (\emph{i.e.}
for the case that $X^{0}$ is simply a photon). The \emph{FeynCalc}
\cite{Mertig:an} package was used to assist Dirac traces and reduction
of the \emph{Passarino Veltman} \emph{functions} \cite{Passarino:1978jh}.
The universal leading log term was compared to the non-universal terms
in this case and was found to dominate the sub-leading terms by a
factor of at least $13$ across the $\cos\theta$ region under study.

\section{Summary and Conclusions}

We have seen that the radiator functions describing ISR, the dominant
correction, are merely the result of an intrinsically factorized DGLAP
type analysis with no additional non-factorizable components and that
these describe the total cross section to better than the 1\% level.
In addition we have established that even ISR{*}FSR interference contributions
to the differential cross section are universal to better than the
leading log approximation (note that equation \ref{eq:5.5.2.47} contains
sub-leading terms which are nonetheless universal). With this in mind
we feel justified in using the numerical radiator functions obtained
from the \texttt{ZFITTER} program or its relatives to correct the
tree-level differential cross sections of processes involving the
exchange of some new particle. This method is valid within the context
of the leading log approximation (and also somewhat beyond).

We would like to stress however that though we conclude that to good
approximation the radiative corrections we have discussed are independent
of the new physics in the diagrams they are not independent of the
topology of the diagrams. The \texttt{ZFITTER} package and this analysis
is only strictly valid for the case of $s$-channel difermion processes.
\texttt{ZFITTER} treats the $t$-channel processes \emph{i.e.} $e^{+}e^{-}\rightarrow e^{+}e^{-}$
differently to the other difermion final states. Our analysis of the
DGLAP electron structure function approach to ISR will still hold
for $t$-channel processes, these corrections pertain only to the
incoming particles. On the other hand we expect that our analysis
of the universality of ISR{*}FSR interference as discussed in section
\ref{sub:Oa3Virtual} will be modified, certainly one should expect
the angular dependence of the radiative corrections to be different.
Nevertheless a large logarithmic, universal, factorizable term will
result. This term will be, by analogy to the $s$-channel result,
of the form $\frac{2\alpha}{\pi}\log\left(\frac{t}{4\omega^{2}}\right)\log\left(...\right)$
\emph{i.e.} the $t$-channel result should still involve a logarithm
of the ratio of the hard scale physics scale to the soft scale $\left(\omega^{2}\right)$.
Provided that the cuts on the $\cos\theta$ region are not too loose
\emph{i.e.} provided all the events in the sample are such that $t\gg\omega^{2}$,
the resulting enhancement for the universal term $\left(\log\left(\frac{t}{4\omega^{2}}\right)\right)$
will still be large thus making the use of radiative corrections to
$e^{+}e^{-}\rightarrow e^{+}e^{-}$ in the standard model valid, to
good approximation, for new physics $e^{+}e^{-}\rightarrow e^{+}e^{-}$
processes.

\section{Acknowledgments}

Thanks to P.Renton, P.J.Holt and O.Vives for helpful discussions. 

\appendix
\renewcommand{\theequation}{A.\arabic{equation}}

\setcounter{equation}{0}

\section*{Appendix A\label{sec:Appendix-A}}

\begin{figure}
\begin{center}\includegraphics[%
  width=0.27\paperwidth,
  height=0.18\paperwidth]{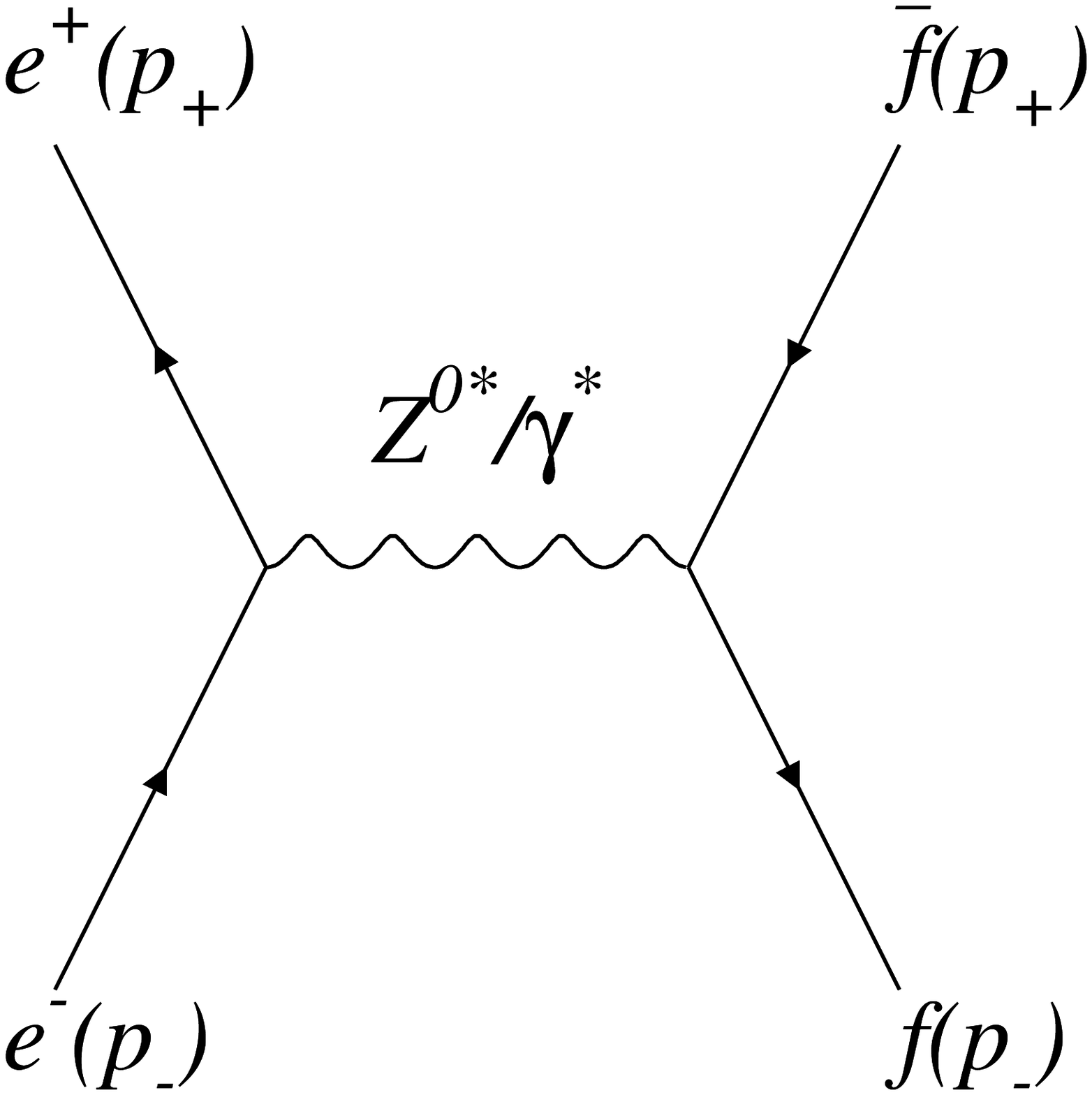}\includegraphics[%
  width=0.27\paperwidth,
  height=0.18\paperwidth]{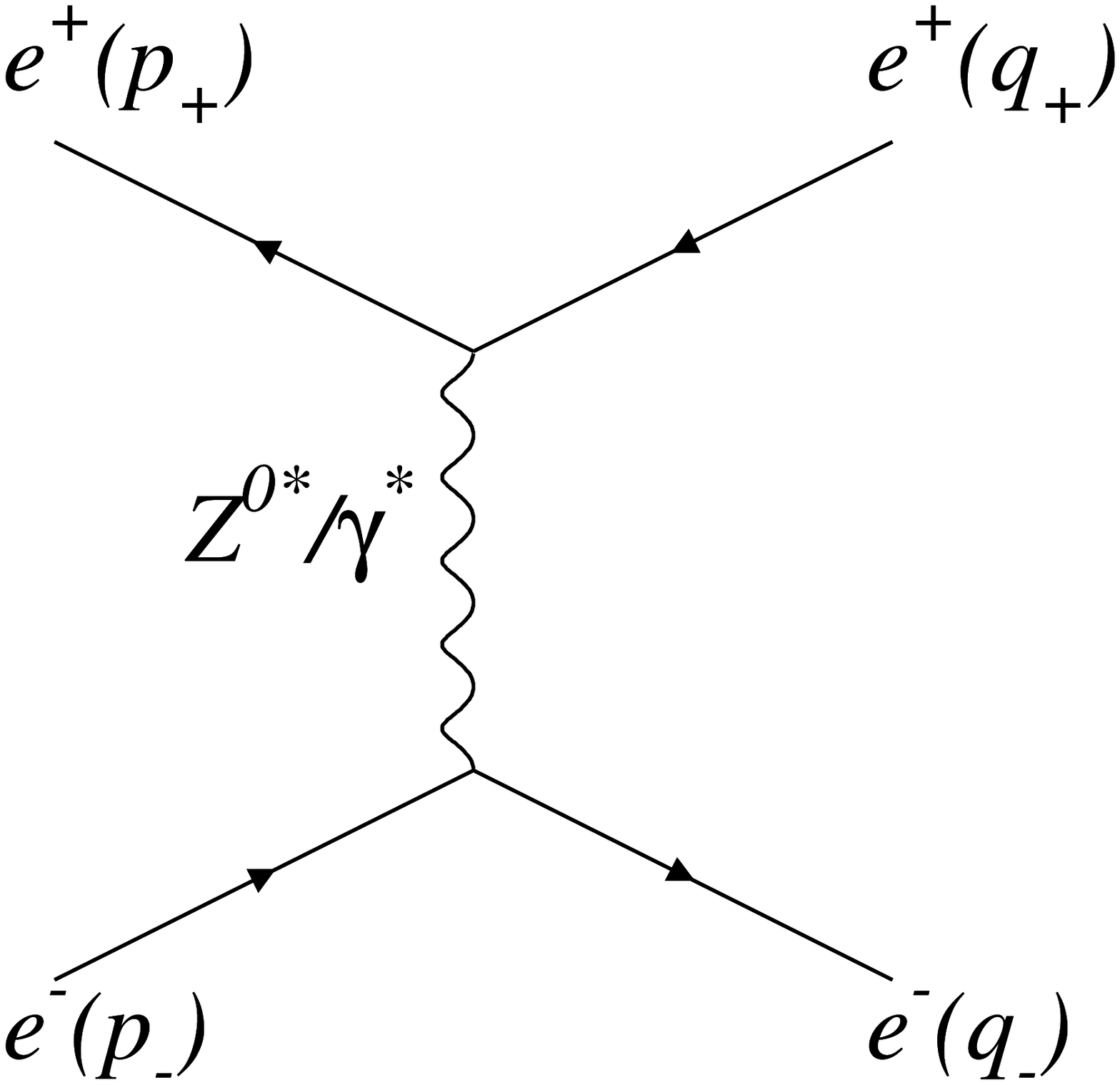}\end{center}

\caption{\label{cap:tree level difermion}The diagrams for tree level difermion
processes. From left to right these are $s$-channel and $t$-channel
processes respectively. Only electron-positron final states may result
from $t$-channel processes. }
\end{figure}

A difermion event is basically an event where an electron and positron
in the initial state exchange a $Z^{0*}$ or a photon $\gamma^{*}$
and leaves a fermion and an anti-fermion in the final state. This
can happen via $s$-channel $t$-channel exchange, these are depicted
in figure \ref{cap:tree level difermion} at tree level. $t$-channel
processes are only possible for a dielectron final state. %

Radiative corrections to these processes are significant in the experiment
and the definition of a difermion process is modified. More generally
we define a difermion event as an event where an electron and positron
emit photons in the initial state then undergo an interaction which
produces a fermion anti-fermion pair and photons, finally the fermion
and anti-fermion may emit additional bremsstrahlung photons. Generally
this can be reduced to three types of diagram, the tree level diagrams
with bremsstrahlung from the external legs and vertices replaced by
effective vertices (one-particle-irreducible diagrams with possible
photons emitted from them) and one with photons radiated off the external
legs and a single 4-point effective vertex, to lowest order this is
a simple \emph{box diagram}. In addition it is possible that photons
are radiated off the one-particle-irreducible diagrams representing
the vertices.

\begin{figure}
\begin{center}\includegraphics[%
  width=0.27\paperwidth,
  height=0.18\paperwidth]{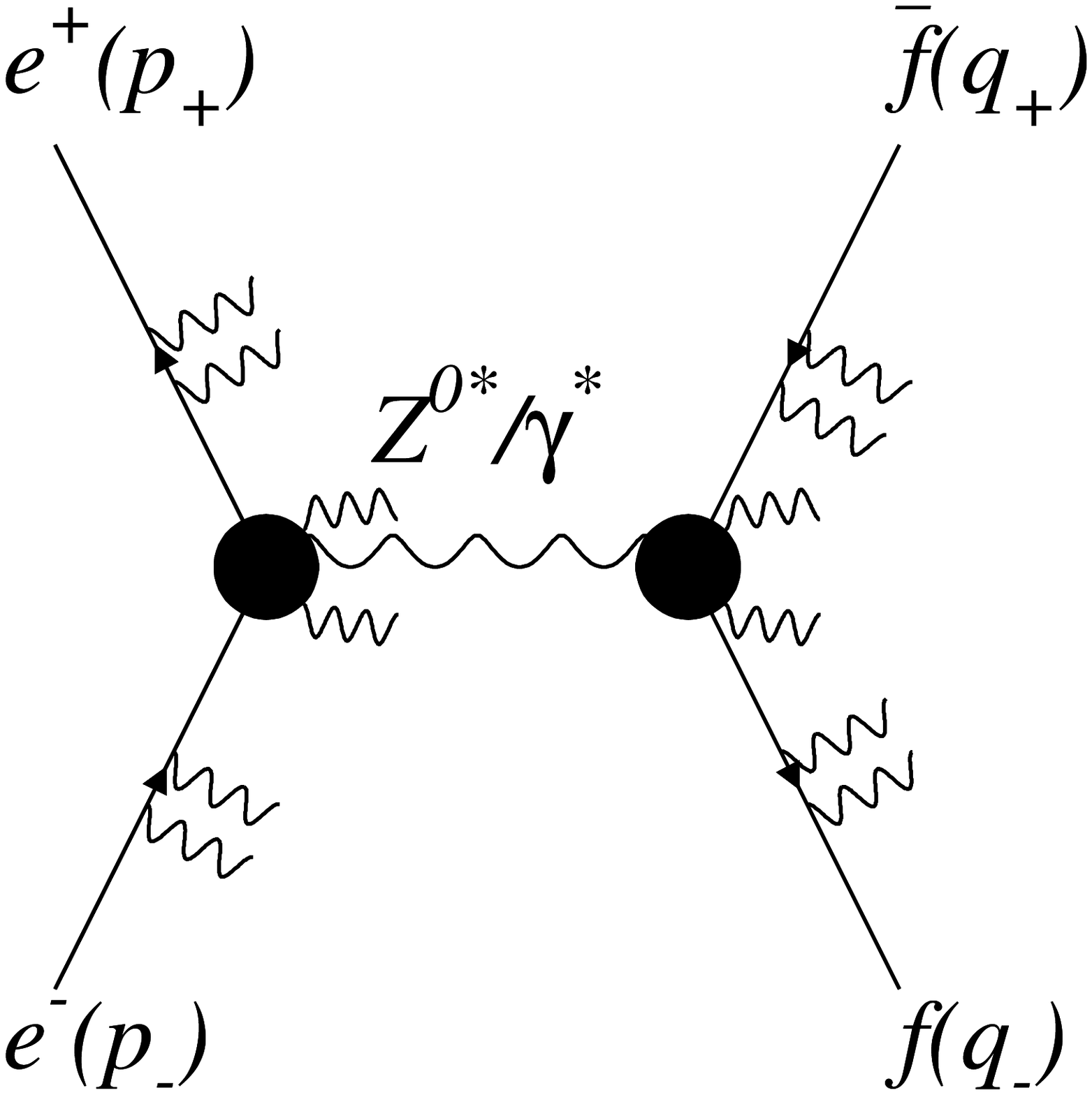}\includegraphics[%
  width=0.27\paperwidth,
  height=0.18\paperwidth]{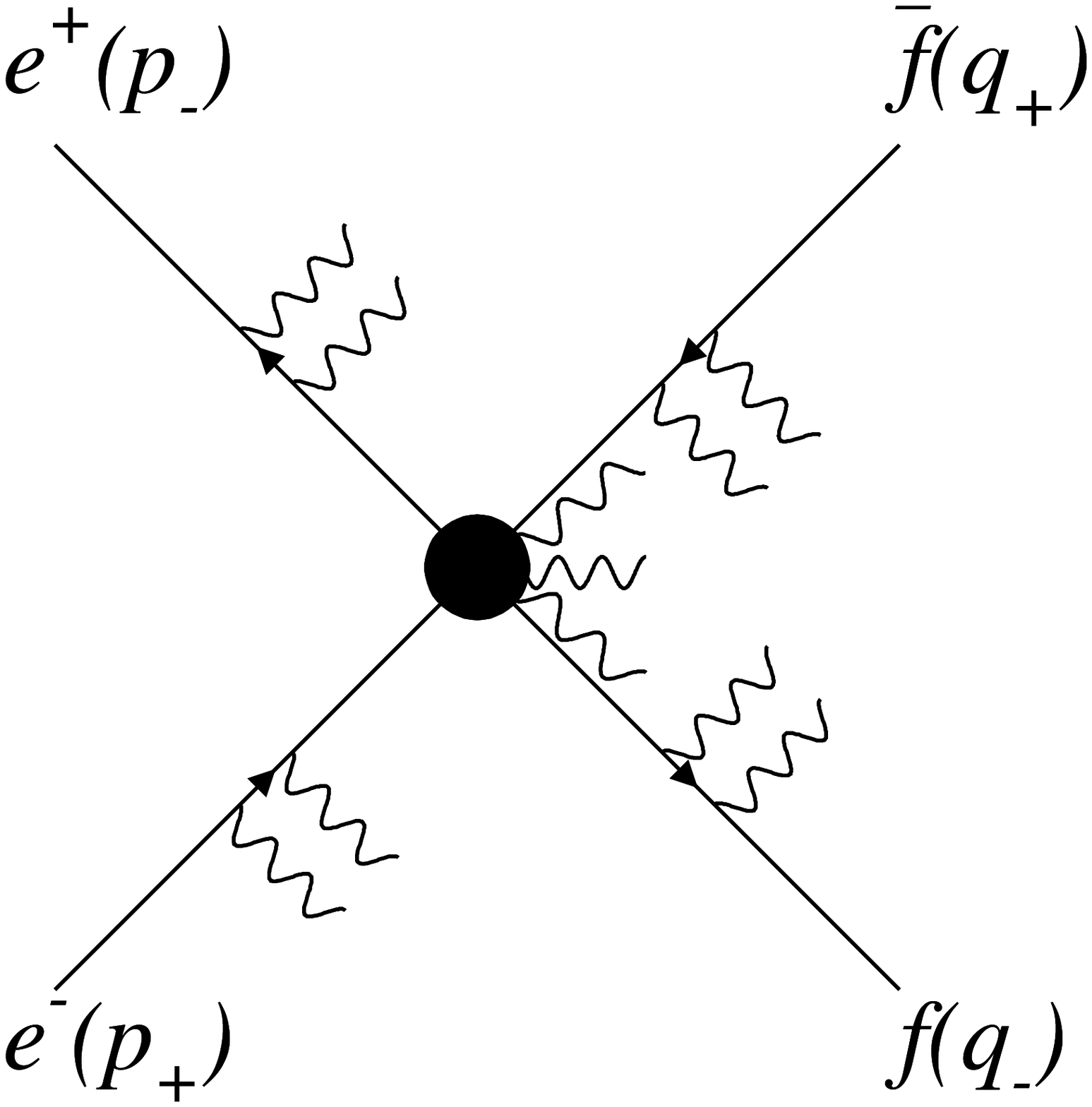}\end{center}

\caption{\label{cap:s-chan boxes}Diagrams corresponding to difermion events
involving bremsstrahlung. The black circles denote all possible one-particle-irreducible
diagrams. The arrows denote the flow of fermion number, momenta $p_{+}$,
$p_{-}$ are flowing inward, $q_{+}$,$q_{-}$ are flowing outward. }
\end{figure}

We define the Mandelstam variables $s$, $t$ , $u$ as%
\footnote{Momenta $p_{i}$ are flowing inward, $q_{i}$ flow outward.%
}\begin{equation}
\begin{array}{lclcl}
s & = & \left(p_{+}+p_{-}\right)^{2} & = & \left(q_{+}+q_{-}\right)^{2}\\
t & = & \left(p_{-}-q_{-}\right)^{2} & = & \left(p_{+}-q_{+}\right)^{2}\\
u & = & \left(p_{-}-q_{+}\right)^{2} & = & \left(p_{+}-q_{-}\right)^{2}\end{array}.\label{eq:A.1}\end{equation}
 Neglecting the mass of the electron it has four momentum $p_{-}=\left(E_{b},0,0,E_{b}\right)$
prior to any bremsstrahlung, this defines the $z$-axis for the experiment,
the positron has four momentum $p_{-}=\left(E_{b},0,0,-E_{b}\right)$,
consequently \begin{equation}
E_{b}=\frac{1}{2}\sqrt{s}.\label{eq:A.2}\end{equation}
 The \emph{initial state radiation} (\emph{ISR}) is defined as that
which is emitted from the external legs of the diagrams. We denote
the combined \emph{outgoing} momentum of the radiation emitted from
the external legs of momentum $p_{i}$ by $k_{j}$ $\left(i=+,-,\textrm{ }j=1,2\right)$
and from the external legs of momentum $q_{i}$ by $k_{j}$ $\left(i=+,-,\textrm{ }j=3,4\right)$.
Primed variables are defined to enable us to discuss the difermion
process in the presence of radiative corrections where the invariant
mass is reduced\begin{equation}
\begin{array}{rcl}
p_{-}^{\prime} & = & p_{-}-k_{1}\\
p_{+}^{\prime} & = & p_{+}-k_{2}\\
q_{-}^{\prime} & = & q_{-}+k_{3}\\
q_{+}^{\prime} & = & q_{+}+k_{4}\end{array}.\label{eq:A.3}\end{equation}
 In the presence of such radiative corrections there is some ambiguity
in the definition of $s$, $t$, $u$ as \begin{equation}
\left(p_{-}+p_{+}\right)^{2}\ne\left(q_{-}+q_{+}\right)^{2}\label{eq:A.4}\end{equation}
 \emph{etc}, hence for the case of radiative corrections we use \begin{equation}
\begin{array}{lclcl}
s & = & \left(p_{-}+p_{+}\right)^{2}\\
t & = & \left(p_{-}-q_{-}\right)^{2}\\
u & = & \left(p_{+}-q_{-}\right)^{2}\\
s^{\prime} & = & \left(p_{-}^{\prime}+p_{+}^{\prime}\right)^{2} & = & \left(q_{-}^{\prime}+q_{+}^{\prime}\right)^{2}\\
t^{\prime} & = & \left(p_{-}^{\prime}-q_{-}^{\prime}\right)^{2} & = & \left(p_{+}^{\prime}-q_{+}^{\prime}\right)^{2}\\
u^{\prime} & = & \left(p_{-}^{\prime}-q_{+}^{\prime}\right)^{2} & = & \left(p_{+}^{\prime}-q_{-}^{\prime}\right)^{2}\end{array}.\label{eq:A.5}\end{equation}

\appendix
\renewcommand{\theequation}{B.\arabic{equation}}\setcounter{equation}{0}

\section*{Appendix B\label{sec:Appendix-B}}

In this appendix we show briefly the integration of the soft photon
phase space factor associated with the soft bremsstrahlung discussed
in subsection \ref{sub:Oa3brem}. The integral to be performed is

\begin{equation}
\frac{1}{4\pi}\left(\frac{1}{\mu}\right)^{\epsilon^{\prime}}\int_{\left|\vec{k}\right|<\omega}\frac{\textrm{d}^{\epsilon^{\prime}+3}\vec{k}}{k_{0}}\frac{1}{\left(k.p_{i}\right)\left(k.p_{j}\right)}.\label{eq:5.5.15b}\end{equation}
 Following 't Hooft and Veltman \cite{'tHooft:1978xw} we introduce
a parameter $\kappa$ and define\begin{equation}
\begin{array}{rcl}
p & = & \kappa p_{i}\\
q & = & p_{j}\end{array},\label{eq:5.5.16}\end{equation}
 where $\kappa$ is defined to be the solution of $\left(p-q\right)^{2}=0$
which gives $q_{0}$ with the same sign as $p_{0}-q_{0}$. With these
definitions \ref{eq:5.5.15b} becomes,\begin{equation}
\frac{\kappa}{4\pi}\left(\frac{1}{\mu}\right)^{\epsilon^{\prime}}\int_{\left|\vec{k}\right|<\omega}\frac{\textrm{d}^{\epsilon^{\prime}+3}\vec{k}}{k_{0}}\frac{1}{\left(k.p\right)\left(k.q\right)}.\label{eq:5.5.17}\end{equation}
 now combine the denominators with a Feynman parameter $\left(ab\right)^{-1}=\int_{0}^{1}\textrm{d}x\textrm{ }\left[ax+b\left(1-x\right)\right]^{-2}$.
This easily gives,\begin{equation}
\begin{array}{rl}
 & \frac{\kappa}{4\pi}\left(\frac{1}{\mu}\right)^{\epsilon^{\prime}}\int_{\left|\vec{k}\right|<\omega}\frac{{\rm {d}}^{\epsilon^{\prime}+3}\vec{k}}{k_{0}}\int_{0}^{1}\textrm{d}x\textrm{ }\frac{1}{\left(k.P\right)^{2}}\\
= & \frac{\kappa}{4\pi}\left(\frac{1}{\mu}\right)^{\epsilon^{\prime}}\int_{0}^{1}\textrm{d}x\int_{\left|\vec{k}\right|<\omega}\frac{{\rm {d}}^{\epsilon^{\prime}+3}\vec{k}}{k_{0}}\textrm{ }\frac{1}{k_{0}^{2}\left(P_{0}-\left|\vec{P}\right|\cos\tilde{\theta}\right)^{2}}\end{array}\label{eq:5.5.18}\end{equation}
 where we have defined $P=q+x\left(p-q\right)$.

The integration measure can be written explicitly in terms of its
angular components and momentum component in the usual way. The angular
integrations are simplified by essentially redefining the photon momentum
space axes such that $\tilde{\theta}$ is the polar angle, in this
case all but the polar angle integration is trivial. This is analogous
to how integration over the azimuthal angle in three dimensional problems
gives a factor $2\pi$ when the integrand has rotational symmetry
about one axis, the $d$-dimensional analogy is a well known result
see \emph{e.g.} \cite{Bardin:ak}\begin{equation}
\int\textrm{d}\Omega_{d}=\frac{2\pi^{\frac{d}{2}}}{\Gamma\left(\frac{d}{2}\right)}\int_{0}^{\pi}\textrm{d}\tilde{\theta}\textrm{ }\sin^{d-1}\tilde{\theta}.\label{eq:5.5.19}\end{equation}
 Equation \ref{eq:5.5.16} gives the integration measure $\textrm{d}\Omega_{d}$
on a $d$-sphere, in our case the phase space is a $d=n-2$ dimensional
sphere of radius $\left|\vec{k}\right|$ embedded in $n-1$ dimensions.
Hence we decompose $\textrm{d}^{\epsilon^{\prime}+3}\vec{k}=\left|\vec{k}\right|^{\epsilon^{\prime}+2}\textrm{d}\left|\vec{k}\right|\textrm{d}\Omega_{\epsilon^{\prime}+2}$
giving \ref{eq:5.5.15b} as\begin{equation}
\frac{\kappa}{4\pi}\left(\frac{1}{\mu}\right)^{\epsilon^{\prime}}\frac{2\pi^{1+\frac{\epsilon^{\prime}}{2}}}{\Gamma\left(1+\frac{\epsilon^{\prime}}{2}\right)}\int_{0}^{1}\textrm{d}x\int_{0}^{\pi}\textrm{d}\tilde{\theta}\textrm{ }\frac{\sin^{\epsilon^{\prime}+1}\tilde{\theta}}{\left(P_{0}-\left|\vec{P}\right|\cos\tilde{\theta}\right)^{2}}\int_{\left|\vec{k}\right|<\omega}\textrm{d}\left|\vec{k}\right|k_{0}^{\epsilon^{\prime}-1},\label{eq:5.5.20}\end{equation}
 making the replacement $\sin\tilde{\theta}\textrm{d}\tilde{\theta}\rightarrow-\textrm{d}\cos\tilde{\theta}$
this becomes,\begin{equation}
\frac{1}{4\pi}\left(\frac{1}{\mu}\right)^{\epsilon^{\prime}}\int_{\left|\vec{k}\right|<\omega}\frac{\textrm{d}^{\epsilon^{\prime}+3}\vec{k}}{k_{0}}\frac{1}{\left(k.p_{i}\right)\left(k.p_{j}\right)}=\frac{\kappa}{2}\frac{\pi^{\frac{\epsilon^{\prime}}{2}}}{\Gamma\left(1+\frac{\epsilon^{\prime}}{2}\right)}\frac{1}{\epsilon^{\prime}}\left(\frac{\omega}{\mu}\right)^{\epsilon^{\prime}}\int_{0}^{1}\textrm{d}x\int_{-1}^{+1}\textrm{d}\cos\tilde{\theta}\textrm{ }\frac{\left(1-\cos^{2}\tilde{\theta}\right)^{\frac{\epsilon^{\prime}}{2}}}{\left(P_{0}-\left|\vec{P}\right|\cos\tilde{\theta}\right)^{2}}.\label{eq:5.5.21}\end{equation}
 Expanding in $\epsilon^{\prime}$, dropping terms ${\mathcal{{O}}}\left(\epsilon^{\prime}\right)$
and above this becomes,\begin{equation}
\frac{\kappa}{2}\int_{0}^{1}\textrm{d}x\int_{-1}^{+1}\textrm{d}C\textrm{ }\frac{1}{\left(P_{0}-\left|\vec{P}\right|C\right)^{2}}\left(\frac{1}{\epsilon^{\prime}}+\log\frac{\omega}{\mu}+\frac{\gamma}{2}+\frac{1}{2}\log\pi+\frac{1}{2}\log\left(1-C^{2}\right)\right)\label{eq:5.5.22}\end{equation}
 with $\cos\theta$ relabeled as $C$. The angular integrations maybe
safely carried out using a symbolic computer algebra package giving
for integral \ref{eq:5.5.22}\begin{equation}
\frac{\kappa}{2}\int_{0}^{1}\textrm{d}x\textrm{ }\frac{1}{P^{2}}\left(\frac{1}{\hat{\epsilon}}+\log\left(\frac{\omega}{\mu}\right)^{2}+2\log2+\frac{P_{0}}{\left|\vec{P}\right|}\log\frac{P_{0}-\left|\vec{P}\right|}{P_{0}+\left|\vec{P}\right|}\right),\label{eq:5.5.23}\end{equation}
 (recall $P=q+x\left(p-q\right)$) where we have introduced the infrared
regulator\begin{equation}
\frac{1}{\hat{\epsilon}}=\frac{2}{\epsilon^{\prime}}+\gamma+\log\pi.\label{eq:5.5.24}\end{equation}
 Again by analogy to \cite{'tHooft:1978xw}, the integration is transformed
such that it is over $m=P_{0}-\left|\vec{P}\right|$ using\begin{equation}
\begin{array}{lcl}
v & = & \frac{1}{2}\frac{p^{2}-q^{2}}{p_{0}-q_{0}}\\
P^{2} & = & q^{2}+\frac{P_{0}-q_{0}}{p_{0}-q_{0}}\left(p^{2}-q^{2}\right)\\
 & = & q^{2}+2vP_{0}-2vq_{0}\\
\left|\vec{P}\right|^{2} & = & P_{0}^{2}-P^{2}\\
 & = & P_{0}^{2}-q^{2}-2vP_{0}+2vq_{0}\end{array},\label{eq:5.5.25}\end{equation}
 where we have used the fact that $\kappa$ was defined so that $\left(p-q\right)^{2}=0$
in determining $P^{2}$ and $\left|\vec{P}\right|^{2}$ above. In
terms of the new variables we have\begin{equation}
\log\left(\frac{P_{0}-\left|\vec{P}\right|}{P_{0}+\left|\vec{P}\right|}\right)=\log\left(\frac{m-v}{v}.\frac{m}{m-\frac{1}{v}\left(2vq_{0}-q^{2}\right)}\right).\label{eq:5.5.26}\end{equation}
 The rest of the integrand and the integration measure transforms
as\begin{equation}
\textrm{d}x\textrm{ }\frac{1}{P^{2}}\frac{P_{0}}{\left|\vec{P}\right|}=\frac{\textrm{d}m}{2vl}\textrm{ }\left(\frac{2}{P^{2}}.\frac{vP_{0}}{\left|\vec{P}\right|}\frac{\textrm{d}P_{0}}{\textrm{d}m}\textrm{ }\right).\label{eq:5.5.27}\end{equation}
 Differentiating \ref{eq:5.5.26} with respect to $m$ we find\begin{equation}
\frac{\textrm{d}}{\textrm{d}m}\log\left(\frac{P_{0}-\left|\vec{P}\right|}{P_{0}+\left|\vec{P}\right|}\right)=\frac{2}{P^{2}}\left(\frac{\left|\vec{P}\right|^{2}-P_{0}^{2}}{\left|\vec{P}\right|}\right)\frac{\textrm{d}P_{0}}{\textrm{d}m}+\frac{2}{P^{2}}\frac{vP_{0}}{\left|\vec{P}\right|}\frac{\textrm{d}P_{0}}{\textrm{d}m},\label{eq:5.5.28}\end{equation}
 this enables us to rewrite \ref{eq:5.5.27} as\begin{equation}
\textrm{d}x\textrm{ }\frac{1}{P^{2}}\frac{P_{0}}{\left|\vec{P}\right|}=\frac{\textrm{d}m}{2vl}\textrm{ }\left(\frac{\textrm{d}}{\textrm{d}m}\log\left(\frac{P_{0}-\left|\vec{P}\right|}{P_{0}+\left|\vec{P}\right|}\right)-\frac{2}{m-v}\right).\label{eq:5.5.29}\end{equation}
 Substituting these transformed quantities \ref{eq:5.5.27} and \ref{eq:5.5.29}
into the phase space integral \ref{eq:5.5.21} gives\begin{equation}
\begin{array}{rl}
 & \frac{1}{4\pi}\left(\frac{1}{\mu}\right)^{\epsilon^{\prime}}\int_{\left|\vec{k}\right|<\omega}\frac{{\rm {d}}^{\epsilon^{\prime}+3}\vec{k}}{k_{0}}\frac{1}{\left(k.p_{i}\right)\left(k.p_{j}\right)}\\
= & \frac{\kappa}{2}\left(\frac{1}{\hat{\epsilon}}+\log\left(\frac{\omega}{\mu}\right)^{2}+2\log2\right)\int_{0}^{1}\textrm{d}x\textrm{ }\frac{1}{P^{2}}\\
+ & \frac{\kappa}{2}\int_{q_{0}-\left|\vec{q}\right|}^{p_{0}-\left|\vec{p}\right|}\frac{{\rm {d}}m}{2vl}\textrm{ }\left(\frac{{\rm {d}}}{{\rm {d}}m}\log\left(\frac{m-v}{v}.\frac{m}{m-\frac{1}{v}\left(2vq_{0}-q^{2}\right)}\right)-\frac{2}{m-v}\right)\log\left(\frac{m-v}{v}.\frac{m}{m-\frac{1}{v}\left(2vq_{0}-q^{2}\right)}\right)\end{array}.\label{eq:5.5.30}\end{equation}
 The term $\log\left(...\right)\frac{\textrm{d}}{\textrm{d}m}\log\left(...\right)$
can be rewritten as $\frac{1}{2}\frac{\textrm{d}}{\textrm{d}m}\log^{2}\left(...\right)$
in which case the integration is trivial. The first integral is also
easy, \begin{equation}
\int_{0}^{1}\textrm{d}x\textrm{ }\frac{1}{P^{2}}=\frac{1}{2vl}\log\left(1+\frac{2vl}{q^{2}}\right).\label{eq:5.5.31}\end{equation}
 This leaves one integral\begin{equation}
\begin{array}{rl}
 & \frac{1}{4\pi}\left(\frac{1}{\mu}\right)^{\epsilon^{\prime}}\int_{\left|\vec{k}\right|<\omega}\frac{{\rm {d}}^{\epsilon^{\prime}+3}\vec{k}}{k_{0}}\frac{1}{\left(k.p_{i}\right)\left(k.p_{j}\right)}\\
= & \frac{\kappa}{4vl}\left(\frac{1}{\hat{\epsilon}}+\log\left(\frac{\omega}{\mu}\right)^{2}+2\log2\right)\log\left(1+\frac{2vl}{q^{2}}\right)\\
+ & \frac{\kappa}{2vl}\textrm{ }\left[\frac{1}{4}\log^{2}\left(\frac{m\left(m-v\right)}{q^{2}+mv-2vq_{0}}\right)\right]_{q_{0}-\left|\vec{q}\right|}^{p_{0}-\left|\vec{p}\right|}-\frac{\kappa}{2vl}\int_{q_{0}-\left|\vec{q}\right|}^{p_{0}-\left|\vec{p}\right|}\textrm{d}m\textrm{ }\frac{1}{m-v}\log\left(\frac{m-v}{v}.\frac{m}{m-\frac{1}{v}\left(2vq_{0}-q^{2}\right)}\right)\end{array}.\label{eq:5.5.30}\end{equation}
 The final integral is awkward and is done with the help of the following
two identities\begin{equation}
\begin{array}{lcl}
\log\left(m-a\right) & = & \log\left(1-\frac{m-v}{a-v}\right)+\log\left(v-a\right)\\
\frac{\textrm{d}}{\textrm{d}m}\textrm{Li}_{2}\left(A\left(m-v\right)\right) & = & -\frac{1}{m-v}\log\left(1-A\left(m-v\right)\right)\end{array},\label{eq:5.5.31}\end{equation}
 {\small \begin{equation}
\begin{array}{rl}
 & \int_{q_{0}-\left|\vec{q}\right|}^{p_{0}-\left|\vec{p}\right|}{\rm {d}}m\textrm{ }\frac{1}{m-v}\log\left(\frac{m-v}{v}.\frac{m}{m-\frac{1}{v}\left(2vq_{0}-q^{2}\right)}\right)\\
= & \left[-\textrm{Li}_{2}\left(1-\frac{m}{v}\right)+\textrm{Li}_{2}\left(\frac{v\left(m-v\right)}{\left(2vq_{0}-q^{2}\right)-v^{2}}\right)\right]_{q_{0}-\left|\vec{q}\right|}^{p_{0}-\left|\vec{p}\right|}+\int_{q_{0}-\left|\vec{q}\right|}^{p_{0}-\left|\vec{p}\right|}{\rm {d}}m\textrm{ }\frac{1}{m-v}\log\left(\frac{\left(m-v\right)v}{v^{2}-2vq_{0}+q^{2}}\right)\end{array}.\label{eq:5.5.32}\end{equation}
} Finally we abbreviate $v\left(v^{2}-2vq_{0}+q^{2}\right)^{-1}=c$
and perform a transformation of variables $x=\left(m-v\right)c$,
$\textrm{d}m=\frac{1}{c}\textrm{d}x$ hence\begin{equation}
\begin{array}{rcl}
\int_{q_{0}-\left|\vec{q}\right|}^{p_{0}-\left|\vec{p}\right|}\textrm{d}m\textrm{ }\frac{c}{\left(m-v\right)c}\log\left(\left(m-v\right)c\right) & = & \int_{x_{1}}^{x_{2}}\textrm{d}x\textrm{ }\frac{1}{x}\log\left(x\right)\\
 & = & \left[\frac{1}{2}\log^{2}\left(\frac{v\left(m-v\right)}{v^{2}-2vq_{0}+q^{2}}\right)\right]_{q_{0}-\left|\vec{q}\right|}^{p_{0}-\left|\vec{p}\right|}\end{array}.\label{eq:5.5.33}\end{equation}
 For the phase space integral \ref{eq:5.5.14} we now have\begin{equation}
\begin{array}{rl}
\Rightarrow & \frac{1}{4\pi}\left(\frac{1}{\mu}\right)^{\epsilon^{\prime}}\int_{\left|\vec{k}\right|<\omega}\frac{{\rm {d}}^{\epsilon^{\prime}+3}\vec{k}}{k_{0}}\frac{1}{\left(k.p_{i}\right)\left(k.p_{j}\right)}\\
= & \frac{\kappa}{4vl}\left(\frac{1}{\hat{\epsilon}}+\log\left(\frac{4\omega^{2}}{\mu^{2}}\right)\right)\log\left(1+\frac{2vl}{q^{2}}\right)\\
+ & \frac{\kappa}{2vl}\textrm{ }\left[\frac{1}{4}\log^{2}\left(\frac{m\left(m-v\right)}{q^{2}+mv-2vq_{0}}\right)-\frac{1}{2}\log^{2}\left(\frac{v\left(m-v\right)}{q^{2}+v^{2}-2vq_{0}}\right)+\textrm{Li}_{2}\left(1-\frac{m}{v}\right)-\textrm{Li}_{2}\left(\frac{v\left(m-v\right)}{2vq_{0}-q^{2}-v^{2}}\right)\right]_{q_{0}-\left|\vec{q}\right|}^{p_{0}-\left|\vec{p}\right|}\end{array}.\label{eq:5.5.34}\end{equation}
 Using the identity for the dilogarithm identity $\textrm{Li}_{2}\left(\frac{1}{z}\right)=-\textrm{Li}_{2}\left(z\right)-\frac{1}{2}\log^{2}\left(-z\right)-\zeta\left(2\right)$
this becomes\begin{equation}
\begin{array}{rl}
\Rightarrow & \frac{1}{4\pi}\left(\frac{1}{\mu}\right)^{\epsilon^{\prime}}\int_{\left|\vec{k}\right|<\omega}\frac{{\rm {d}}^{\epsilon^{\prime}+3}\vec{k}}{k_{0}}\frac{1}{\left(k.p_{i}\right)\left(k.p_{j}\right)}\\
= & \frac{\kappa}{4vl}\left(\frac{1}{\hat{\epsilon}}+\log\left(\frac{4\omega^{2}}{\mu^{2}}\right)\right)\log\left(1+\frac{2vl}{q^{2}}\right)\\
+ & \frac{\kappa}{2vl}\textrm{ }\left[\frac{1}{4}\log^{2}\left(\frac{m\left(m-v\right)}{q^{2}+mv-2vq_{0}}\right)+\textrm{Li}_{2}\left(1-\frac{m}{v}\right)+\textrm{Li}_{2}\left(\frac{2vq_{0}-q^{2}-v^{2}}{v\left(m-v\right)}\right)\right]_{q_{0}-\left|\vec{q}\right|}^{p_{0}-\left|\vec{p}\right|}\end{array}.\label{eq:5.5.34}\end{equation}
 Retracing the algebra of \ref{eq:5.5.26} backward and using the
relation $p^{2}-2vp_{0}=q^{2}-2vq_{0}$ we can write\begin{equation}
\begin{array}{rcl}
\frac{\left(q_{0}-\left|\vec{q}\right|\right)\left(q_{0}-\left|\vec{q}\right|-v\right)}{q^{2}+\left(q_{0}-\left|\vec{q}\right|\right)v-2vq_{0}} & = & \frac{q_{0}-\left|\vec{q}\right|}{q_{0}+\left|\vec{q}\right|}\\
\frac{\left(p_{0}-\left|\vec{p}\right|\right)\left(p_{0}-\left|\vec{p}\right|-v\right)}{p^{2}+\left(p_{0}-\left|\vec{p}\right|\right)v-2vp_{0}} & = & \frac{p_{0}-\left|\vec{p}\right|}{p_{0}+\left|\vec{p}\right|}\\
\frac{2vq_{0}-q^{2}-v^{2}}{v\left(q_{0}-\left|\vec{q}\right|-v\right)} & = & \frac{v-q_{0}-\left|\vec{q}\right|}{v}\\
\frac{2vp_{0}-p^{2}-v^{2}}{v\left(p_{0}-\left|\vec{p}\right|-v\right)} & = & \frac{v-p_{0}-\left|\vec{p}\right|}{v}\end{array}.\label{eq:5.5.35}\end{equation}
 Also for any momentum four vector $V$, in the limit of small masses,
$V^{2}\ll V_{0}^{2}$ and we can approximate, \begin{equation}
\begin{array}{lclcl}
V_{0}-\left|\vec{V}\right| & = & V_{0}-V_{0}\left(1-\frac{V^{2}}{V_{0}^{2}}\right)^{\frac{1}{2}} & = & \frac{V^{2}}{2V_{0}}\\
V_{0}+\left|\vec{V}\right| & = & V_{0}+V_{0}\left(1-\frac{V^{2}}{V_{0}^{2}}\right)^{\frac{1}{2}} & = & 2V_{0}\end{array}.\label{eq:5.5.36}\end{equation}
 Specifically in the case of initial state-final state interference
contributions to $\textrm{d}\sigma_{Soft}$ we require four phase
space integrals corresponding to all possible permutations of a bremsstrahlung
photon from an initial state leg and a bremsstrahlung photon from
a final state leg \emph{i.e.} all combinations of $p_{i}$ and $p_{j}$
where $p_{i}=p_{-},\textrm{ }p_{+}$ and $p_{j}=q_{-},\textrm{ }q_{+}$.
Working also in the limit that the photon carries off no energy $p_{i,0}=p_{j,0}$,
the phase space integral can now be written\begin{equation}
\begin{array}{rl}
\Rightarrow & \frac{1}{4\pi}\left(\frac{1}{\mu}\right)^{\epsilon^{\prime}}\int_{\left|\vec{k}\right|<\omega}\frac{{\rm {d}}^{\epsilon^{\prime}+3}\vec{k}}{k_{0}}\frac{1}{\left(k.p_{i}\right)\left(k.p_{j}\right)}\\
= & \frac{\kappa}{4vl}\left(\frac{1}{\hat{\epsilon}}+\log\left(\frac{4\omega^{2}}{\mu^{2}}\right)\right)\log\left(1+\frac{2vl}{q^{2}}\right)+\frac{\kappa}{8vl}\log^{2}\left(\frac{m_{e}^{2}}{s}\right)-\frac{\kappa}{8vl}\log^{2}\left(\frac{m_{f}^{2}}{s}\right)\\
+ & \frac{\kappa}{2vl}\textrm{ }\left[\textrm{Li}_{2}\left(\frac{v-m_{0}+\left|\vec{m}\right|}{v}\right)+\textrm{Li}_{2}\left(\frac{v-m_{0}-\left|\vec{m}\right|}{v}\right)\right]_{m=p_{j}}^{m=\kappa p_{i}}\end{array}.\label{eq:5.5.37}\end{equation}
 \textcolor{black}{Returning to equation \ref{eq:5.5.12} we see
that the soft photon contribution to ISR{*}FSR bremsstrahlung is{\small \begin{equation}
{\textrm{{d}}}\sigma_{Soft,Int}=-\frac{\alpha}{4\pi^{2}}\int_{\left|\vec{k}\right|<\omega}\frac{{\rm {d}}^{3}\vec{k}}{k_{0}}\left(\frac{t}{\left(k.p_{+}\right)\left(k.q_{+}\right)}+\frac{t}{\left(k.p_{-}\right)\left(k.q_{-}\right)}-\frac{u}{\left(k.p_{+}\right)\left(k.q_{-}\right)}-\frac{u}{\left(k.p_{-}\right)\left(k.q_{+}\right)}\right){\textrm{{d}}}\sigma_{Born}.\label{eq:5.5.38}\end{equation}
}  }Finally we must determine $\kappa$ for each phase space integral
where $\kappa$ was defined earlier as the solution to \begin{equation}
\left(\kappa p_{i}-p_{j}\right)^{2}=0,\label{eq:5.5.39}\end{equation}
 for which $\kappa p_{i,0}-p_{j,0}$ and $p_{j,0}$ have the same
sign. We shall demonstrate how to obtain $\kappa$ for the case $p_{i}=p_{+}$
$p_{j}=q_{+}$, the results generalize easily to the other combinations
of momenta. In this case expanding \ref{eq:5.5.39} gives a simple
quadratic equation for $\kappa$. Working in the limit $t\gg m_{f}^{2}$
gives the two solutions $\left(\kappa_{+},\kappa_{-}\right)$ for
$\kappa$ to first order as\begin{equation}
\begin{array}{lcl}
\kappa_{+} & = & -\frac{t}{m_{e}^{2}}\\
\kappa_{-} & = & -\frac{m_{f}^{2}}{t}\end{array}.\label{eq:5.5.40}\end{equation}
 In the massless limit the Mandelstam variable $t$ is\begin{equation}
t=-\frac{s}{2}\left(1-\cos\theta_{++}\right),\label{eq:5.5.41}\end{equation}
 where $\theta_{++}$ is the angle between $\left|\vec{p_{+}}\right|$
and $\left|\vec{q_{+}}\right|$, therefore $t$ is a negative quantity.
We require the solution for which $\textrm{sign}\left(\kappa p_{+,0}-q_{+,0}\right)=\textrm{sign}\left(q_{+,0}\right)$
\emph{i.e.}\begin{equation}
\textrm{sign}\left(\frac{1}{2}\sqrt{s}\left(\kappa-1\right)\right)=\textrm{sign}\left(\frac{1}{2}\sqrt{s}\right)\label{eq:5.5.42}\end{equation}
 so we require $\kappa>1$. Again working in the limit $t\gg m_{f}^{2}$
this means taking solution $\kappa_{+}=-\frac{t}{m_{e}^{2}}$. Exactly
the same mathematics and the same $\kappa$ are obtained using $p_{i}=p_{-}$,
$p_{j}=q_{-}$ and only marginal differences in working give $\kappa_{+}=-\frac{u}{m_{e}^{2}}$
for the other two combinations of momenta. Noting that $\kappa$ is
large we have that to lowest order in small things \begin{equation}
\begin{array}{rcl}
v & = & \frac{1}{2\kappa p_{0}}\kappa^{2}p^{2}\left(1-\frac{q^{2}}{\kappa^{2}p^{2}}\right)\left(1-\frac{q_{0}}{\kappa p_{0}}\right)^{-1}\\
 & \approx & \frac{\kappa m_{e}^{2}}{\sqrt{s}}\end{array}.\label{eq:5.5.43}\end{equation}
 Now considering solely the dilogarithm terms in the small mass approximation
\ref{eq:5.5.36}, we have for $\kappa=-\frac{t}{m_{e}^{2}}$\begin{equation}
\left[\textrm{Li}_{2}\left(1-\frac{m_{0}-\left|\vec{m}\right|}{v}\right)+\textrm{Li}_{2}\left(1-\frac{m_{0}+\left|\vec{m}\right|}{v}\right)\right]_{m=p_{j}}^{m=\kappa p_{i}}=-\frac{1}{2}\log\left(\frac{m_{e}^{2}}{s}\right)-\textrm{Li}_{2}\left(1+\frac{s}{t}\right)-\frac{\pi^{2}}{3},\label{eq:5.5.44}\end{equation}
 where we have used $\textrm{Li}_{2}\left(-\frac{s}{m_{e}^{2}}\right)=-\textrm{Li}_{2}\left(-\frac{m_{e}^{2}}{s}\right)-\frac{1}{2}\log\left(\frac{m_{e}^{2}}{s}\right)-\zeta\left(2\right)$
and $\textrm{Li}_{2}\left(1\right)=\zeta\left(2\right)$. The same
term for the second solution $\kappa=-\frac{u}{m_{e}^{2}}$ takes
the same form as above but with the replacement $t\rightarrow u$.
Substituting all of this into \ref{eq:5.5.38} gives finally\begin{equation}
\textrm{d}\sigma_{Soft,Int}=\frac{2\alpha}{\pi}\left[\left(\frac{1}{\hat{\epsilon}}+\log\left(\frac{4\omega^{2}}{\mu^{2}}\right)\right)\log\left(\frac{t}{u}\right)-\textrm{Li}_{2}\left(1+\frac{s}{t}\right)+\textrm{Li}_{2}\left(1+\frac{s}{u}\right)\right]\textrm{d}\sigma_{Born}.\label{eq:5.5.45}\end{equation}

\end{document}